\documentclass[a4paper,12pt]{article}
\usepackage{epsfig}
\usepackage{amssymb}
\usepackage{graphicx}
\usepackage{color}
\definecolor{rosso}{cmyk}{0,1,1,0.4}
\definecolor{rossos}{cmyk}{0,1,1,0.55}
\definecolor{rossoc}{cmyk}{0,1,1,0.2}
\definecolor{blu}{cmyk}{1,1,0,0.3}
\definecolor{blus}{cmyk}{1,1,0,0.6}
\definecolor{bluc}{cmyk}{1,1,0,0.1}
\definecolor{verde}{cmyk}{0.92,0,0.59,0.25}
\definecolor{verdec}{cmyk}{0.92,0,0.59,0.15}
\definecolor{verdes}{cmyk}{0.92,0,0.59,0.4}

\def\Jo#1#2#3#4{{\it #1} {\bf #2}, #3 (#4)}

\def\circa#1{\,\raise.3ex\hbox{$#1$\kern-.75em\lower1ex\hbox{$\sim$}}\,}
\def\NPB{{\rm Nucl. Phys. B}}

\def\PLB{{\rm Phys. Lett. B}}
\def\PRL{\rm Phys. Rev. Lett.}
\def\PRD{{\rm Phys. Rev.  D}}

\def\JHEP{JHEP}

\newcommand{\eV}{\,{\rm eV}}

\makeatletter
\newcommand{\beq}{\begin{equation}}
\newcommand{\eeq}{\end{equation}}
\newcommand{\bea}{\begin{eqnarray}}
\newcommand{\eea}{\end{eqnarray}}
\newcommand{\ba}{\begin{array}}
\newcommand{\ea}{\end{array}}
\newcommand{\bn}{\begin{enumerate}}
\newcommand{\en}{\end{enumerate}}
\newcommand{\bc}{\begin{center}}
\newcommand{\ec}{\end{center}}

\newcommand{\ds}{\displaystyle}

\newcommand{\gsim}{\lower.7ex\hbox{$\;\stackrel{\textstyle>}{\sim}\;$}}
\newcommand{\lsim}{\lower.7ex\hbox{$\;\stackrel{\textstyle<}{\sim}\;$}}
\newcommand{\baz}{\begin{array}{cc}}
\newcommand{\bad}{\begin{array}{ccc}}
\def\gtap{\mathrel{ \rlap{\raise 0.511ex \hbox{$>$}}{\lower 0.511ex
   \hbox{$\sim$}}}} 
\def\ltap{\mathrel{ \rlap{\raise 0.511ex
   \hbox{$<$}}{\lower 0.511ex \hbox{$\sim$}}}}
   \newcommand{\deltaatm}{\mbox{$\Delta m^2_{\mathrm{A}}$}}
   \newcommand{\deltasol}{\mbox{$ \Delta m^2_{\odot}$}}
\newcommand{\dmsol}{\mbox{$\Delta m^2_{\odot}$}}
\newcommand{\dma}{\mbox{$\Delta m^2_{\rm A}$}}
   
   \newcommand{\betabeta}{\mbox{$(\beta \beta)_{0 \nu}$}}

  \newcommand{\meff}{\mbox{$ \left| \langle   m  \rangle \right| $}}

   \newcommand{\hbeta}{$\mbox{}^3 {\rm H}$ $\beta$-decay }

\newcommand{\pmns}{\mbox{$ U$}}

\begin{document}

\tolerance=100000
\thispagestyle{empty}
\setcounter{page}{0}
\begin{flushright}
\texttt{DCPT/06/166}\\
\texttt{CERN-PH-TH/2006-213}\\
\texttt{IPPP/06/83}\\
\texttt{SISSA 71/2006/EP}
\end{flushright}
\begin{center}
{\large \bf Leptogenesis and 
Low Energy CP Violation in Neutrino Physics \\[0.3cm]
}
\vskip 15pt
{\bf S. Pascoli $^{a)}$, S.T. Petcov$^{b)}$
\footnote{Also at: Institute
of Nuclear Research and Nuclear Energy, Bulgarian Academy 
of Sciences, 1784 Sofia, Bulgaria.} 
 and A. Riotto$^{c,d)}$}
 
\vskip 5pt
$^{a)}${\it IPPP, Dept. of Physics, University of Durham, DH1 3LE,UK}\\
$^{b)}${\it SISSA and INFN-Sezione di Trieste, Trieste I-34014, Italy}\\
$^{c)}${\it CERN Theory Division, Geneve 23, CH-1211, Switzerland}\\
$^{d)}${\it INFN, Sezione di Padova, Via Marzolo 8, Padova I-35131, Italy}\\
\vspace{0.3cm}
{\large\bf Abstract}
\end{center}
\begin{quote}
{\noindent
Taking into account the recent progress
in the understanding of the 
lepton flavour effects in leptogenesis,
we investigate in detail the
possibility that the CP-violation necessary 
for the generation of the baryon asymmetry of the Universe
is due exclusively to the Dirac and/or Majorana CP-violating 
phases in the PMNS neutrino mixing matrix $U$, 
and thus is directly related to the low energy 
CP-violation in the lepton sector
(e.g., in neutrino oscillations, etc.).
We first derive the conditions of CP-invariance
of the neutrino   Yukawa couplings $\lambda$
in the see-saw Lagrangian, and of the 
complex orthogonal matrix $R$ in the ``orthogonal''
parametrisation of $\lambda$.
We show, e.g. that under certain
conditions i) real $R$ and
specific CP-conserving values
of the Majorana and Dirac phases 
can imply CP-violation, and 
ii) purely imaginary $R$ does not necessarily
imply breaking of CP-symmetry. 
We study in detail the case of hierarchical 
heavy Majorana neutrino mass spectrum, 
presenting results for three possible 
types of light neutrino mass spectrum:
i) normal hierarchical, ii) inverted hierarchical, 
and iii) quasi-degenerate.
Results in the alternative case of  
quasi-degenerate in mass heavy Majorana neutrinos,
are also derived. The
minimal supersymmetric extension of the 
Standard Theory with right-handed Majorana 
neutrinos and see-saw mechanism of neutrino mass 
generation is discussed as well. 
We illustrate the possible correlations
between the baryon asymmetry of the Universe
and i) the rephasing invariant $J_{\rm CP}$ 
controlling the magnitude of 
CP-violation in neutrino oscillations,
or ii) the effective Majorana mass 
in neutrinoless double beta decay, 
in the cases when the only source of 
CP-violation is respectively the Dirac 
or the Majorana phases in the neutrino 
mixing matrix.
}
\end{quote}
\setcounter{page}{1}

\section{Introduction}
\noindent
Baryogenesis through Leptogenesis \cite{FY} is 
a simple  mechanism to explain the observed 
baryon asymmetry of the Universe. 
A lepton asymmetry is dynamically generated
and then  converted into a baryon asymmetry
due to $(B+L)$-violating sphaleron interactions \cite{kuzmin}
which exist within the Standard Model (SM).
A simple scheme in which this mechanism can be implemented is the 
``seesaw''(type I) model of neutrino mass generation \cite{seesaw}.
In its minimal version it 
includes the Standard Model (SM) plus two or three   
right-handed (RH) heavy Majorana neutrinos.
Thermal leptogenesis \cite{lept,ogen,work} 
can take place, for instance, 
in the case of hierarchical spectrum of the heavy RH 
Majorana neutrinos. The lightest of the RH Majorana 
neutrinos is produced by thermal 
scattering after inflation. It subsequently
decays out-of-equilibrium in a lepton number and CP-violating way, 
thus satisfying Sakharov's conditions \cite{sakharov}.
In grand unified theories (GUT) the masses of the heavy 
RH Majorana neutrinos are typically by a few to several 
orders of magnitude smaller than the
scale of unification of the electroweak and strong interactions,
$M_{\rm GUT} \cong 2\times 10^{16}$ GeV.
This range coincides with the range of values of the
heavy Majorana neutrino masses, required 
for a successful thermal leptogenesis.

\indent Compelling evidence for existence 
non-zero neutrino masses and non-trivial neutrino mixing 
have been obtained during the last several years 
in the experiments studying oscillations of 
solar, atmospheric, reactor and accelerator 
neutrinos~\cite{sol,SKsolaratm,SNO123,KamLAND,K2K}.
The currently existing data imply
the presence of 3-neutrino mixing 
in the weak charged-lepton
current (see, e.g.,~\cite{STPNu04}):
\begin{equation}
\nu_{l \mathrm{L}}  = \sum_{j=1}^{3} U_{l j} \, \nu_{j \mathrm{L}},~~
l  = e,\mu,\tau,
\label{3numix}
\end{equation}
%
\noindent where 
$\nu_{lL}$ are the flavour neutrino fields, $\nu_{j \mathrm{L}}$ is
the field of neutrino $\nu_j$ having a mass $m_j$ and $U$ is the
Pontecorvo-Maki-Nakagawa-Sakata (PMNS) mixing
matrix~\cite{BPont57}.  The existing data, 
including the data from the \hbeta 
experiments~\cite{MoscowH3Mainz}, show that
the massive neutrinos $\nu_j$ are 
significantly lighter than the charged leptons and quarks
\footnote{The data from the \hbeta experiments~\cite{MoscowH3Mainz} 
imply $m_{j} < 2.3$ eV (95\%C.L.). More stringent upper limit 
on $m_j$ follows from the constraints on the sum of neutrino 
masses obtained from cosmological/astrophysical observations,
namely, from the CMB data of the WMAP experiment 
\cite{WMAPnu} combined with data from large scale 
structure surveys (2dFGRS, SDSS)~(see, e.g., \cite{Teg2}).
}.
The see-saw mechanism of neutrino 
mass generation~\cite{seesaw}
provides a natural explanation of the smallness of 
neutrino masses: integrating out 
the heavy RH Majorana neutrinos  
generates a mass term of Majorana type for the 
left-handed flavour neutrinos, which is inversely 
proportional to the large mass of the RH ones. 

  Establishing a connection between the low 
energy neutrino mixing parameters 
and high energy leptogenesis parameters 
has received much attention in recent years 
\cite{comb,ich,others,others2,nolowCP,23,231,triangular,lissac,PPR03,PRST05,antusch}. 
These studies showed, in particular, that 
the number of phenomenological 
parameters of the seesaw mechanism is 
significantly larger than  the 
number of quantities measurable 
in the ``low energy'' neutrino experiments.

    In the present article we investigate
the  link between the high energy 
CP-violation responsible for
the generation of baryon asymmetry
through  leptogenesis and the leptonic CP violation 
at low energy, which can manifest itself 
in non-zero  CP-violating asymmetry in 
neutrino oscillations and, in an indirect way, 
in the effective Majorana mass in neutrinoless 
double beta ($\betabeta$-) decay, $\meff$ 
(see, e.g., \cite{BiPet87,BPP1}). 

   It was concluded in a large number of studies of leptogenesis 
performed in the so-called ``one-flavor approximation'', 
that no direct link exists between the leptogenesis 
CP-violating parameters  and the CP-violating Dirac 
and Majorana phases in the PMNS neutrino mixing matrix, 
measurable at low energies. In particular, 
an observation of leptonic low energy 
CP-violating phases would not automatically 
imply a non-vanishing baryon asymmetry 
through leptogenesis in the one-flavour case. 
This conclusion, however, does not universally hold 
\cite{davidsonetal,nardietal,davidsonetal2}. 
Moreover, as was shown in \cite{PPRio106} 
(see also \cite{brancoj})
and will be discussed in detail in the present article,
the low-energy Dirac and/or Majorana CP-violating phases
in $\pmns$, which enter into the expressions respectively
of the leptonic CP-violating rephasing invariant 
$J_{\rm CP}$ \cite{PKSP3nu88}, controlling the magnitude of the
CP-violation effects 
in neutrino oscillations, 
and of the effective Majorana mass $\meff$ 
\cite{BGKP96,BPP1,STPFocusNu04}
can be the CP-violating  parameters responsible 
for the generation of the baryon asymmetry of the Universe. 
Consequently, the leptogenesis mechanism 
can be maximally connected to the low energy 
CP-violating phases in $\pmns$:
within the scenario under discussion, 
the observation of CP-violation in the lepton (neutrino) 
sector would generically ensure the 
existence of a baryon asymmetry.

    The possibility of a direct connection between
the high-energy leptogenesis and the low-energy leptonic 
CP-violation
is based on the fact that a new ingredient
has been recently accounted for in the leptogenesis 
mechanism, namely, the lepton flavour effects 
\cite{davidsonetal,nardietal,davidsonetal2}. 
As we have indicated above, 
the dynamics of leptogenesis was usually  
addressed within the `one-flavour' approximation.
In the latter, the Boltzmann equations
are written for the abundance of the 
lightest RH Majorana neutrino, $N_1$, responsible
for the out of equilibrium and CP-asymmetric decays, 
and  for the {\it total lepton charge} asymmetry. 
However, this 
approximation is rigourously correct 
only when the interactions mediated by 
charged lepton Yukawa couplings are out of equilibrium. 
Supposing that leptogenesis takes place at 
temperatures $T\sim M_1$, where $M_1$
is the mass of $N_1$,
the `one-flavour' approximation 
holds only for $T\sim M_1\gsim 10^{12}$ GeV. 
For $M_1\gsim 10^{12}$ GeV, i.e., at
temperatures higher than $10^{12}$ GeV,
all lepton  flavours are indistinguishable 
- there is no notion of flavour.
The lepton asymmetry generated in $N_1$
decays is effectively ``stored'' in one 
lepton flavour. 
However, for $T\sim M_1\sim 10^{12}$ GeV, 
the interactions mediated by the
$\tau$-lepton Yukawa couplings 
come into equilibrium, followed by those mediated
by the muon Yukawa couplings at $T\sim M_1\sim 10^{9}$ GeV, 
and the notion of lepton flavour becomes physical.

  The impact of flavour in thermal leptogenesis
has been recently investigated in detail in
\cite{Barbieri99,davidsonetal,nardietal,davidsonetal2} and in 
\cite{antusch,dibari}
including the quantum oscillations/correlations
of the asymmetries in lepton flavour space
\cite{davidsonetal}.
The Boltzmann equations 
describing the asymmetries in flavour space 
have  additional terms which can  
significantly affect the result for the
final baryon asymmetry. The ultimate reason is that 
realistic leptogenesis is a dynamical
process, involving  the  production and 
destruction of the heavy RH Majorana neutrinos, 
and of a lepton asymmetry  that is distributed among
{\it distinguishable} lepton flavours.  
Contrary to what is generically assumed 
in the one-single flavour 
approximation, the $\Delta L=1$
inverse decay processes which
wash-out the net lepton number
are flavour dependent, that is  
the lepton asymmetry carried by, 
say, electrons can be 
washed out only by the inverse decays 
involving the electron flavour. 
The  asymmetries in each lepton flavour,
are  therefore washed out differently,
and will appear with different 
weights in the final formula
for the baryon asymmetry. 
This is physically inequivalent
to the treatment of wash-out in the one-flavour approximation,
where the flavours are taken indistinguishable, 
thus obtaining  the unphysical
result that,  e.g., an asymmetry stored in 
the electron lepton charge may be washed
out by inverse decays involving the 
muon  or the tau charges.

  When flavour effects are accounted for, 
the final value of the baryon asymmetry is 
the sum of three contributions. Each term is 
given by the CP asymmetry in a given lepton 
flavour $l$, properly weighted by a wash-out factor 
induced by the same  lepton number violating processes. 
The wash-out factors are also flavour dependent.
In the present article
we perform a detailed analysis of the indicated
flavour effects in leptogenesis. We show that
the low energy Dirac and/or Majorana$ CP$-violating 
phases in $\pmns$ can be responsible for the 
generation of the baryon asymmetry of the Universe. 
We study also in detail 
the possible correlations between the physical 
low energy observables which depend on the 
CP-violating phases in $U_{\rm PMNS}$ - 
the rephasing invariant $J_{\rm CP}$ and the
effective Majorana mass in neutrinoless 
double beta decay $\meff$, 
and the baryon asymmetry.

  The paper is organized as follows.
In Section 2 we briefly review the existing data on 
the neutrino mixing parameters and the 
phenomenology of the low energy Dirac and Majorana 
CP-violation in the lepton sector. We note, in particular,
that  searching for CP-violation effects in $\nu$-
oscillations is the only practical way to
get information about the CP-violation due to the
Dirac phase in the neutrino mixing matrix $\pmns$ 
(Dirac CP-violation), and that  
the only feasible experiments which
at present have the potential of
establishing the Majorana nature 
of light neutrinos $\nu_j$ and of
providing information on the Majorana 
CPV phases in $\pmns$ are the 
the neutrinoless double beta
decay experiments searching for the process
$(A,Z) \rightarrow (A,Z+2) + e^- + e^-$.
In Sections 3-7 we present a detailed discussion of the 
possible connection between the
CP-violation at high energy in leptogenesis and 
the CP-violation in the lepton sector at low energies.
We first derive the conditions of CP-invariance
of the see-saw Lagrangian, i.e. of the neutrino    
Yukawa couplings, taking into account that
the light and heavy neutrinos with definite mass are 
Majorana particles, and thus have definite CP-parities
in the case of exact CP-symmetry (Section 3). 
We review briefly the arguments,
based on the ``one-flavour'' approximation,
leading to the conclusion that the connection
between leptogenesis CP-violating parameters 
and the CP-violating phases in the PMNS matrix
generically does not hold.
We next discuss the conditions under which
the lepton flavour effects in leptogenesis
become important and present a brief summary
of the results obtained recently on these effects, 
which are used in our analysis (Section 4).
In Sections 5, 6 and 7 we investigate the
possibility that the CP-violation necessary 
for the generation of the baryon asymmetry of the Universe
is due exclusively to the Dirac and/or Majorana CP-violating 
phases in the PMNS matrix, and thus is directly related
to the low energy CP-violation in the lepton sector
(e.g., in neutrino oscillations, etc.).    
The case of hierarchical heavy Majorana neutrino 
mass spectrum is studied in detail in Section 5, 
where we present results for three possible 
types of light neutrino mass spectrum:
i) normal hierarchical, ii) inverted hierarchical, 
and iii) quasi-degenerate.
In Section 6 results in the alternative case of  
quasi-degenerate in mass heavy Majorana neutrinos
are derived, while in Section 7 we 
discuss how the results on leptogenesis 
in Section 4 - 6 will be modified in 
the minimal supersymmetric extension of the 
Standard Theory with right-handed Majorana 
neutrinos and see-saw mechanism of neutrino mass 
generation. Section 8 represents Conclusions.

%
\section{Neutrino Mixing Parameters and Low Energy
Dirac and Majorana CP-Violation}
%
\noindent
We will use the standard parametrisation of the PMNS matrix:
\bea 
\label{eq:Upara}
\pmns = \left( \bad 
c_{12} c_{13} & s_{12} c_{13} & s_{13}e^{-i \delta}  \\[0.2cm] 
 -s_{12} c_{23} - c_{12} s_{23} s_{13} e^{i \delta} 
 & c_{12} c_{23} - s_{12} s_{23} s_{13} e^{i \delta} 
& s_{23} c_{13}  \\[0.2cm] 
 s_{12} s_{23} - c_{12} c_{23} s_{13} e^{i \delta} & 
 - c_{12} s_{23} - s_{12} c_{23} s_{13} e^{i \delta} 
 & c_{23} c_{13} \\ 
                \ea   \right) 
~{\rm diag}(1, e^{i \frac{\alpha_{21}}{2}}, e^{i \frac{\alpha_{31}}{2}})
\eea
\noindent where 
$c_{ij} \equiv \cos\theta_{ij}$,
$s_{ij} \equiv \sin\theta_{ij}$, 
$\theta_{ij} = [0,\pi/2]$,
$\delta = [0,2\pi]$ is the 
Dirac CP-violating (CPV) phase and 
and $\alpha_{21}$ and $\alpha_{31}$
two Majorana  CPV phases \cite{BHP80,Doi81SchValle80}.
We standardly identify  
$\dmsol$=$\Delta m^2_{21} \equiv m^2_2-m^2_1 > 0$, 
where $\dmsol$ drives the solar neutrino oscillations.
In this case 
$|\dma|$=$|\Delta m^2_{31}|\cong |\Delta m^2_{32}|$,
$\theta_{23} = \theta_{\rm A}$ and
$\theta_{12} = \theta_{\odot}$,
$|\dma|$, $\theta_{\rm A}$ 
and $\theta_{\odot}$  being
the $\nu$-mass squared difference
and mixing angles responsible respectively 
for atmospheric and solar neutrino oscillations,
while $\theta_{13}$ is the CHOOZ angle
\cite{CHOOZPV}.
The existing neutrino oscillation data,
including the recent results of the MINOS
experiment \cite{MINOS0706},
allow us to determine 
$\dmsol$, $|\dma|$, 
$\sin^2\theta_{12}$ and
$\sin^22\theta_{23}$ with a relatively 
good precision and to obtain
rather stringent limits on $\sin^2\theta_{13}$ (see,
e.g.,~\cite{BCGPRKL2,TSchwSNOW06,Fogli06}). The best fit values 
and the 95\% C.L. allowed ranges 
of $\dmsol$, $\sin^2\theta_{12}$
$|\dma|$ and $\sin^22\theta_{23}$
read:
\begin{eqnarray}
\label{deltaatmvalues}
(|\deltaatm|)_{\rm BF} = 2.5 \times 10^{-3} \ \eV^2, 
& 2.1  \times 10^{-3} \ \eV^2 \leq |\deltaatm| 
\leq 2.9  \times 10^{-3} \ \eV^2 ,  \\
\label{deltasolvalues}
(\deltasol)_{\rm BF} = 8.0 \times 10^{-5} \ \eV^2, 
& 7.3 \times 10^{-5} \ \eV^2 \leq \deltasol \leq 8.5 \times 10^{-5} \ \eV^2, \\
\label{th12th23}
(\sin^2 \theta_{12})_{\rm BF} = 0.31,
& 0.26 \leq \sin^2 \theta_{12} \leq 0.36\,\\ 
(\sin^2 2\theta_{23})_{\rm BF} = 1,
& \sin^2 2\theta_{23} \geq 0.90\,.
\end{eqnarray}
%
A combined 3-$\nu$ oscillation analysis of 
the global neutrino oscillation data
gives~\cite{BCGPRKL2,TSchwSNOW06}
\beq
\sin^2\theta_{13} < 0.027~(0.041)
\quad\mbox{at}\quad 95\%~(99.73\%)~{\rm C.L.}
\label{th13}
\eeq
%
 The neutrino oscillation parameters 
discussed above can (and very
likely will) be measured with much higher 
accuracy in the future (see, e.g.,~\cite{STPNu04}).

  Depending on the sign of 
$\deltaatm \equiv \Delta m^2_{31} \cong \Delta m^2_{32}$ 
which cannot be determined from presently existing
data, the $\nu$-mass spectrum can be of two types:\\
-- {\it with normal ordering} 
$m_1 < m_2 < m_3$, $\dma=\Delta m^2_{31} >0$, and \\
-- {\it with inverted ordering} 
$m_3 < m_1 < m_2$, $\dma =\Delta m^2_{32}< 0$, $m_j > 0$,\\
\noindent where we have employed the standardly 
used convention to define the two spectra. 
Depending on the sign of $\dma$, 
${\rm sgn}(\dma)$, and on the value of 
the lightest neutrino mass (i.e., absolute 
neutrino mass scale), ${\rm min}(m_j)$, 
the $\nu$-mass  spectrum can be\\
-- {\it Normal Hierarchical}: $m_1{\small \ll m_2 \ll }m_3$,
$m_2{\small \cong }(\dmsol)^ {1\over{2}}{\small \sim}$ 0.009 eV,
$m_3{\small \cong }|\dma|^{1\over{2}}{\small \sim}$ 0.05 eV;\\
-- {\it Inverted Hierarchical}: $m_3 \ll m_1 < m_2$,
with $m_{1,2} \cong |\dma|^{1\over{2}}\sim$ 0.05 eV; \\
-- {\it Quasi-Degenerate}: $m_1 \cong m_2 \cong m_3 \cong m$,
$m_j^2 \gg |\dma|$, $m \gtap 0.10$~eV.

  Determining the nature of massive 
neutrinos, obtaining information
on the type of $\nu$-mass spectrum, absolute 
$\nu$-mass scale
and on the status of CP-symmetry 
in the lepton sector
are among the fundamental problems in
the studies of neutrino mixing \cite{STPNu04}.

\vspace{-0.35cm}
\subsection{CP and T Violation in Neutrino Oscillations}
\noindent
 Searching for CP-violation effects in $\nu$-
oscillations is the only practical way to
get information about Dirac CP-violation in 
the lepton sector, associated with the phase 
$\delta$ in $\pmns$.
A measure of  CP- and $T$- violation is provided
by the asymmetries \cite{Cabibbo78,BHP80,Barger80,PKSP3nu88}:
\beq
A^{(l,l')}_{\rm CP} = P({\nu_l \rightarrow \nu_{l'}}) 
- P({\bar{\nu}_l \rightarrow \bar{\nu}_{l'}})\,,~~~
A^{(l,l')}_{\rm T} = P(\nu_l \rightarrow \nu_{l'}) -
P(\nu_{l'} \rightarrow \nu_{l})\,,~~l\neq l'=e,\mu,\tau\,.
\eeq
%
\noindent For 3-$\nu$ oscillations in vacuum,
which respect the CPT-symmetry, one has \cite{PKSP3nu88}:
\begin{eqnarray}
A^{(e,\mu)}_{\rm T} &=& A^{(\mu,\tau)}_{\rm T} = - A^{(e,\tau)}_{\rm T} = 
J_{\rm CP}~F^{\rm vac}_{\rm osc}\,,~~~A^{(l,l')}_{\rm CP} = A^{(l,l')}_{\rm T}\,,\\
J_{\rm CP} &=& {\rm Im}\left\{ U_{e1} U_{\mu 2} U_{e 2}^\ast 
U_{\mu 1}^\ast \right\} = \frac{1}{4}\sin2\theta_{12} \sin2\theta_{23}
\cos^2\theta_{13}\sin\theta_{13}\sin\delta\,,\\
F^{\rm vac}_{\rm osc} &=&
\sin\left(\frac{\Delta m^2_{21}}{2E}L\right) +
\sin\left(\frac{\Delta m^2_{32}}{2E}L\right) +
\sin\left(\frac{\Delta m^2_{13}}{2E}L\right)\,.
\end{eqnarray}
%
\noindent
Thus, the magnitude of CP-violation effects
in neutrino oscillations is controlled by
the rephasing invariant associated with the Dirac
phase $\delta$, $J_{\rm CP}$  \cite{PKSP3nu88}.
The existence of Dirac CP-violation
in the lepton sector would be established 
if, e.g., some of the vacuum oscillation 
asymmetries 
$A^{(e,\mu)}_{\rm CP(T)}$, etc. are proven
experimentally to be nonzero.
This would imply that $J_{\rm CP} \neq 0$,
and, consequently, that 
$\sin\theta_{13}\sin\delta \neq 0$
\footnote{
Let us note that 
the oscillations in matter, e.g., in the Earth, 
are neither CP- nor CPT- invariant \cite{Lang87}
as a consequence of the fact 
that the Earth matter 
is not charge-symmetric 
(it contains $e^{-},~p$ and $n$, 
but does not contain their antiparticles). 
This complicates the studies of
CP-violation due to the Dirac phase $\delta$ 
in $\nu$-oscillations in matter (Earth) 
(see, e.g., \cite{Future}).
The matter effects in $\nu$-oscillations 
in the Earth to a good precision 
are not T-violating \cite{PKSP3nu88},
however. In matter with constant density, 
e.g., Earth mantle, one has \cite{PKSP3nu88}: 
$A^{(e,\mu)}_{\rm T} = J^{\rm m}_{\rm CP}F^{\rm m}_{\rm osc}$,
$J^{\rm m}_{\rm CP} = J_{\rm CP}~R_{\rm CP}$, where 
the function $R_{\rm CP}$ does not
depend on $\theta_{23}$ and $\delta$.
}.
One of the major goals of the future 
experimental studies of neutrino 
oscillations is the searches 
for CP-violation effects 
due to the Dirac phase in $\pmns$ 
(see, e.g., \cite{STPNu04,Future}).
  
\subsection{Majorana CP-Violating Phases and $\betabeta$-Decay}
\noindent
%
  As is well-known, the theories with 
see-saw mechanism of neutrino mass 
generation~\cite{seesaw} of interest for our 
discussion, predict the massive neutrinos 
$\nu_j$ to be Majorana particles. 
We will assume in what follows that the 
fields $\nu_j(x)$ satisfy the Majorana 
condition:
\begin{equation}
C(\bar{\nu}_{j})^{T} = \nu_{j},~j=1,2,3,
\label{Majnu}
\end{equation}
%
\noindent where $C$ is the charge conjugation matrix:
$C^{-1} \gamma_{\mu} C = - \gamma_{\mu}^{T}$,
$C^{T} = - C$, $C^{\dagger} = C^{-1}$.  

 Determining the nature of massive 
neutrinos is one of the most formidable 
and pressing problems in today's neutrino physics
(see, e.g.,~\cite{STPNu04,APSbb0nu}). 
If $\nu_j$ are proven to be Majorana fermions,
getting experimental information about the  
Majorana CPV phases in $\pmns$,
$\alpha_{21}$ and $\alpha_{31}$,
would be a remarkably difficult problem. 
The oscillations of flavour neutrinos, 
$\nu_{l} \rightarrow \nu_{l'}$ and 
$\bar{\nu}_{l} \rightarrow \bar{\nu}_{l'}$,
$l,l'=e,\mu,\tau$, are insensitive to the phases
$\alpha_{21,31}$~\cite{BHP80,Lang87}.
The Majorana phases of interest
$\alpha_{21,31}$ can affect 
significantly the predictions for the 
rates of (LFV) decays $\mu \rightarrow e + \gamma$,
$\tau \rightarrow \mu + \gamma$, etc.
in a large class of supersymmetric theories
with see-saw mechanism of $\nu$-mass generation 
(see, e.g., \cite{PPY03}).

  In the case of 3-$\nu$ mixing under discussion
there are, in principle, three independent 
CP violation rephasing invariants.
The first is $J_{\rm CP}$ - the Dirac one, 
associated with the Dirac phase $\delta$,
we have discussed in the preceding
subsection. 
The existence of two additional 
invariants, $S_1$ and $S_2$
is related to the two Majorana 
CP violation phases in $\pmns$.
The invariants $S_1$ and $S_2$ can be chosen as
\footnote{The expressions for the invariants
$S_{1,2}$ we give and will use further 
correspond to the Majorana 
conditions (\ref{Majnu}) for the fields of 
neutrinos $\nu_j$, which do not contain 
phase factors, see, e.g., \cite{BPP1}.} 
\cite{JMaj87,ASBranco00,BPP1}:
\beq
S_1 = {\rm Im}\left\{ U_{\tau 1}^\ast \, U_{\tau 2} \right\}~,~~~~
S_2 = {\rm Im}\left\{ U_{\tau 2}^\ast \, U_{\tau 3} \right\}~.
\label{eq:SCP}
\eeq
%
The rephasing invariants associated 
with the Majorana phases are 
not uniquely determined. Instead of $S_1$ 
defined above we could have chosen, e.g.,
$S'_1 = {\rm Im}\left\{ U_{e 1} U_{e 2}^\ast \right\}$ 
or $S''_1 = {\rm Im}\left\{ U_{\mu 1} U_{\mu 2}^\ast \right\}$, 
while instead of $S_2$ we could have used 
$S'_2 = {\rm Im}\left\{ U_{e2} \, U_{e 3}^\ast \right\}$, 
or $S''_2 = {\rm Im}\left\{ U_{\mu 2} \, U_{\mu 3}^\ast \right\}$.
The Majorana phases $\alpha_{21}$ and $\alpha_{31}$, 
or $\alpha_{21}$ and $\alpha_{32} \equiv (\alpha_{31} - \alpha_{21})$,
can be expressed in terms of the rephasing invariants
thus introduced \cite{BPP1}:
$\cos \alpha_{21} = 1 - (S'_1)^2/(|U_{e1}U_{e2}|^2)$, etc.
The expression for $\cos \alpha_{21}$ in terms of 
$S_1$ is somewhat more cumbersome (it involves also 
$J_{\rm CP}$) and we will not give it here.
Note that CP-violation due to the Majorana phase
$\alpha_{21}$ requires that both
$S_1 = {\rm Im}\left\{ U_{\tau 1} U_{\tau 2}^\ast \right\}\neq 0$ and
${\rm Re}\left\{ U_{\tau 1} U_{\tau 2}^\ast \right\}\neq 0$.
Similarly,
$S_2 = {\rm Im}\left\{ U_{\tau 2}^\ast U_{\tau 3} \right\} \neq 0$
would imply violation of the CP-symmetry only 
if in addition ${\rm Re}\left\{ U_{\tau 2}^\ast U_{\tau 3} \right\} \neq 0$.

  The only feasible experiments which
at present have the potential of
establishing the Majorana nature 
of light neutrinos $\nu_j$ and of
providing information on the Majorana 
CPV phases in $\pmns$ are the experiments 
searching for the neutrinoless double beta
($\betabeta$-) decay, 
$(A,Z) \rightarrow (A,Z+2) + e^- + e^-$ (see, 
e.g.,~\cite{BiPet87,APSbb0nu,STPFocusNu04}). 
The $\betabeta$-decay effective Majorana mass, $\meff$ 
(see, e.g., \cite{BiPet87}), which contains all the 
dependence of the $\betabeta$-decay amplitude on 
the neutrino mixing parameters, 
is given by the following expressions 
for the normal hierarchical (NH),  
inverted hierarchical (IH) 
and quasi-degenerate (QD) neutrino mass spectra
(see, e.g., \cite{STPFocusNu04}): 
\begin{eqnarray}
\meff&\cong&\left|\sqrt{\deltasol}\sin^2 \theta_{12}e^{i\alpha_{21}}
+\sqrt{\deltaatm}\sin^2\theta_{13}
e^{i(\alpha_{31}-2\delta)} \right|\;,~m_1\ll m_2 \ll m_3~{\rm (NH)},
\nonumber\\
&&\\
\label{meffNH2}
\meff &\cong& \sqrt{|\deltaatm|}
\left|\cos^2\theta_{12}  + 
e^{i\alpha_{21}}~\sin^2 \theta_{12} \right|\;,
~~m_3 \ll m_1< m_2~{\rm (IH)},\\ [0.2cm] 
\label{meffIH1}
\meff &\cong& m \left|\cos^2\theta_{12}
 +  e^{i \alpha_{21}}~\sin^2 \theta_{12} 
\right|\;,~~m_{1,2,3} \cong m \gtap 0.10~{\rm eV}\;~{\rm (QD)}\,. 
\label{meffQD0} 
\end{eqnarray}
%
Obviously, $\meff$ depends strongly 
on the Majorana CPV phase(s):
the CP-conserving values of 
$\alpha_{21} = 0,\pm\pi$ \cite{LW81}, 
for instance, determine the range of 
possible values of $\meff$ in the 
cases of IH and QD spectrum, while
the CP-conserving values of
$(\alpha_{31} - \alpha_{21})
\equiv \alpha_{32} = 0,\pm \pi$,
can be important in the
case of NH spectrum. 
As is well-known, 
in the case of CP-invariance 
the phase factors 
\begin{equation}
\eta_{21} \equiv e^{i\alpha_{21}}= \pm 1\,,~ 
\eta_{31}\equiv e^{i\alpha_{31}} = \pm 1\,,~
\eta_{32}\equiv e^{i\alpha_{32}} = \pm 1\,,~   
\label{eta2131}
\end{equation}
%
\noindent have a very simple physical 
interpretation \cite{LW81,BiPet87}:
$\eta_{ik}$ is the
relative CP-parity of Majorana neutrinos 
$\nu_{i}$ and $\nu_k$,
$\eta_{ik} = (\eta^{\nu \mathrm{CP}}_i)^{\ast} \eta^{\nu \mathrm{CP}}_k$,
$\eta^{\nu \mathrm{CP}}_{i(k)} \pm i$ being the CP-parity of
$\nu_{i(k)}$.

   As can be shown, in the general case of
arbitrary ${\rm min}(m_j)$,
$\meff$ depends on the two invariants 
$S_1$ and $S_2$ \cite{BPP1}. 
In the chosen parametrisation of the
PMNS matrix, Eq.~(\ref{eq:Upara}), 
$\meff$ depends also on $J_{\rm CP}$.

   The $\betabeta$-decay experiments
of the next generation, which are under 
preparation at present (see, e.g., \cite{APSbb0nu}), 
are aiming to probe the QD and IH ranges of $\meff$.
If the $\betabeta$-decay will be 
observed in these experiments, 
the measurement of the $\betabeta$-decay 
half-life might allow to obtain constraints on 
the Majorana phase $\alpha_{21}$ 
\cite{BPP1,BGKP96,PPW}. 

%
%
\section{The CP-Invariance Constraints}
%
\noindent
In the next Section, we 
will summarize the arguments leading to the
conclusions that the 
leptogenesis CPV phases, 
responsible for the generation of the 
baryon asymmetry, can indeed be directly 
connected to the low energy CPV phases in 
$\pmns$ and, correspondingly, to 
CP violating phenomena, e.g., 
in neutrino physics. 
The starting point of our discussion 
is the Lagrangian of the Standard Model (SM)
with the addition of three heavy right-handed  
Majorana neutrinos $N_{i}$ ($i=1,2,3$) with 
masses $M_{3}> M_{2}> M_{1} > 0$
and Yukawa couplings 
$\lambda_{i l}$. 
It will be assumed (without loss of generality) 
that the fields $N_j$ satisfy the Majorana 
condition:
\begin{equation}
C(\overline{N}_{j})^{T} = N_{j},~j=1,2,3.
\label{Majcondnu}
\end{equation}
%
\noindent We will work in the basis in which 
the Yukawa couplings for the
charged leptons are flavour-diagonal. 
In this basis the leptonic part of the 
Lagrangian of interest reads:
\beq
{\cal L}^{\rm lep}(x) = {\cal L}_{\rm CC}(x) + 
{\cal L}_{\rm Y}(x) + {\cal L}_{\rm M}^{\rm N}(x)\,, 
\label{Llep}
\eeq
%
\noindent where ${\cal L}_{\rm CC}(x)$ and
${\cal L}_{\rm Y}(x)$
are the charged current (CC) weak interaction
and Yukawa coupling Lagrangians, while 
${\cal L}_{\rm M}^{\rm N}(x)$ is the mass term
of the heavy Majorana neutrinos $N_i$:
\begin{eqnarray}
\label{CC}
{\cal L}_{\rm CC} &=& - ~\frac{g}{\sqrt{2}}\,
\overline{l_L}(x)\, \gamma_{\alpha}\,\nu_{lL}(x)\,
W^{\alpha \dagger}(x) + \hbox{h.c.}\,,\\
\label{Ynu}
{\cal L}_{\rm Y}(x) &=& 
\lambda_{i l}\,\overline{N_{iR}}(x)\, H^{\dagger}(x)\,\psi_{l L}(x) +  
h_l\,H^c(x) \,\overline{l_{R}}(x)\,\psi_{lL}(x) + \hbox{h.c.}\,,\\ 
\label{MN}
{\cal L}_{\rm M}^{\rm N}(x) &=& -\,\frac{1}{2}\,M_{i}\,\overline{N_i}(x)\, N_i(x)\,.
\end{eqnarray}
%
\noindent 
Here $\psi_{lL}$ and $l_{R}$ denote respectively
the left-handed (LH) lepton doublet and 
right-handed  lepton singlet fields of flavour 
$l=e,\mu,\tau$, $\psi^{\rm T}_{lL} = (\nu_{lL}~l_L)$,   
$W^{\alpha}(x)$ is the $W^{\pm}$-boson field, 
and $H$ is the Higgs doublet field whose neutral 
component has a vacuum expectation value equal to
$v=174$ GeV. Obviously, 
${\cal L}_{\rm Y}(x) + {\cal L}_{\rm M}^{\rm N}(x)$ 
includes all the necessary ingredients of
the see-saw mechanism. At energies below 
the heavy Majorana neutrino mass scale $M_1$, 
the heavy Majorana neutrino fields 
are integrated out and after the breaking 
of the electroweak symmetry, a Majorana mass 
term for the LH flavour neutrinos 
is generated:
\begin{equation}
\label{Rm}
m_{\nu} =  v^2 \lambda^T\, M^{-1}\, \lambda = U^{\ast}\, m \,U^{\dagger}\;,
\label{mnusees}
\end{equation} 
%
\noindent where $M$ and $m$ are diagonal matrices
formed by the masses of $N_j$ and $\nu_j$,
$M \equiv {\rm Diag}(M_1,M_2,M_3)$,
$m \equiv {\rm Diag}(m_1,m_2,m_3)$,
$M_j > 0$, $m_k \geq 0$, 
and $U$ is the PMNS matrix. The 
diagonalisation of the Majorana 
mass matrix $m_{\nu}$, Eq.~(\ref{mnusees}), leads to 
the relation (\ref{3numix}) between the 
LH flavour neutrino fields $\nu_{lL}$ 
and the fields $\nu_{jL}$ of neutrinos 
with definite mass and, correspondingly,
to the appearance of the PMNS 
neutrino mixing matrix in the charged current
weak interaction Lagrangian 
${\cal L}_{\rm CC}(x)$.

   In what follows we will often use 
the well-known ``orthogonal 
parametrisation`` of the matrix 
of neutrino Yukawa couplings \cite{Casas:2001sr}:
\begin{equation}
\label{R}
\lambda = \frac{1}{v} \,  \sqrt{M} \, R\, \sqrt{m}\, U^{\dagger}\;,
\end{equation} 
%
\noindent where $R$ is, in general, 
a complex orthogonal matrix, 
$R~R^T = R^T~R = {\bf 1}$.

  Before discussing leptogenesis and the 
violation of CP-symmetry associated with it,
it proves useful to analyze the constraints 
which the requirement  CP-invariance 
imposes on the Yukawa couplings 
$\lambda_{j l}$, on the PMNS matrix 
$U$ and on the matrix $R$.
If the CP-symmetry is unbroken,  
the Majorana fields $N_j$ and $\nu_k$ 
have definite CP-parities 
(see, e.g., \cite{BiPet87,LW81})
$\eta^{N \mathrm{CP}}_j = \pm i$, 
$\eta^{\nu \mathrm{CP}}_k = \pm i$,
and transform in the following way
under the CP-symmetry transformation:
\begin{eqnarray}
\label{CPN}
U_{\rm CP}\, N_j(x)\, U^{\dagger}_{\rm CP} 
&=& \eta^{N \mathrm{CP}}_j\, \gamma_0\, N_j(x')\,,
~~\eta^{N \mathrm{CP}}_j = i\rho^N_j = \pm i\,, \\
U_{\rm CP}\, \nu_k(x)\, U^{\dagger}_{\rm CP} 
&=& \eta^{\nu \mathrm{CP}}_k\, \gamma_0\, \nu_k(x')\,,~~\eta^{\nu \mathrm{CP}}_k = 
i\rho^{\nu}_k = \pm i\,.
\label{CPnu}
\end{eqnarray}
%
\noindent The sign factors $\rho^N_j$ and $\rho^{\nu}_k$, 
and correspondingly, the CP-parities of the heavy 
and light Majorana neutrinos $N_j$ and $\nu_k$,
are determined by the properties of the 
corresponding RH and LH flavour neutrino Majorana
mass matrices \cite{BiPet87}. They will be treated 
as free discrete parameters in what follows.

 Obviously, the mass term 
${\cal L}_{\rm M}^{\rm N}(x)$, Eq.~ (\ref{MN}), 
is CP-invariant. The requirement of 
CP-invariance of the Lagrangian ${\cal L}_{\rm Y}(x)$, 
Eq.~(\ref{Ynu}), leads, as can be shown, 
to the following constraint on the neutrino 
Yukawa couplings:
\begin{equation}
\lambda^{\ast}_{j l} = \lambda_{j l}\,
(\eta^{N \mathrm{CP}}_j)^{\ast}\, \eta^{l}\, \eta^{H\ast}\,,~~j=1,2,3,~l=e,\mu,\tau, 
\label{YCPinv1}
\end{equation}
%
\noindent where $\eta^{l}$ and $\eta^{H}$ are 
unphysical phase factors which appear in the
CP-transformations of the LH lepton doublet  
and Higgs doublet fields, $\psi_{lL}(x)$ 
and $H(x)$, respectively. These phase factors 
do not affect any of the physical observables and, 
for convenience, we will set them to
$i$ and 1: $\eta^{l} = i$,
$\eta^{H} = 1$. Thus, in what follows we will
work with the constraint (\ref{YCPinv1}) in the form 
\begin{equation}
\lambda^{\ast}_{j l} = \lambda_{j l}\,\rho^{N}_j\,,~\rho^{N}_j = \pm 1\,,
~~~j=1,2,3,~l=e,\mu,\tau, 
\label{YCPinv2}
\end{equation}
%
\noindent where we have used Eq.~(\ref{CPN}). Note that
if the CP-parity of a given heavy Majorana neutrino $N_j$
is equal to $(-i)$, i.e., if $\rho^{N}_j = -1$,
the elements $\lambda_{j l}$, $l=e,\mu,\tau$,
of the matrix of neutrino Yukawa couplings $\lambda$
will be {\it purely imaginary}.

  It follows from Eqs.~(\ref{mnusees}) and (\ref{YCPinv2}) 
that if CP-invariance holds, the Majorana mass matrix
of the LH flavour neutrinos generated by the see-saw 
mechanism is real (in the convention for the various
unphysical phase factors employed by us): 
$m_{\nu \ast} = m_{\nu}$. This leads to    
the following CP-invariance constraint 
on the PMNS matrix $U$ \cite{BiPet87}:
\begin{equation}
U^{\ast}_{l j} = U_{l j}\,\rho^{\nu}_j,,~~j=1,2,3,~l=e,\mu,\tau \,. 
\label{UCPinv1}
\end{equation}
%
\noindent In the parametrisation (\ref{eq:Upara}) 
we are using these conditions 
imply that the Dirac phase $\delta = \pi q$, $q=0,1,2,...$,
and that the Majorana phases should satisfy:
$\alpha_{21}= \pi q'$, $\alpha_{31}= \pi q''$, 
$q',q''=0,1,2,...$.   

   Using Eqs.~(\ref{Rm}), (\ref{YCPinv2}) 
and (\ref{UCPinv1}) it is easy to 
derive the constraints on the matrix $R$
following (in the convention we are using)
from the requirement of
CP-invariance of the ``high-energy'' 
Lagrangian of interest (\ref{Llep}): 
\begin{equation}
R^{\ast}_{jk} = R_{jk}\,\rho^{N}_j \,\rho^{\nu}_k,~~j,k=1,2,3\,. 
\label{RCPinv1}
\end{equation}
%
\noindent Obviously, this would be a condition
of reality of the matrix $R$ only if 
$\rho^{N}_j \rho^{\nu}_k = 1$ for any $j,k=1,2,3$.
However, we can also have $\rho^{N}_j \rho^{\nu}_k = - 1$
for some $j$ and $k$ and in that case 
the corresponding elements of $R$ will be 
{\it purely imaginary}. 

  The preceding results 
lead  to the following 
(perhaps obvious) conclusions.\\
i) If CP-invariance holds at ``high'' energy, i.e., if 
the neutrino Yukawa couplings satisfy 
(in the convention used) the constraints 
(\ref{YCPinv2}), it will also hold at 
``low'' energy in the lepton sector, 
i.e., the elements of  the PMNS matrix 
will satisfy (in the same convention) 
Eq.~(\ref{UCPinv1}).\\
ii) If the CP-symmetry is violated 
at ``low'' energy, i.e., if the PMNS matrix  
does not satisfy conditions (\ref{UCPinv1}) 
and the CC weak interaction is not CP-invariant, 
it will also be violated at ``high'' energy,
i.e., the neutrino Yukawa couplings
will not satisfy Eq.~(\ref{YCPinv2})
and the Lagrangian  ${\cal L}_{\rm Y}(x)$ 
will not be CP-invariant.
Obviously, it is not 
possible to use the matrix $R$ 
in order to render the Yukawa 
couplings CP-conserving.\\
iii) If CP-invariance holds 
at ``low'' energy, i.e., if 
the CC weak interaction (\ref{CC}) 
is CP-invariant and the PMNS 
matrix satisfy conditions (\ref{UCPinv1}), 
it can still be violated at ``high'' energy,
i.e., the neutrino Yukawa couplings
will not necessarily satisfy Eq.~(\ref{YCPinv2})
and the Lagrangian  ${\cal L}_{\rm Y}(x)$ 
will not necessarily be CP-invariant.
The CP-violation in this case can be 
due to the matrix $R$.

   As we will see, of interest 
for our further analysis is, 
in particular, the product 
\begin{equation}
P_{jkml}\equiv R_{j k}\,R_{j m} \, U^{\ast}_{lk} \, U_{lm}\,,~ k\neq m. 
\label{P1}
\end{equation}
%
\noindent If CP-invariance holds, we find from 
Eqs. (\ref{UCPinv1}) and (\ref{RCPinv1})
that $P_{jkml}$ is real:
\begin{equation}
P^{\ast}_{jkml} = P_{jkml}\,(\rho^{N}_j)^2\,
(\rho^{\nu}_k)^2\, (\rho^{\nu}_m)^2 = P_{jkml}\,,~~
{\rm Im}(P_{jkml}) = 0\,. 
\label{P2}
\end{equation}
%

   Let us consider next the case
when conditions (\ref{UCPinv1}) are satisfied and 
${\rm Re}(U^{\ast}_{\tau k} U_{\tau m})=0$, $k<m$, i.e.,
$U^{\ast}_{\tau k}U_{\tau m}$ is {\it purely imaginary}.
This can be realised, as Eqs. (\ref{eq:Upara}) and 
(\ref{UCPinv1}) show, for $\delta = \pi q$, 
$q=0,1,2,...$, and 
$\rho^{\nu}_k \rho^{\nu}_m = -1$, i.e., 
if the relative CP-parity of the light Majorana
neutrinos $\nu_k$ and  $\nu_m$ is equal to $(-1)$,
or, correspondingly, the Majorana phase 
$\alpha_{mk} = \pi (2q' +1)$, $q'=0,1,2,...$.
If $J_{\rm CP} = 0$ and both  
${\rm Re}(U^{\ast}_{\tau 1} U_{\tau 2})=0$
and ${\rm Re}(U^{\ast}_{\tau 2} U_{\tau 3})=0$,
CP-invariance holds in the lepton sector 
at low energies. In order for CP-invariance to hold 
at ``high'' energy, i.e., for $P_{jkml}$ to be real,
the product $R_{j k} R_{j m}$,
in the case under consideration,
has also to be {\it purely imaginary},
${\rm Re}(R_{j k} R_{j m}) = 0$.
Thus, {\it purely imaginary} $U^{\ast}_{\tau k}U_{\tau m}\neq 0$
and {\it purely real} $R_{j k} R_{j m} \neq 0$, i.e.,
${\rm Re}(U^{\ast}_{\tau k} U_{\tau m})=0$,
${\rm Im}(R_{j k} R_{j m}) = 0$,
in particular, {\it imply violation of the
CP-symmetry at ``high'' energy  by the matrix $R$}.
 
%
\section{Leptogenesis and the Baryon Asymmetry}
\noindent
\noindent
In this Section
we will repeat briefly the arguments leading to the
conclusion that the connection
between leptogenesis CP-violating parameters 
and the CP-violating phases in the PMNS matrix
generically does not hold.
We will therefore first restrict 
our discussion to the ``one-flavour'' approximation.
Then, we will review the recent results 
on leptogenesis which 
take into account the lepton flavour effects~
\cite{davidsonetal,nardietal,davidsonetal2}.
In this case, the baryon asymmetry depends
explicitly on the parameters of the lepton mixing matrix $U$. 

%
\subsection{The One Flavour Case}
\noindent
\noindent
   We assume that 
the heavy Majorana neutrinos $N_i$ have a 
hierarchical mass spectrum,
$M_{2,3}\gg M_1$, so that studying the 
evolution of the number density of  $N_1$ suffices.  
The  final amount of $(B-L)$ 
asymmetry can be parametrized as 
$Y_{B-L}=n_{B-L}/s$, 
where $s=2\pi^2 g_* T^3/45$ 
is the entropy density and $g_*$ counts 
the effective number of
spin-degrees of freedom in 
thermal equilibrium ($g_*=217/2$ in the SM
with a single generation of right-handed
neutrinos).  After reprocessing by sphaleron 
transitions, the baryon asymmetry is related to the $(B-L)$
asymmetry by \cite{Harvey:1990qw} 
$Y_B=(12/37) \left(Y_{B - L}\right)$.

     One  defines the CP asymmetry 
generated by $N_1$ decays  as
\begin{equation}
\epsilon_{1} \equiv 
\frac{\sum_l[\Gamma(N_1\rightarrow H \ell_l)-\Gamma(
N_1\rightarrow \overline{H} \overline{\ell}_l)]}{
\sum_l[\Gamma(N_1\rightarrow H
\ell_l)+\Gamma
(N_1\rightarrow \overline{H} \overline{\ell}_l)]}=
\frac{1}{8\pi}\sum_{j\neq 1}
\frac{\textrm{Im}
\left[ (\lambda \lambda^{\dagger})_{j 1}^2\right] }{\left[\lambda
\lambda^{\dagger}\right]_{11}}
g\left(\frac{M_{j}^2}{M_{1}^2}\right)\, .
\end{equation}
%
The wavefunction plus vertex contributions are included 
giving $g(x)\simeq - \frac{3}{2
\sqrt{x}}$ for $x\gg 1$.
Notice, in particular, that $\epsilon_1$ denotes
the CP asymmetry in the total lepton charge (i.e., trace over the lepton 
flavour index).

   Besides  the CP parameter $\epsilon_1$, the final baryon asymmetry
depends on a single  parameter,
\begin{equation}
\label{oldK}
\left(\frac{\widetilde{m_1}}{\widetilde{m_*}}\right )\equiv
\frac{\sum_l\Gamma(N_1\rightarrow
H\ell_l)}{H(M_1)}\, ,
\end{equation}
%
where $H(M_1)$ denotes the value of the Hubble rate
evaluated at a temperature $T=M_1$, $\widetilde{m_*}\sim 3\times 10^{-3}\,
{\rm eV}$ and 
\begin{equation}
\widetilde{m_1}\equiv \frac{(\lambda \lambda^\dagger)_{11}
v^2}{M_{1}}\, 
\end{equation} 
is proportional to
the total decay rate of the heavy RH
Majorana neutrino $N_1$. 
Notice again that $\widetilde{m_1}$ is obtained by computing the
total decay rate of the $N_1$, that is by summing over all the 
flavours. 

The asymmetry in the total lepton charge is then given by
\begin{equation}
Y_L\simeq  \frac{\epsilon_1}{g_*}\eta\left(\widetilde{m_1}\right)\, ,
\end{equation}
%
where $\eta\left(\widetilde{m_1}\right)$ accounts for the 
washing out of the total lepton asymmetry due to inverse decays.

The final baryon asymmetry in the ``one flavour
approximation'' depends always upon the trace of the CP asymmetries
over flavours, $\epsilon_1$, times  a function of the trace over flavours
of the decay rate of the RH neutrino $N_1$. 
This is an unexpected result because it suggests, in particular, 
that the lepton asymmetry, say, in 
the electron lepton number 
can be washed out by inverse decays 
involving the second and/or the
third family (which erase only the muon and/or tau lepton charges).

    Let us turn now 
to the issue of the connection between the CP
violating parameters in leptogenesis  and the low energy 
CP-violating phases in the 
lepton sector, i.e., in the PMNS matrix $\pmns$.
The CP asymmetry $\epsilon_1$ can be written in terms
of the diagonal light and heavy Majorana 
neutrino mass matrices introduced earlier,
$m={\rm Diag}(m_1,m_2,m_3)$ and
$M={\rm Diag}(M_1,M_2,M_3)$,  
and the orthogonal complex matrix $R$: 
\begin{eqnarray}
\epsilon_{1}&=& -\frac{3 M_1}{16\pi v^2}\, \frac{{\rm Im}\left(
\sum_{l \beta \rho} 
m_\beta^{1/2}m_\rho^{3/2} 
U^*_{l \beta}U_{l \rho}
R_{1\beta }R_{1\rho}\right)}{\sum_\beta m_\beta\left|R_{1\beta}\right|^2}
\nonumber\\
&=&-\frac{3 M_1}{16\pi v^2}\, \frac{{\rm Im}\left(
\sum_{\rho} m_\rho^{2} 
R^2_{1 \rho }\right)}{\sum_\beta m_\beta\left|R_{1\beta}\right|^2}\,,
\end{eqnarray}
%
\noindent where we have summed over all lepton 
flavours $l=e,\mu,\tau$. 
In particular, if the elements of the matrix $R$ satisfy 
CP-invariance condition (\ref{RCPinv1}), i.e., 
if $R_{1 \rho }$, $\rho=1,2,3$, 
is {\it real or purely imaginary},
the leptogenesis CP-asymmetry $\epsilon_{1} = 0$. 
In the top-down approach, working  in  the 
basis in which the matrix of charged lepton Yukawa 
couplings and the RH neutrino Majorana 
mass matrix are diagonal, the matrix of neutrino
Yukawa couplings can be written as 
$\lambda=V_R^\dagger{\rm Diag}(\lambda_1,
\lambda_2,\lambda_3)V_L$ and  the 
low energy leptonic phases can arise from the
phases in $V_L$, i.e. in the left-handed (LH) sector,
in $V_R$, i.e. in the RH sector, or from both $V_L$ and $V_R$.
However, from $\lambda\lambda^\dagger=
V_R^\dagger \, {\rm Diag}(\lambda_1^2,\lambda_2^2,\lambda_3^2) \, V_R=
M^{1/2}R\,  m\,  R^\dagger M^{1/2}/v^2$, one
sees that the phases of $R$ are related 
to those of $V_R$. This means that
in the ``one-flavour'' approximation a 
future possible observation of CP-violating
phases at low energies in the neutrino sector 
does not imply the existence
of a baryon asymmetry. Indeed, low
energy CP-violating phases might stem  
entirely from the LH sector, i.e. $V_L$, 
and hence be irrelevant for leptogenesis 
which would be  driven by the phases in $R$, i.e. of the RH sector. 

%
\subsection{Taking the Flavour Effects 
into Account}
\noindent
%
\noindent
As we mentioned in the Introduction, the  `one-flavour' approximation
rigorously holds only when the interactions mediated by the charged lepton
Yukawas are out of equilibrium, that is at $T\sim M_1\gsim 10^{12}$ GeV.
In this regime, flavours are indistinguishable and one may indeed
perform a rotation is flavour space to store all the
asymmetry in  a single flavour.
At smaller temperatures, though, this operation is not 
possible. The $\tau$ ($\mu$) lepton doublet is  
a distinguishable mass eigenstate
for $T\sim M_1\lsim 10^{12}\,(10^9)$ GeV. 
Flavoured Boltzmann equations
may be written down for the asymmetry $Y_{\Delta_l}$ in each flavour 
$\Delta_l=B/3-L_l$ \cite{davidsonetal,nardietal,davidsonetal2}. For the
case of hierarchical RH neutrinos, they read 
\footnote{The equations
of motion should include the terms accounting for the quantum oscillations
among the different flavours \cite{davidsonetal}. They become  
relevant only  in the
proximity of the transition between the 
one-single flavour and the two-flavour
state when the $\tau$ flavour becomes distinguishable.}
\begin{eqnarray}
\label{1a}
\!\!\!\!\!\!\!\!\!\!\!\frac{\mathrm{d} Y_{N_1}}{\mathrm{d} z} \!\!&=&\!\! 
  \frac{-z}{s H(M_1)} (\gamma_D + \gamma_{\mathrm{S},\Delta L = 1})
\left(  \frac{Y_{N_1}}{Y^\mathrm{eq}_{N_1}} - 1  \right)\!,\\
\!\!\!\!\!\!\!\!\!\!\!\frac{\mathrm{d} Y_{\Delta_l}}{\mathrm{d} z} 
\!\!&=&\!\! 
 \frac{-z}{s H(M_1)}  \!\left[
\epsilon_{l}   (\gamma_D + \gamma_{\mathrm{S},\Delta L = 1}) 
\left(  \frac{Y_{N_1}}{Y^\mathrm{eq}_{N_1}} - 1  \right) -
\left(\frac{\gamma^{l}_D}{2} + 
\gamma^{l}_{\mathrm{W},\Delta L = 1}\right)\, 
\frac{\sum_\beta A_{lm} Y_{\Delta_m}}{Y^\mathrm{eq}_\ell}
\right]\!,
\label{2a}
\end{eqnarray} 
%
where $z = M_1/T$, $T$ being the temperature, and
there is no sum over $l$ in the last term in
the right-side of Eq.~(\ref{2a}). In the
above equation
$Y_{N_1}$ is the density of the 
lightest right-handed neutrino $N_1$ with mass $M_1$.  
$Y_{\Delta_l}$ are defined as $Y_{\Delta_l}\equiv Y_B/3 - 
Y_{L_l}$, where $Y_{L_l}$ are the total 
lepton number densities for the flavours 
$l = e,\mu,\tau$ and $Y_B$ is the total baryon density. 
One has to solve the Boltzmann equations for $Y_{\Delta_l}$ 
instead of for the number densities $Y_{l}$ of the lepton doublets 
$\ell_l$, since $\left(\Delta_l\equiv B/3 - L_l\right)$ is conserved 
by sphalerons and by the other SM interactions. 
Furthermore, all number densities $Y$, normalization 
to the entropy density $s$ is understood and 
$Y^\mathrm{eq}_{N_1}$ and $Y^\mathrm{eq}_\ell$ 
stand for the corresponding equilibrium number densities;  
$\gamma_D$ is the thermally averaged total 
decay rate of $N_1$ and 
 $\gamma_{\mathrm{S},\Delta L = 1}$ 
represents the rates for the $\Delta L = 1$ 
scattering processes in the thermal bath. 
Notice, in particular, that 
$\gamma_{\mathrm{S},\Delta L = 1}$ contributes 
to the asymmetry, as was recently pointed out in 
\cite{davidsonetal2}. 
The corresponding flavour-dependent rates for wash-out 
processes involving 
the lepton flavour $l$ are $\gamma^{l}_D$ 
(from inverse decays involving 
leptons $\ell_l$) and $\gamma^{l}_{\mathrm{W},\Delta L = 1}$, 
while $\Delta L = 2$ scatterings may be safely neglected. 
The matrix $A$, which appears in the wash-out term connects
the asymmetries in the lepton doublets to the asymmetries in the 
charges $\Delta_l$  by $Y_l = \sum_m A_{lm} \,Y_{\Delta_m}$. 
The values of its elements depend on which interactions, in addition to the 
weak and strong sphalerons, are in thermal equilibrium at the temperatures 
where leptogenesis takes place. 
Below $10^9$ GeV in the SM, $A$ is given by \cite{davidsonetal2}
\begin{eqnarray}
A=\left(
\begin{array}{ccc}
-151/179 & 20/179 & 20/179 \\
25/358 & -344/537 & 14/537 \\
25/358 & 14/537 & -344/537
\end{array}\right)\, .
\end{eqnarray} 
%
In what regards the charged leptons, 
in the SM only the interaction mediated by 
the $\tau$ Yukawa coupling is in equilibrium  
between $10^9$ and $10^{12}$ GeV, 
and the lepton asymmetries and $B/3 - L_l$ 
asymmetries in the $e$ and $\mu$ flavour can be 
combined to $Y_2 \equiv Y_{e + \mu}$ and $Y_{\Delta_2} \equiv 
Y_{\Delta_e+\Delta_\mu}$. In this temperature range, $A$ is given by 
\cite{davidsonetal2}
\begin{eqnarray}
A=\left(
\begin{array}{cc}
-920/589 & 120/589  \\
30/589 & -390/589 
\end{array}
\right)\, .
\end{eqnarray} 
Above $10^{12}$ GeV, all asymmetries can be combined 
to $Y_{B-L}$, and $A$ is given by $A = - 1$.

The asymmetry in each flavour is given by
\begin{equation}
\epsilon_{l} = -\frac{3 M_1}{16\pi v^2}\, \frac{{\rm Im}\left(
\sum_{\beta\rho} 
m_\beta^{1/2}m_\rho^{3/2} U^*_{l \beta}U_{l \rho}
R_{1\beta}R_{1\rho}\right)}{\sum_\beta m_\beta\left|R_{1\beta}\right|^2}\,,~~
l=e,\mu,\tau\,.
\label{epsa1}
\end{equation}
%
Obviously, the trace over the flavours of  
$\epsilon_{l}$ coincides with $\epsilon_1$. 
It should be clear from Eq.~(\ref{epsa1}) 
that we can have 
$\epsilon_{l} \neq 0$
even if, e.g., $R$ is a real 
matrix and $R\neq {\bf 1}$. If, however, 
$R$ is a diagonal matrix, e.g., if 
$R = {\bf 1}$, all three lepton number asymmetries
vanish: 
$\epsilon_{l} = 0$, $l=e,\mu,\tau$.

Similarly, one has to define a ``wash-out mass 
parameter'' for each flavour  $l$  \cite{davidsonetal,davidsonetal2}:
\begin{eqnarray}
\left(
\frac{\widetilde{m_l}}{3\times 10^{-3}\,{\rm eV}}\right)
&\equiv& \frac{\Gamma(N_1\rightarrow
H\, l)}{H(M_1)}\,,\nonumber\\
\widetilde{m_l} \equiv \frac{|\lambda_{1l}|^2\,v^2}{M_{1}}&=&
\left|\sum_{k}R_{1k}m_k^{1/2}U_{lk}^*\right|^2
\, , ~~~l=e,\mu,\tau\,.
\label{tildmal1}
\end{eqnarray}
%
The quantity $\widetilde{m_l}$ parametrizes 
the decay rate of $N_1$ to the 
leptons of flavour $l$. The trace  
$\sum_l \widetilde{m_l}$ 
coincides with the $\widetilde{m_1}$ 
parameter defined in the previous section.

%
What is more relevant is that the total baryon asymmetry
is the sum of each individual lepton asymmetry. In the rest of the 
paper we will be concerned with  temperatures
$(10^9\lsim T\sim M_1\lsim 10^{12})$ GeV. In  this range 
 only the interactions
mediated by the $\tau$ Yukawa coupling are in equilibrium and the 
final baryon asymmetry is well approximated by  \cite{{davidsonetal2}}
\begin{equation}
\label{interm}
Y_B\simeq -\frac{12}{37 g_*}\left(\epsilon_2\,\eta\left(
\frac{417}{589}\,\widetilde{m_2}\right)+
\epsilon_\tau\,\eta\left(\frac{390}{589}
\,\widetilde{m_\tau}\right)\right)\, ,
\label{BAUM1large}
\end{equation}
%
where $\epsilon_2 =\epsilon_{e}+\epsilon_{\mu}$, 
$\widetilde{m_2}=\widetilde{m_e}+
\widetilde{m_\mu}$ and
\begin{equation}
\eta\left(\widetilde{m_l}\right)\simeq \left(
\left(\frac{\widetilde{m_l}}{8.25\times
10^{-3}\,{\rm eV}}\right)^{-1}+
\left(\frac{0.2\times
10^{-3}\,{\rm eV}}{\widetilde{m_l}}
\right)^{-1.16}\
\right)^{-1}\, .
\label{eta1}
\end{equation}
Notice that the wash-out
masses $\widetilde{m}_2$ and $\widetilde{m_\tau}$ in Eq.~(\ref{BAUM1large}) 
are multiplied by
some numerical coefficients which account for the dynamics involving
the lepton doublet asymmetries and the asymmetries stored in the
charges $\Delta_l=(1/3)B-L_l$ \cite{davidsonetal2}. 

From the generic expression for the baryon asymmetry, 
we deduce that the CP asymmetry in each
flavour is weighted by the corresponding wash-out parameter. Therefore, 
the total baryon number is
generically not proportional  to $\epsilon_1$.

The dependence of $\epsilon_l$ 
on the PMNS matrix elements is such that  
nonvanishing low energy leptonic CP-violating 
phases imply, in the context of
leptogenesis and barring accidental cancellations,  a nonvanishing  
baryon asymmetry. 
In the general case of complex matrix $R$, the low energy
phases in the neutrino mixing matrix $\pmns$ could stem 
from the phases in the left-handed sector, 
in the RH sector, or  in both sectors. 
This result only follows when the lepton flavour effects 
are correctly taken into account in the Boltzmann 
equations. In previous analyses
of leptogenesis ignoring the flavour effects, 
the observation of low energy CP violation
did not automatically imply the existence of a baryon asymmetry,
since the possibility existed that the low energy phases 
could stem exclusively from the left-handed sector 
and hence be irrelevant for leptogenesis. 

This conclusion is by 
itself already quite
interesting. 
However, we can go even further. 
Let us consider, for instance,  the case
in which the CP parities of the heavy and light Majorana neutrinos
are such that $\rho_i^N=\rho^\nu_j=1$ for all $i,j=1,2,3$. In such a case,
CP invariance  corresponds to having all the elements of the matrix $R$ real,
see Eq.~(\ref{RCPinv1}) and $\delta=\alpha_{21}=\alpha_{31}=0$ 
(mod $2\pi$)\footnote{Alternatively, one may consider the case in which 
the CP parities of the heavy and light neutrinos are such that CP 
invariance is realized for the elements of the matrix $R$ purely imaginary
and $\delta=0$, $\alpha_{21}=\alpha_{31}=\pi$ 
(mod $2\pi$).}. In the top-down approach, a real  
matrix $R$ would  correspond to the class
of models where CP is an exact symmetry in the
RH neutrino sector. The reason for this 
can be more easily understood working in the basis where the 
charged lepton Yukawa coupling and the 
RH mass matrix are diagonal, so that 
the neutrino Yukawa matrix is the only
coupling in the leptonic Lagrangian that violates CP. 
More specifically, since the neutrino Yukawa coupling
can be written in its singular value decomposition, 
$\lambda= V^{\dagger}_R {\rm Diag}(\lambda_1,\lambda_2,\lambda_3) V_L$, 
CP violation in the RH neutrino sector 
is encoded in the  phases in $V_R$, that can be extracted from
diagonalizing the combination $\lambda \lambda^{\dagger}=  V^{\dagger}_R
{\rm Diag}(\lambda^2_1,\lambda^2_2,\lambda^2_3) V_R$. On the other
hand, as we have seen, 
using the parametrization of the Yukawa coupling (\ref{R}), 
this same combination of matrices 
can be written as 
 $\lambda \lambda^{\dagger}=M^{1/2} R m R^{\dagger}M^{1/2}/v^2$.
Comparing the two expressions it is apparent that $R$ is real
if and only if $V_R$ is real, i.e. when there is no CP violation
in the RH sector. It
has been recently pointed out that the case of a real $R$ matrix 
is naturally realized within  the class of models based on sequential
dominance \cite{sd}. 
In the case of $R$ real,  the flavour CP asymmetries and the baryon asymmetry
depend exclusively on the phases of the left-handed sector, that are
in turn uniquely determined by the low energy phases. Consequently, 
for real matrix $R \neq {\bf 1}$, 
the leptogenesis mechanism 
is directly connected to the 
low energy CP-violating phases in $U_{\rm PMNS}$. 
This connection is more apparent
from the expression of the flavour CP asymmetries in the
parametrization (\ref{R}):
\begin{eqnarray}
\label{epsa2}
\epsilon_{l}&=& - \frac{3M_1}{16\pi v^2}
\frac{{\rm Im}\left(\sum_\beta\sqrt{m_\beta}R_{1\beta} U^*_{l\beta}
\sum_\rho \sqrt{m^3_\rho}R_{1\rho} U_{l \rho}\right)}
{\sum_\gamma m_\gamma |R_{1\gamma}|^2} \nonumber \\
&=& -\frac{3M_1}{16\pi v^2} 
\frac{\sum_\beta \sum_{\rho>\beta}\sqrt{m_\beta m_\rho}
(m_\rho-m_\beta)R_{1\beta}R_{1\rho}
{\rm Im}\,\left(U^*_{l\beta} U_{l \rho}\right)}{
\sum_\gamma m_\gamma |R_{1\gamma}|^2}\,.
\label{epsa3}
\end{eqnarray}
%
If $R_{1\beta}R_{1\rho}$ is purely imaginary,
$R_{1\beta}R_{1\rho} = \pm i\,|R_{1\beta}R_{1\rho}|$,
we would get $\pm (m_\rho+m_\beta)|R_{1\beta}R_{1\rho}|
{\rm Re}(U^*_{l\beta} U_{l \rho})$ instead of
$(m_\rho-m_\beta)R_{1\beta}R_{1\rho}
{\rm Im}(U^*_{l\beta} U_{l \rho})$
in Eq.~(\ref{epsa3}). Purely imaginary 
$R_{1\beta}R_{1\rho}$  and real 
$(U^*_{l\beta} U_{l \rho})$ implies violation of 
CP-invariance. In order for the CP-symmetry to be broken 
at low energies, we should have both
${\rm Re}(U^*_{l\beta} U_{l \rho})\neq 0$ and
${\rm Im}(U^*_{l\beta} U_{l \rho})\neq 0$.

   In the next Section we will study in greater 
detail the possibility that
the baryon asymmetry stems only from the low energy measurable
CP violating phases in the neutrino mixing matrix $\pmns$.

%
\section{Baryon Asymmetry from Low Energy CP-Violating 
Dirac and Majorana Phases in $U_{\rm PMNS}$: 
RH Neutrinos with Hierarchical Mass Spectrum}
%
\noindent
In what follows we shall assume that 
the matrix $R$ has real and/or purely imaginary 
elements and that the heavy Majorana
neutrinos possess a hierarchical mass spectrum,
$M_1 \ll M_2 \ll M_3$, with $M_1$ having a value 
in the interval of interest, 
$10^{9}~{\rm GeV} \ltap M_1 \ltap 10^{12}$ GeV.
We will investigate the case when the RG running of
$m_j$ and of the parameters in $\pmns$ from $M_Z$ to $M_1$ is
relatively small and can be neglected. This possibility is realised 
(in the class of theories under discussion) for sufficiently small values
of the lightest neutrino mass  ${\rm min}(m_j)$~\cite{rad}, e.g., 
for ${\rm  min}(m_j) \ltap 0.05$ eV. The latter condition is fulfilled
for the normal hierarchical (NH) and 
inverted hierarchical (IH) light neutrino mass spectrum.
Under the indicated condition
$m_j$, and correspondingly $\deltaatm$ and $\deltasol$,
and $U$ in Eqs.~(\ref{epsa3}) and Eqs.~(\ref{tildmal1})  can be
taken at the scale $\sim M_Z$, at which the neutrino 
mixing parameters are measured.

  Taking into account that in the case under discussion
$\epsilon_2 =\epsilon_{e}+\epsilon_{\mu} = -\epsilon_{\tau}$, 
we can cast Eq.~(\ref{BAUM1large}) 
in a somewhat more convenient form:
\begin{equation}
\label{interm1}
Y_B=-\frac{12}{37}~\frac{\epsilon_{\tau}}{g_*}\,
\left(\eta\left(\frac{390}{589}\widetilde{m_{\tau}}\right) - 
\eta\left(\frac{417}{589}\widetilde{m_2}\right) \right)\,,
\label{BAUM11large}
\end{equation}
%
where $\widetilde{m_2} =\widetilde{m_e} + \widetilde{m_\mu}$ and 
$\widetilde{m_l}$ and $\eta(\widetilde{m_l})$
are given in Eqs. (\ref{tildmal1}) and (\ref{eta1}).

%
\subsection{\label{sec:NHNH}
\large{Normal Hierarchical Light Neutrino Mass Spectrum}}
%
\indent\
Given the inequalities $m_1 \ll m_2 \ll m_3$, 
we will assume 
that the terms $\propto \sqrt{m_1}$ 
give sub-leading contributions
to the lepton flavour asymmetries 
$\epsilon_{l}$, 
Eq.~(\ref{epsa3}), and to the mass parameters
$\widetilde{m_l}$ related to the wash-out
effects, and we neglect them with respect to those
$\propto \sqrt{m_{2,3}}$ 
giving the dominant contribution.
This requires that 
$\sqrt{m_1}|R_{11}| \ll 
\sqrt{m_2}|R_{12}|,\sqrt{m_3}|R_{13}|$.
In the indicated approximation, we 
find using Eq.~(\ref{epsa3})
and the fact that $m_2 \cong \sqrt{\deltasol}$,
$m_3 \cong \sqrt{\deltaatm}$:
\begin{eqnarray} 
\epsilon_{l} &\simeq& - 
~\frac{3M_1\sqrt{\deltaatm}}{16\pi v^2}
\left(1 - \frac{\sqrt{\deltasol}}{\sqrt{\deltaatm}}\right)\;
\left (\frac{\deltasol}{\deltaatm} \right )^{\frac{1}{4}}\,
\frac{\left | R_{12}R_{13} \right | }
{\left (\frac{\deltasol}{\deltaatm} \right )^{\frac{1}{2}}\,|R_{12}|^2 
+ |R_{13}|^2}\, \nonumber \\
& & \times  
\left [ {\rm Im}\,\left (e^{i\beta_{23}}\, U^*_{l 2}U_{l 3}\right )
+ \frac{2\sqrt{\deltasol}}{\sqrt{\deltaatm} - \sqrt{\deltasol}}\,
{\rm Im}\,\left (e^{i\beta_{23}}\,
{\rm Re}\left ( U^*_{l 2}U_{l 3}\right ) \right ) \right ]\, \\ 
&=& - 
~\frac{3M_1\sqrt{\deltaatm}}{16\pi v^2}
\left (\frac{\deltasol}{\deltaatm} \right )^{\frac{1}{4}}\,
\frac{\left | R_{12}R_{13} \right | }
{\left (\frac{\deltasol}{\deltaatm} \right )^{\frac{1}{2}}\,|R_{12}|^2 
+ |R_{13}|^2}\, \nonumber \\
& & \times 
{\rm Im}\left [ \!
\left(\! 1\!  - \! \frac{\sqrt{\deltasol}}{\sqrt{\deltaatm}}\right)
e^{i(\beta_{23} + \frac{\pi}{2})}
{\rm Im}\left ( U^*_{l 2}U_{l 3}\right )
+ \left(\! 1 \! + \frac{\sqrt{\deltasol}}{\sqrt{\deltaatm}}\right)
e^{i\beta_{23}}{\rm Re}\left ( U^*_{l 2}U_{l 3} \right ) \!
\right ] \! ,
\label{epsl1}
\end{eqnarray}
%
where $\beta_{23}\equiv \widetilde{\beta_{12}} + 
\widetilde{\beta_{13}} \equiv
 {\rm arg}(R_{12}R_{13})$, $\widetilde{\beta_{1j}} \equiv
{\rm arg}(R_{1j})$.
The phase $\beta_{23}$ parametrises 
the effect of CP-violation due to the matrix $R$
in the asymmetry $\epsilon_{l}$.
In the case of CP-invariance, $\beta_{23} = 0$ or $\pi/2$
\footnote{More precisely, in the case of CP-invariance
we have $\beta_{23} = \pi q$ or $(2q + 1)\pi/2 $,
$q=0,1,2,...$.}
depending on whether ${\rm Im}(U^*_{l 2}U_{l 3})=0$
or ${\rm Re}(U^*_{l 2}U_{l 3}) =0$, respectively,
and $\epsilon_{l} = 0$, $l=e,\mu,\tau$.

   From Eq.~(\ref{eq:Upara}), it is 
straightforward to obtain 
${\rm Im}(e^{i\beta_{23}}U^*_{l 2}U_{l 3})$ and
${\rm Im}(e^{i\beta_{23}}{\rm Re}( U^*_{l 2}U_{l 3}))$.
The expressions, e.g., for ${\rm Im}(e^{i\beta_{23}}U^*_{l 2}U_{l 3})$
read:
\begin{eqnarray}
\label{Ime2e3}
{\rm Im} \left(e^{i\beta_{23}}\,U^*_{e 2}U_{e 3}\right) &=& 
- s_{12}c_{13}s_{13}\sin \left (\delta - 
\left(\frac{\alpha_{32}}{2} + \beta_{23}\right) \right)\,,\\
{\rm Im} \left(e^{i\beta_{23}}\,U^*_{\mu 2}U_{\mu 3}\right) &=& 
- c_{13}\left [- c_{23}s_{23}c_{12}
\sin \! \left(\frac{\alpha_{32}}{2} + \beta_{23}\right)
- s^2_{23}s_{12}s_{13}\sin  \! \left (\delta \!-\! 
\left(\frac{\alpha_{32}}{2} + \beta_{23}\right)\right) \right ]\,,\nonumber
\\
&&
\label{Imm2m3}\\
{\rm Im}\left(e^{i\beta_{23}}\,U^*_{\tau 2}U_{\tau 3}\right) &=& 
- c_{13}\left [ c_{23}s_{23}c_{12}\sin
\left(\frac{\alpha_{32}}{2} + \beta_{23}\right)
- c^2_{23}s_{12}s_{13}\sin \! \left (\delta \!-\! 
\left(\frac{\alpha_{32}}{2} + \beta_{23}\right)\right) \right ]\,,\nonumber\\
&& \label{Imt2t3}
\end{eqnarray}
%
where $\alpha_{32} \equiv \alpha_{31} - \alpha_{21}$.
Thus, $\epsilon_{l}$
depend on the same Majorana phase difference
(or phase)  the effective Majorana mass 
in $\betabeta$-decay, $\meff$, depends (see Eq.~(\ref{meffNH2})) on. 
It follows from Eq.~(\ref{epsl1}) 
that, as could be expected, for real or purely imaginary
$R_{12}$ and $R_{13}$ we have 
$\epsilon_{e}+ \epsilon_{\mu} + \epsilon_{\tau} = 0$.

   We are interested in the case in which the 
CP-violating phases in $\pmns$ play the role
of leptogenesis CP-violating parameters.
It follows from the preceding discussions that 
one has to consider two cases:
i) $\beta_{23} = 0$ (or more generally, 
$\beta_{23} = \pi q$, $q=0,1,2,...$), and  
ii) $\beta_{23} = \pi/2$  
(or more generally, 
$\beta_{23} = (2q + 1)\pi/2$, $q=0,1,2,...$).
We have to remember
that if ${\rm Im}(U^*_{l 2}U_{l 3})\neq 0$ 
(${\rm Re}(U^*_{l 2}U_{l 3}) \neq 0$)
but ${\rm Re}(U^*_{l 2}U_{l 3}) = 0$
(${\rm Im}(U^*_{l 2}U_{l 3}) = 0$), the case of
$\beta_{23} = 0,\pi$ ($\beta_{23} = \pi/2,3\pi/2$) 
{\it corresponds to violation of the
CP-symmetry by the matrix $R$}.

  Consider first the possibility of $\beta_{23} = 0~(\pi)$.
It is quite remarkable that, 
as it follows from Eqs.~(\ref{epsl1}) 
and (\ref{Imt2t3}), for $\beta_{23} = 0~(\pi)$ 
the asymmetry $\epsilon_{\tau}$ depends 
on the rephasing invariant $S_2$,
Eq.~(\ref{eq:SCP}): we have 
$\epsilon_{\tau} \propto 
{\rm Im}\left (U^*_{\tau 2}U_{\tau 3}\right )\equiv S_2$.
The requirement of a nonzero asymmetry, 
$\epsilon_{\tau} \neq 0$,
implies $S_2 \neq 0$, and correspondingly
$\alpha_{32} \neq 2\pi k$
and/or $(\delta - \alpha_{32}/2)\neq \pi k'$, 
$k,k'=0,1,...$.
In order to reproduce the observed value
of the baryon asymmetry, $|\sin(\alpha_{32}/2)|$  
and/or $|\sin\theta_{13}\sin(\delta - \alpha_{32}/2)|$ 
have to be sufficiently large (see further). 
Since in the case under study 
$\beta_{23} = 0~{\rm or}~\pi$, the sign of 
$\epsilon_{\tau}$ is not uniquely determined 
by the sign of 
${\rm Im}\left (U^*_{\tau 2}U_{\tau 3}\right )$.

 In the alternative possibility of $\beta_{23} = \pi/2~(3\pi/2)$ 
we have $|\epsilon_{\tau}| \propto 
|{\rm Re}(U^*_{\tau 2}U_{\tau 3})|$.
Now  $\epsilon_{\tau} \neq 0$ provided
$\alpha_{32} \neq \pi(2k + 1)$
and/or $(\delta - \alpha_{32}/2)\neq (\pi/2)(2k' + 1)$, 
$k,k'=0,1,...$.
In this case $|\cos(\alpha_{32}/2)|$  
and/or $|\cos\theta_{13}\cos(\delta - \alpha_{32}/2)|$ 
have to be sufficiently large in order for leptogenesis
to be successful. The sign of 
$\epsilon_{\tau}$ is again not uniquely determined 
by the sign of ${\rm Re}\left (U^*_{\tau 2}U_{\tau 3}\right )$.

   The maximal value of 
$|{\rm Im}(e^{i\beta_{23}}U^*_{\tau 2}U_{\tau 3})|$ 
and of $|\epsilon_{\tau}|$, 
is reached for $\beta_{23} = 0;~\pi$
($\beta_{23} = \pi/2;~3\pi/2$) at
\footnote{Note that for $\beta_{23} = \pi/2;~3\pi/2$
we have $|{\rm Im}(e^{i\beta_{23}}U^*_{\tau 2}U_{\tau 3})| =
|{\rm Re}(U^*_{\tau 2}U_{\tau 3})|$, 
and the maximum of 
$|{\rm Im}(e^{i\beta_{23}}U^*_{\tau 2}U_{\tau 3})|$
coincides with the maximum of 
$|{\rm Re}(U^*_{\tau 2}U_{\tau 3})|$.}
$\alpha_{32} = \pi (2k + 1)$ 
and $\delta = 2\pi k$
($\alpha_{32} = 2\pi k$ 
and $\delta = \pi(2k +1)$),
$k=0,1,...$:
\beq
{\rm max}\left | {\rm Im}\,\left (e^{i\beta_{23}}\,
U^*_{\tau 2}U_{\tau 3}\right )\right |
= c_{23}c_{13}~\left ( s_{23}c_{12} + c_{23}s_{12}s_{13} \right )
\ltap 0.47\,,~~
\beta_{23} = \frac{\pi}{2}\,q\,,q=0,1,2,...,
\label{maxImb23t2t3}
\eeq
%
where we have used the best fit values of
$\sin^22\theta_{23}$ and $\sin^2\theta_{12}$ 
given in Eq.~(\ref{th12th23}) and the upper limit 
$s_{13} \leq 0.2$ (see Eq.~(\ref{th13})).
For $s_{13} = 0$ 
we get 
$|{\rm Im}(e^{i\beta_{23}}U^*_{\tau 2}U_{\tau 3})| 
\ltap 0.42~|\sin(\alpha_{32}/2)| \leq 0.42$
while if $\alpha_{32} = 0$, one obtains 
$|{\rm Im}(e^{i\beta_{23}}U^*_{\tau 2}U_{\tau 3})|
\cong 0.27~|s_{13}\sin\delta| \ltap 0.054$.
Thus, the effect of the term 
$\propto |s_{13}\sin\delta|$ in $|S_2|$, 
and correspondingly in $|\epsilon_{\tau}|$,
can be significant only if 
$|\sin(\alpha_{32}/2)| \ltap 0.20$.
Note that for $\beta_{23} = 0;~\pi$
($\beta_{23} = \pi/2;~3\pi/2$) 
${\rm max}~(|{\rm Im}(e^{i\beta_{23}}
U^*_{\tau 2}U_{\tau 3})|)$
corresponds to 
${\rm Re}(U^*_{l 2}U_{l 3}) = 0$
(${\rm Im}(U^*_{l 2}U_{l 3}) = 0$), $l=e,\mu,\tau$,
i.e., to CP-conserving values of $\alpha_{32}$ and $\delta$.
Nevertheless, in this case we still have 
$|\epsilon_{\tau}| \neq 0$ as a consequence of the
violation of the CP-symmetry by the matrix $R$.

 We turn next to the dependence on the parameters in $R$.
  Taking into account that $R$ is, in general, a complex 
orthogonal matrix, we get for the elements of $R$ of interest,
$R_{1j}$, $j=1,2,3$, obeying the CP-invariance constraints:
\beq
\left ( |R_{11}|^2\, \rho^{\nu}_{1} + |R_{12}|^2\, \rho^{\nu}_{2} 
+ |R_{13}|^2\, \rho^{\nu}_{3} \right )\,\rho^{N}_{1} = 1\,.
\label{CPinvR1j}    
\eeq
%
The CP-asymmetry $|\epsilon_{\tau}|$ is proportional to the factor $r$:
\beq
r \equiv \frac{|R_{12}R_{13}|}
{\left (\frac{\deltasol}{\deltaatm} \right )^{\frac{1}{2}}\,|R_{12}|^2 
+ |R_{13}|^2}\,.
\label{r}
\eeq
%
It is clear that as long as $|R_{12}| \sim |R_{13}|$,
$r$ will not act as a suppression factor for the
asymmetry $|\epsilon_{\tau}|$; the latter can 
be suppressed if, for instance, $|R_{12}| \ll |R_{13}|$
or  $|R_{13}| \ll (\deltasol/\deltaatm)^{\frac{1}{4}}|R_{12}|$. 

We now focus on the wash-out effects. 
The mass parameters 
$\widetilde{m_l}$
related to the wash-out effects for 
the three lepton number asymmetries read:
\begin{equation}
\widetilde{m_l} \cong 
\sqrt{\deltaatm} \left | 
\left (\frac{\deltasol}{\deltaatm} \right )^{\frac{1}{4}}\,
R_{12}\,U^*_{l 2} + R_{13}\,U^*_{l 3}\right |^2\,,~~l=e,\mu,\tau~.
\label{tmal1}
\end{equation}
%
Using the unitarity of the PMNS matrix $U$ we obtain: 
\begin{equation}
\widetilde{m_e} + \widetilde{m_\mu} + \widetilde{m_{\tau}} =
\sqrt{\deltaatm} \left [ 
\left (\frac{\deltasol}{\deltaatm} \right )^{\frac{1}{2}}\,
|R_{12}|^2 + |R_{13}|^2 \right ]\,.
\label{sumtmal1}
\end{equation}
%
Since $\widetilde{m_l} \geq 0$, $l=e,\mu,\tau$,
each of the individual wash-out mass parameters
is limited from above by the expression in the
right-hand side of Eq.~(\ref{sumtmal1}).

Except for strong fine tuning, the contribution due to $\sin \theta_{13}$ 
is subdominant and can be neglected at first order. Therefore,
for simplicity, we take $\sin \theta_{13} =0$ in the following analysis
and we get:
\begin{eqnarray}
\label{m2NH}
\widetilde{m_2}  \simeq  & \hspace{-2truecm} \sqrt{\deltaatm} 
\Big( \left(\frac{\deltasol}{\deltaatm}\right)^{\frac{1}{2}}
|R_{12}|^2 (1 - c_{12}^2 s_{23}^2 ) + |R_{13}|^2  s_{23}^2 \\ \nonumber
 & + \ 2 \ \left( \frac{\deltasol}{\deltaatm}  \right)^{\frac{1}{4}} 
|R_{12} R_{13} | \ s_{23} c_{23}
c_{12} \cos ( (\widetilde{\beta_{13}} - 
\widetilde{\beta_{12}}  - \ds\frac{\alpha_{32}}{2} )\Big) ~, \\
\label{mtNH}
\widetilde{m_\tau} \simeq &  \hspace{-2.5truecm} \sqrt{\deltaatm}  \Big( \left(\frac{\deltasol}{\deltaatm}\right)^{\frac{1}{2}}
|R_{12}|^2 c_{12}^2 s_{23}^2  + |R_{13}|^2 c_{23}^2 \\ \nonumber
 & - 2 \left( \frac{\deltasol}{\deltaatm}  \right)^{\frac{1}{4}}  |R_{12} R_{13} | \ s_{23} c_{23}
c_{12} \cos ( (\widetilde{\beta_{13}} - \widetilde{\beta_{12}}  - 
\ds\frac{\alpha_{32}}{2} ) \Big) ~.
\end{eqnarray}
%

 It follows from Eq.~(\ref{BAUM1large}) that 
the baryon asymmetry $|Y_B|$ can be zero if 
$|\eta(0.66\widetilde{m_\tau}) - \eta(0.71\widetilde{m_2})| = 0$,
although $|\epsilon_{\tau}| \neq 0$ and 
$|\epsilon_{e} + \epsilon_{\mu}| \neq 0$.
This corresponds to the physical 
case when the asymmetries generated 
in the lepton doublet charges $\tau$
and $(e + \mu)$ are equal in 
magnitude, but have opposite signs.
Such a possibility can occur if the
following relation holds:
\begin{equation}
\label{etaeq}
\eta\left( \frac{390}{589} \widetilde{m_\tau} \right) = \eta \left(
\frac{417}{589} \widetilde{m_2} \right)~.
\end{equation}
%
One solution is given by:
\beq
\widetilde{m_2} = \frac{390}{417}\,\widetilde{m_{\tau}}\,
\cong 0.935\,\widetilde{m_{\tau}}\,.
\label{YB0MajCP0}
\eeq
%
For fixed $|R_{12}|^2$ and 
$|R_{13}|^2$, Eqs. (\ref{sumtmal1}) and  (\ref{YB0MajCP0}) 
determine the value of $\widetilde{m_{\tau}}$
for which we can have $|Y_B| = 0$:
\begin{equation}
\widetilde{m_{\tau}} =
\frac{\sqrt{\deltaatm}}{1 + 0.935}\, 
 \left [ 
\left (\frac{\deltasol}{\deltaatm} \right )^{\frac{1}{2}}\,
|R_{12}|^2 + |R_{13}|^2 \right ]\,.
\label{tmalYB0}
\end{equation}
%
Since $\widetilde{m_{\tau}}$ is calculated 
from Eq.~(\ref{tmal1}), given 
the neutrino oscillation parameters
and $|R_{12}|$ and $|R_{13}|$, 
the requirement that the value of 
$\widetilde{m_{\tau}}$ obtained from 
Eq.~(\ref{tmal1}) coincides with 
that determined by Eq.~(\ref{tmalYB0}) 
leads to a combined constraint on 
the quantities $|U_{\tau 2}|^2$ and 
$2\,{\rm Re}(R_{12}U^*_{\tau 2}R^*_{13}\,U_{\tau 3})$
which depend on the phases $\delta$  and $\alpha_{32}$.
The phase factors associated with the latter
vary only between $(-1)$ and 1.
Therefore it is not guaranteed that 
condition (\ref{YB0MajCP0})
can be satisfied and we can have 
$|Y_B| = 0$ with $|\epsilon_{\tau}| \neq 0$ and 
$|\epsilon_{e} + \epsilon_{\mu}| \neq 0$
for any possible 
values of $R_{12}$ and $R_{13}$.  
Taking for simplicity $(\widetilde{\beta_{13}} - \widetilde{\beta_{12}} 
- \alpha_{32}/2) = \pi/2$, a solution is given by:
\begin{equation}
|R_{12}|^2 = 0.94 |R_{13}|^2 \frac{\cos 2 \theta_{23} - 0.069 s_{23}^2}{1 - 2.069 c_{12}^2 s_{23}^2}~,
\end{equation}
which holds for $2.069 \,  c_{12}^2 s_{23}^2 \neq 1$ and 
$(\cos2\theta_{23} - 0.069s_{23}^2)/(1 - 2.069 c_{12}^2 s_{23}^2)\geq 0$. 
For example, if $c_{12}^2 =0.69$,
it can be satisfied for $s_{23}^2 \ltap 0.48$.

   A general solution of Eq.~(\ref{etaeq}) can also 
be found and is given by:
\begin{equation}
\widetilde{m_2}  \, \widetilde{m_\tau} \simeq \frac{589^2}{390 \ 417} \, 10^{-6} ~\mathrm{eV}^2
\left( \frac{y-1}{y^{1.16} -1} y^{0.08} \right)^{\frac{2}{2.16}}
\label{yb0NH}
\end{equation}
%
where $y \equiv (417/390) \widetilde{m_2}/\widetilde{m_\tau}$.
If, for instance, $\widetilde{\beta_{13}} - \widetilde{\beta_{12}} 
- \alpha_{32}/2 = \pi/2$ and
$|R_{13}|^2$ dominates in
$\widetilde{m_{\tau}}$ and $\widetilde{m_{2}}$, 
$y$ does not depend on the parameters in $R$
and we can solve Eq.~(\ref{yb0NH}):
\begin{equation}
|R_{13}| \simeq \left( \frac{10^{-6} \ \mathrm{eV}^2}{\deltaatm} 
\frac{1}{c^2_{23} s^2_{23}} 
\frac{589}{390}\, \frac{589}{417}
\Big( \frac{y-1}{y^{1.16} -1} y^{0.08} \Big)^{\frac{2}{2.26}} \right)^{\frac{1}{4}}
\simeq 0.25 ~.
\end{equation}
%
 
 We will analyse next the dependence of the baryon 
asymmetry on the parameters in $R$.
For simplicity, we take the values of $\alpha_{32}$ and of
$\widetilde{\beta_{12}},\widetilde{\beta_{13}} = 0, \pi/2$, 
which maximize the CP-asymmetry in Eq.~(\ref{epsl1}).
The wash-out parameters read:
\begin{eqnarray}
\label{m2NH1}
\widetilde{m_2} & \simeq & \sqrt{\deltaatm} \,  \Big( \sqrt{\frac{\deltasol}{\deltaatm}} 
|R_{12}|^2 (1 - c_{12}^2 s_{23}^2 ) + |R_{13}|^2 s_{23}^2 \Big) ~, \\
\label{mtNH1}
\widetilde{m_\tau} & \simeq & \sqrt{\deltaatm}  \,  \Big( \sqrt{\frac{\deltasol}{\deltaatm}} 
|R_{12}|^2 c_{12}^2 s_{23}^2  + |R_{13}|^2 c_{23}^2 \Big) ~.
\end{eqnarray}
%

{\em Case A: Strong wash-out}. 
This case is realized if $\widetilde{m_{2, \tau}} \gg 2 \times 10^{-3} ~$eV. 
The latter condition is satisfied for 
$|R_{12}|^2 \gg 
2 \times 10^{-3} \mathrm{eV}/
(\sqrt{\deltasol} c^2_{12} s^2_{23}) \cong 0.64$, and/or
for  
$|R_{13}|^2 \gg 2\times 10^{-3}~\mathrm{eV} / 
(\sqrt{\deltaatm} c^2_{23}) \cong 0.08$.
The baryon asymmetry can be approximated as:
\begin{equation}
|Y_B| \sim {\cal C} \frac{|R_{12}| |R_{13}|} { 
\Big(\frac{\deltasol}{\deltaatm}\Big)^{\frac{1}{2}} |R_{12}|^2 + |R_{13}|^2} 
\left| \Big(\frac{589}{390} \frac{2 \times 10^{-4}~\mathrm{eV}}{
\widetilde{m_\tau}} \Big)^{1.16} \!\!\! - \Big( \frac{589}{417} \  
\frac{2 \times 10^{-4} \ \mathrm{eV}}{\widetilde{m_2}} \Big)^{1.16} \right|~.
\label{ybSW}
\end{equation}
%
The constant ${\cal C}$ is defined as 
\begin{equation}
{\cal C} \equiv 
\frac{9 M_1}{74 \pi g_\ast v^2} \, \sqrt{\deltaatm}  \,
\Big( \frac{\deltasol}{\deltaatm} \Big)^{\frac{1}{4}}\,  
c_{23}\,s_{23}\, c_{12} \simeq 
1.1 \times 10^{-8} \, \left ( \frac{M_1}{10^{11}~\mathrm{GeV}} \right)\,, 
\end{equation}
%
where we have used the present best fit values of the 
neutrino oscillation parameters.
If the $|R_{13}|^2$ term in the wash-out factors dominates, 
i.e. if $|R_{12}|^2 / |R_{13}|^2  \ll  
\Big(\deltaatm / \deltasol \Big)^{1/2} s_{23}^2/ (1-c_{12}^2 s_{23}^2)
\cong 4.3$, the dependence on the $R$ parameters 
becomes:
\begin{equation}
|Y_B| \sim 2.7 \times 10^{-3}  \ {\cal C} \ 
\frac{|R_{12}| } {|R_{13}|^{3.32} c_{23}^{2.32}} 
 \left| 1 -
\Big(  \frac{390}{417} \ \frac{c_{23}^2}{ s_{23}^2} \Big)^{1.16} \right|  ~.
\end{equation}
%
For maximal atmospheric neutrino mixing there is a strong cancellation 
and the resulting baryon asymmetry 
has an additional suppression factor $\sim 0.075$:
\begin{equation}
|Y_B| \sim  4.5 \times 10^{-4} \  {\cal C} \ 
\frac{|R_{12}| } {|R_{13}|^{3.32} }\,.
\end{equation}
%
Taking into account the dependence of $\epsilon_\tau$ on 
$\sin 2 \theta_{23}$,
we find that for $\sin^2 \theta_{23} = 0.34 \ (0.66)$
$ |Y_B|$ is larger by a factor of 
9 \ (11). Notice that the asymmetry changes 
sign when $\sin^2 \theta_{23}$ increases 
from 0.34  to 0.66. Thus, there is  also 
a value of $\sin^2\theta_{23}$ in the interval
[0.34,0.66], for which $Y_B = 0$ (up to corrections
$\sim s_{13}$). The asymmetry decreases very rapidly 
with $|R_{13}|$ and therefore the maximal asymmetry 
should correspond to relatively small $|R_{13}|$.

In the alternative case when $|R_{12}|^2$ dominates 
in $\widetilde{m_{\tau, 2}}$, which is realised for
$|R_{12}|^2 \gg 7.9 \  |R_{13}|^2$, 
the asymmetry $|Y_B|$ is proportional to:
\begin{equation}
|Y_B| \sim 2.7 \times 10^{-3}  \  {\cal C} \  
\Big( \frac{\deltaatm}{\deltasol}\Big)^{1.08} 
\frac{|R_{13}| } {|R_{12}|^{3.32} }
\left| \Big( c_{12}^2 s_{23}^2  \Big)^{-1.16} -
\Big(  \frac{417}{390} (1- c_{12}^2 s_{23}^2)  \Big)^{-1.16} \right| ~.
\end{equation}
%
In this case, $|Y_B|$ is not 
suppressed
for maximal atmospheric neutrino 
mixing. Varying the solar and atmospheric 
angles within their 95\%~C.L.
ranges induces a factor of a few 
in the predicted baryon asymmetry.
Also in this case the baryon asymmetry is a
decreasing function of $|R_{12}|$.

{\em Case B: weak wash-out.} For sufficiently 
small values of $|R_{12}|$ and $|R_{13}|$,
one enters the weak wash-out regime. 
The asymmetry is given approximately by:
\begin{equation}
|Y_B| \sim {\cal C} \  \frac{|R_{12}| |R_{13}|} { \Big(\protect\frac{\deltasol}{\deltaatm}\Big)^{\frac{1}{2}} |R_{12}|^2 + |R_{13}|^2} 
\left| \frac{390}{589} \, \frac{\widetilde{m_\tau}}{8.25 \times 10^{-3}~\mathrm{eV}}  -  \frac{417}{589} \,  \frac{\widetilde{m_2}}{8.25 \times 10^{-3}~\mathrm{eV}}  \right| ~.
\label{ybWW}
\end{equation}
%
If the term $\propto |R_{13}|^2$ dominates in $\widetilde{m_{\tau, 2}}$,
we can have again
a partial cancellation between the two 
leading terms in  $|Y_B|$
for maximal atmospheric 
neutrino mixing, $\theta_{23} = \pi/4$:
\begin{equation}
|Y_B| \simeq 4.0 \ {\cal C} \  |R_{12}| |R_{13}| \left| \cos {2 \theta_{23}} - \frac{27}{390} s_{23}^2 \right|
\simeq 0.12 \ {\cal C} \, |R_{12}| |R_{13}|~.
\end{equation}
%
The asymmetry $|Y_B|$ is bigger 
by approximately a factor of 10 if the atmospheric 
mixing angle assumes the values $\sin^2 \theta_{23} = 0.34,~0.66$.
The asymmetry has opposite sign for these two values,
indicating that $|Y_B|$ goes through zero
(up to corrections $\sim s_{13}$) 
when  $\sin^2 \theta_{23}$ is varied 
from 0.34 to 0.66.

  If the term $\propto |R_{12}|^2$ is the dominant one in
$\widetilde{m_{\tau, 2}}$, one finds:
\begin{equation}
|Y_B| \sim 4.0 \ {\cal C } \, \frac{\deltasol}{\deltaatm}  |R_{12}| |R_{13}| \left( \frac{417}{390} + \frac{27}{390} c_{12}^2 s_{23}^2 \right)
\simeq 0.14 \, {\cal C} \,  |R_{12}| |R_{13}| ~.
\end{equation}
We notice that in the case of weak wash-out one finds
that the asymmetry increases with the values of $|R_{12}|$
or $|R_{13}|$.

On the basis of the above 
discussion, we can conclude
that more than one local maximum of $|Y_B|$ is present 
and that the absolute one
is found, in general, in a regime
which interpolates between the weak and strong wash-out
regimes. However, 
it is possible to show that in this region 
$Y_B$ can be even zero if Eq.~(\ref{etaeq}) is satisfied.
%
%
\subsection{The case of $N_3$ decoupling}
%
%
   We will assume further for simplicity that
$|R_{11}|^2 \ll ||R_{12}|^2 \pm |R_{13}|^2|$.
This corresponds to the case of 
``decoupling'' of the heaviest 
Majorana neutrino $N_3$ 
\cite{IR041,JE2RHN04,PRST05}. 
For $\beta_{23} = 0;~\pi$ 
($\beta_{23} = \pi/2;~3\pi/2)$, 
which implies 
$\rho^{\nu}_{2}\rho^{\nu}_{3}=1~(-1)$,
we get:
\beq
|R_{12}|^2~ ^{~+}_{(-)}|R_{13}|^2 = \rho^{\nu}_{2}\,\rho^{N}_{1}\,.
\label{CPinvR1213}   
\eeq
%
Obviously, if $\beta_{23} = 0;~\pi$ we have 
$|R_{12}|^2+|R_{13}|^2 =1$, while
for $\beta_{23} = \pi/2;~3\pi/2$,
one finds $|R_{12}|^2 - |R_{13}|^2 = \pm 1$. 
It is not difficult to find the maximal value of $r$
for $\rho^{\nu}_{2}\,\rho^{N}_{1} = 1$:
\beq
\label{b230pi}
{\rm max}\,r = \frac{1}{2} 
\left (\frac{\deltaatm}{\deltasol} \right )^{\frac{1}{4}} \cong 1.2\,,~~
\rho^{\nu}_{2}\,\rho^{N}_{1} = 1\,,~~
\beta_{23} = \frac{\pi}{2} k\,, k=0,1,2,... 
\label{maxr1}
\eeq 
%
The maximum is reached when
$\beta_{23} = 0;~\pi$ at 
$|R_{12}|^2 = (1 +  
\sqrt{\deltasol}/\sqrt{\deltaatm})^{-1} \cong 0.85$,
while for $\beta_{23} = \pi/2;~3\pi/2$ it corresponds to
$|R_{12}|^2 = (1 -  
\sqrt{\deltasol}/\sqrt{\deltaatm})^{-1} \cong 1.22$.
In the latter case $|R_{12}|^2 - |R_{13}|^2 = 1$ and
$|R_{12}|$, $|R_{13}|$, in general, 
are not limited from above
\footnote{Note that if  $|R_{12}|^2 + |R_{13}|^2 = 1$,
one can obviously use the parametrisation
$R_{12} = \cos\omega$, $R_{13} = \sin\omega$,
while in the case of $|R_{12}|^2 - |R_{13}|^2 = 1$,
we have $R_{12} = \cos(i\omega) = \cosh \omega$, 
$R_{13} = \sin(i\omega) = i\sinh\omega$,
$\omega$ being a real parameter.
}.
However, it is not difficult to convince oneself
that we can have successful leptogenesis
only if $|R_{12}|$ and $|R_{13}|$ are not
exceedingly large, namely for  
$|R_{12}|,|R_{13}| \ltap 10$.
Indeed, barring accidental cancellations,
the wash-out mass parameters
$\widetilde{m_l}$, $l=e,\mu,\tau$,
increase with the increasing of 
$|R_{12}|$ ($|R_{13}|^2 = |R_{12}|^2 - 1$),
and for $|R_{12}|\sim |R_{13}| \sim 10$,
one would typically have 
$\widetilde{m_l} \sim 1$ eV.
The corresponding efficiency factors 
would be exceedingly small and
$|\eta\left(\frac{390}{589}\widetilde{m_{\tau}}\right) - 
\eta\left(\frac{417}{589}\widetilde{m_2}\right)| \sim 10^{-4}$
(extremely strong wash-out regime),
which makes it impossible to reproduce the
observed value of the baryon asymmetry
for $M_1 \ltap 10^{12}~{\rm GeV}$.
The preceding rather qualitative analysis
can obviously be refined to obtain more 
precise upper limits on $|R_{12}|$ and 
$|R_{13}|$ satisfying $|R_{12}|^2 - |R_{13}|^2 = 1$. 

   If $\beta_{23} = \pi/2;~3\pi/2$ and 
$\rho^{\nu}_{2}\rho^{N}_{1} = - 1$,
one always has
$r < 1$. In this case 
$|R_{12}|^2 < |R_{13}|^2$.
For $|R_{12}| \ll  |R_{13}|$
we have $r \ll 1$ and the asymmetry
$|\epsilon_{\tau}|$ will be suppressed
by the $r$-factor, Eq.~(\ref{r}).
We will not analise this case further.

  Given the fact that 
$\sqrt{\deltaatm} \cong 0.05$ eV,
$\sqrt{\deltasol}/\sqrt{\deltaatm} \cong 0.18$,
$r \ltap 1.2$, Eq.~(\ref{maxr1}), 
and $|{\rm Im}(U^*_{\tau 2}U_{\tau 3})| \ltap 0.47$,
we obtain for the maximal value 
of the asymmetry $|\epsilon_{\tau}|$
in the cases of interest
($|R_{12}|^2~ ^{~+}_{(-)}|R_{13}|^2 = 1$,
$\beta_{23} = (\pi/2)k,~k=0,1,2,...$):
\begin{eqnarray}
|\epsilon_{\tau}| & \leq & 
~\frac{3M_1\sqrt{\deltaatm}}{32\pi v^2}\,
\left(1 ^{~-~}_{(+)} \frac{\sqrt{\deltasol}}{\sqrt{\deltaatm}}\right)\;
\left | {\rm Im}\,\left (e^{i\beta_{23}}\, 
U^*_{l 2}U_{l 3}\right )\right |\, \\
& \ltap &  0.19~(0.27)\, \frac{3M_1\sqrt{\deltaatm}}{16\pi v^2}
\simeq 1.9~(2.7)\times 10^{-8}
\left (\frac{\sqrt{\deltaatm}}{0.05~{\rm eV}}\right )
\left (\frac{M_1}{10^{9}~{\rm GeV}}\right )\,,
\label{maxepstau1}
\end{eqnarray}
%
where the minus (plus) sign and the value 0.19 (0.27) correspond 
to $\beta_{23} = \pi k$, $k=0,1,2,...$
($\beta_{23} = (\pi/2)(2k + 1)$, $k=0,1,2,...$).

 Let us note that, although 
for the values of the parameters we have used 
the asymmetry $|\epsilon_{\tau}|$
has a maximal value, the baryon asymmetry
$|Y_B|$ does not necessarily 
has a maximum since the maximum of $|\epsilon_{\tau}|$
does not correspond, in general, to a maximum
of the effective wash-out factor
$|\eta(0.66\widetilde{m_\tau}) - \eta(0.71\widetilde{m_2})|$ 
in the expression for $|Y_B|$
(see Eq.~(\ref{BAUM1large})).
As can be shown and Figs. \ref{MaxYBDiracMajplus.eps} 
and \ref{MaxYBDiracMajminus.eps} 
illustrate, however, 
for  real $R_{12}$ and $R_{13}$
($\beta_{23} = 0;~\pi$)
satisfying $|R_{12}|^2 + |R_{13}|^2 = 1$, 
the maximum of $|\epsilon_{\tau}|$  
with respect to the parameter $R_{12}$
practically coincides for $\delta = 0$
($\delta = \pi k$, $k=0,1,...$) 
and CP-violation due only to the Majorana 
phase $\alpha_{32}$,  with the maximum of
$|Y_B|$. In the case of 
CP-violation generated only by the 
Dirac phase $\delta$ ($\alpha_{32}= 0$),
the maximum of $|Y_B|$
occurs for $\beta_{23} = 0$
at slightly smaller value of 
$|R_{12}|$, namely at 
$|R_{12}| \cong 0.86$, 
compared to the value of 
$|R_{12}| \cong 0.92$ at which the 
maximum of $|\epsilon_{\tau}|$ takes place.
We will work for convenience
with $R_{12}$ and $R_{13}$ 
maximising $|\epsilon_{\tau}|$,
for which we have simple analytic 
expressions in terms of the 
ratio $\deltasol/\deltaatm$.
In most of the cases we will discuss
these values of $R_{12}$ and $R_{13}$
maximise also $|Y_B|$.

 Consider next the wash-out parameters
$\widetilde{m_l}$, $l=e,\mu,\tau$.
In the case of real $R_{12}$ and $R_{13}$
we have $\beta_{23} = 0;~\pi$
and $|R_{12}|^2 + |R_{13}|^2 = 1$. 
Taking into account this constraint we get from
Eq.~(\ref{sumtmal1}) \cite{PRST05,IR041}:
\begin{equation}
\sqrt{\deltasol} \leq
\widetilde{m_e} + \widetilde{m_\mu} + \widetilde{m_{\tau}} 
\leq \sqrt{\deltaatm}\,. 
\label{sumtmal2}
\end{equation}
%
The maximum and the minimum 
are reached respectively for
$|R_{12}|^2 = 0$ and $|R_{13}|^2 = 0$, and 
in both cases $|\epsilon_{\tau}| = 0$. 
If $\beta_{23} = \pi/2;~3\pi/2$,
we have $|R_{12}|^2 - |R_{13}|^2 = 1$
and for $\widetilde{m_e} + \widetilde{m_\mu} + \widetilde{m_{\tau}}$
we get the same lower bound as in Eq.~(\ref{sumtmal2}), 
but no upper bound, in general, since $|R_{12}|^2$ is not 
limited from above. The requirement of 
successful leptogenesis leads 
to an upper bound roughly of the order
of 1 eV. For $|R_{12}|^2 = (1~ ^{~+}_{(-)}
\sqrt{\deltasol}/\sqrt{\deltaatm})^{-1}$,
corresponding to the maximum of $r$ 
and of $|\epsilon_{\tau}|$,  we find
\begin{equation}
\widetilde{m_e} + \widetilde{m_\mu} + \widetilde{m_{\tau}} = 
2\,\sqrt{\deltasol} \left [1~ ^{~+}_{(-)}
\left ( \frac{\deltasol}{\deltaatm}\right )^{\frac{1}{2}}\right ]^{-1}
\cong 0.30\,(0.44)\, \sqrt{\deltaatm}\,.
\label{sumtmal3}
\end{equation}
%
 
 In what follows we will analyse 
the case of real $R_{12}$ and $R_{13}$.
For real $R_{12}$ and $R_{13}$ we have $\beta_{23} = 0;~\pi$
corresponding to 
${\rm sgn}(R_{12}R_{13}) = +1;~(-1)$.
%
%
\subsubsection{\label{sec:NHNHMajCP}
\large{Leptogenesis due to Majorana CP-Violation in $U_{\rm PMNS}$ }}
%
\indent\    
  We shall consider first the interesting 
possibility of CP-symmetry being violated 
by the Majorana phase $\alpha_{32}$
in $U$, 
and not by the Dirac phase $\delta$.
To be concrete, we choose
$\delta = 0$
and {\it real} $R_{12}$ and $R_{13}$ 
($\beta_{23} = 0;~\pi$) which maximise $r$, 
$|\epsilon_{\tau}|$ and $|Y_B|$, i.e., 
$|R_{12}|^2 = (1 +  \sqrt{\deltasol}/\sqrt{\deltaatm})^{-1} \cong 0.85$,
$|R_{13}|^2 = 1 -  |R_{12}|^2 \cong 0.15$.
For $\alpha_{32} = 0;~2\pi$, the CP-symmetry is not violated.
If $\alpha_{32}$ takes the CP-conserving value 
of $\alpha_{32} = \pi$, CP-symmetry is violated by $R$ 
and $|\epsilon_{\tau}|\neq 0$. 
It should also be noted that the terms 
$\propto \sin\theta_{13}$ in 
the expressions for  
$\epsilon_{\tau}$ and the wash-out 
mass parameters $\widetilde{m_l}$
are sub-dominant in the case being studied.

  With the choices for $\delta$,   
$R_{12}$ and $R_{13}$ 
made we get using Eqs. (\ref{eq:Upara}), 
(\ref{Imt2t3}) and (\ref{tmal1}):
\beq
\label{Imt2t3MajCPV}
\left | {\rm Im}\,\left (e^{i\beta_{23}}U^*_{\tau 2}U_{\tau 3}\right )\right | 
= c_{23}\, c_{13}\,\left (s_{23}c_{12} + c_{23}s_{12}s_{13}\right )\,
\left | \sin \frac{\alpha_{32}}{2} \right |
\cong 
\frac{1}{2}\, (c_{12} + s_{12}s_{13})\,
\left | \sin \frac{\alpha_{32}}{2} \right |\,,
\eeq
\begin{eqnarray}
\widetilde{m_{\tau}} 
&=& 
\sqrt{\deltaatm} \left | 
\left (\frac{\deltasol}{\deltaatm} \right )^{\frac{1}{4}}\,
|R_{12}|\,\left (s_{23}c_{12} + c_{23}s_{12}s_{13}\right )\,
e^{i\frac{\alpha_{32}}{2}}
 - \kappa \, |R_{13}|\,c_{23}
\right |^2\,  
\label{tmtauMajCP1}\\
&\cong& 
\frac{1}{2}\, \frac{\sqrt{\deltaatm}}
{1+ \left ( \frac{\deltaatm}{\deltasol} \right )^{\frac{1}{2}}}\,
\left [ 1 + (c_{12} + s_{12}s_{13})\,
(c_{12} + s_{12}s_{13} - 2\, \kappa \, \cos\frac{\alpha_{32}}{2})\right ]\,,
\label{tmtauMajCP2}\\
\widetilde{m_{2}} &\equiv& \widetilde{m_{e}} + \widetilde{m_{\mu}}
=  \frac{2\,\sqrt{\deltaatm}}
{1+ \left ( \frac{\deltaatm}{\deltasol} \right )^{\frac{1}{2}}}
- \widetilde{m_{\tau}}\,,
\label{tm2MajCP1}
\end{eqnarray}
%
where $\kappa \equiv e^{i\beta_{23}} = 
{\rm sgn}(R_{12}R_{13}) = \pm 1$
and we have set $s_{23}=c_{23}= 1/\sqrt{2}$,
used $|R_{12}|$ and $|R_{13}|$ specified 
above and neglected the terms $\propto s^2_{13}$ 
in Eqs. (\ref{Imt2t3MajCPV}) and (\ref{tmtauMajCP2}).

   It follows from Eqs. (\ref{epsl1}), 
(\ref{Imt2t3MajCPV}) and (\ref{tmtauMajCP1}) 
that in the case under discussion the following 
relations hold: 
$|\epsilon_{\tau}(\alpha_{32})| = 
|\epsilon_{\tau}(2\pi - \alpha_{32})|$, and
$\widetilde{m_{\tau}}(\alpha_{32},\beta_{23}=0) = 
\widetilde{m_{\tau}}(2\pi - \alpha_{32},\beta_{23}=\pi)$.
As a consequence we have from Eq.~(\ref{BAUM1large}):
$|Y_B(\alpha_{32},\beta_{23}=0)| =
|Y_B(2\pi - \alpha_{32},\beta_{23}=\pi)|$.
Thus, we will consider 
values of $\alpha_{32}$ in the interval
$[0,2\pi]$ and limit our discussion to 
$\beta_{23}=0$ (i.e., $\kappa \equiv {\rm sgn}(R_{12}R_{13}) = 1$).
After fixing the value of $\beta_{23}$
the only free parameter 
left in the problem being studied is 
the Majorana phase $\alpha_{32}$.

  For $\alpha_{32} = 0;~2\pi$, obviously 
$|\epsilon_{\tau}| = 0$ and therefore $|Y_B| = 0$.
We can have $|Y_B| = 0$ also even if 
$|\epsilon_{\tau}| \neq 0$, provided 
the efficiency factor in the expression for 
$|Y_B|$, Eq.~(\ref{BAUM1large}), 
is zero, i.e., if 
the condition given in Eq.~(\ref{YB0MajCP0}) is fulfilled.
 For the values of $|R_{12}|$ and $|R_{13}|$ 
considered and $\alpha_{32}$ having a value 
in the interval $[0,2\pi]$, this condition is 
satisfied for $\alpha_{32} \cong 1.2\pi$. 
Thus, in this case we have 
$|\epsilon_{\tau}| \neq 0$ but $|Y_B| = 0$
for $\alpha_{32} \cong 1.2\pi$.

  The absolute maximum of the 
baryon asymmetry $|Y_B|$
as a function of $\alpha_{32}$ is reached 
close to $\alpha_{32} \cong \pi/2$.  
We have at the absolute maximum 
for $s_{13} = 0~(0.2)$:
\begin{eqnarray}
|\epsilon_{\tau}| 
&\cong & 1.2\times 10^{-8}
\left (\frac{\sqrt{\deltaatm}}{0.05~{\rm eV}}\right )
\left (\frac{M_1}{10^{9}~{\rm GeV}}\right )\,,
\label{maxepstMajCP}\\
|Y_B|
&\cong & 2.0\,(2.2)\times 10^{-12}
\left (\frac{\sqrt{\deltaatm}}{0.05~{\rm eV}}\right )
\left (\frac{M_1}{10^{9}~{\rm GeV}}\right )\,.
\label{maxYBMajCP}
\end{eqnarray}
%
As we have already indicated,
the values of $|R_{12}|$ and $|R_{13}|$
we have chosen in this analysis 
maximise not only $\epsilon_{\tau}$,
but also $|Y_B|$.
Thus, the observed baryon asymmetry 
having a value in the interval
$8.0\times 10^{-11} \ltap |Y_B| \ltap 9.2\times 10^{-11}$,
can be reproduced for 
$M_1 \gtap 3.6\times 10^{10}$ GeV.
A second (local) maximum of $|Y_B|$ occurs 
close to $\alpha_{32} = 3\pi/2$
(at $\alpha_{32} \cong 1.7\pi$). 
At this maximum the value of
$|Y_B|$ for $s_{13} = 0~(0.2)$ 
is approximately by a factor of 3 
(1.8) smaller than that at the absolute 
maximum  in Eq.~(\ref{maxYBMajCP}).
Given the fact that the flavour effects 
of interest can be substantial for values of 
$M_1$ up to $M_1 \sim 10^{12}$ GeV,
leptogenesis can be successful even 
for rather small values of the Majorana phase 
$\alpha_{32}$, or more precisely, 
of $|\sin(\alpha_{32}/2)|$. 
If, for instance, $M_1 \sim 5\times 10^{11}$ GeV,
the observed baryon asymmetry can be generated
for $|\sin(\alpha_{32}/2)| \cong 0.15$.

  The results obtained in this subsection 
are illustrated in Figs. 
\ref{MaxYBDiracMajplus.eps}
and \ref{MaxYBDiracMajminus.eps} we have
already discussed, and in 
Fig. \ref{Fig4alp02pis130200plus.eps} 
where the dependence of the baryon asymmetry $|Y_B|$ 
on the Majorana phase $\alpha_{32}$ is shown 
for $M_1 \sim 5\times 10^{10}$ GeV
in the case analised above: $\delta = 0$, 
real $R_{12}$ and $R_{13}$ which maximise $|Y_B|$, 
i.e., $|R_{12}| \cong 0.92$ and $|R_{13}| \cong 0.39$,
and $\beta_{23} = 0$, i.e.,
$\kappa \equiv {\rm sgn}(R_{12}R_{13}) = 1$ 
(Fig. \ref{Fig4alp02pis130200plus.eps}), and
$\beta_{23} = \pi$, i.e., $\kappa = - 1$ 
(Fig. \ref{Fig4balp02piminus.eps}).
The figures have been obtained using the 
best fit values of the oscillation parameters 
$\deltaatm$, $\deltasol$, $\sin^2\theta_{23}$ 
and $\sin^2\theta_{12}$. Results for two values of  
$\sin\theta_{13}$ are presented:
$\sin\theta_{13} = 0;~0.20$.
Figure \ref{Fig4alp02pis130200plus.eps} shows, 
in particular, that in the case being studied, 
the predicted baryon asymmetry $|Y_B|$ 
exhibits very weak dependence on 
$\sin\theta_{13}$ for $\alpha_{32} \ltap \pi$.
If $\sin\theta_{13}$ has a value close to the 
existing upper limit, the effect of $\sin\theta_{13} \sim 0.2$
can be noticeable in the region of the local maximum
of $|Y_B|$ at $\alpha_{32} \cong 1.7\pi$:
it can lead to an increase of $|Y_B|$ by 
a factor of 1.7. For $M_1 = 10^{11}$ GeV, 
for instance, we can have successful leptogenesis for 
$\pi/4 \ltap \alpha_{32} \ltap 0.9\pi$, and, if 
$s_{13} \sim 0.2$, also for 
$1.4\pi \ltap \alpha_{32} \ltap 1.9\pi$.
We note that, as Figs. \ref{MaxYBDiracMajplus.eps}
and \ref{MaxYBDiracMajminus.eps} show, 
that the predicted value of 
$|Y_B|$ exhibits a
relatively strong dependence on 
the elements of the matrix $R$.

 It follows from the results obtained in the 
present subsection that as long as 
$|\sin(\alpha_{32}/2)|$ is 
not exceedingly small
\footnote{Obviously, 
$\alpha_{32}$ should have a value 
sufficiently different from the
``special'' one for which
$|\eta(0.66\widetilde{m_\tau}) - 
\eta(0.71\widetilde{m_2})| = 0$.
}
and $M_1 \gtap 3.5\times 10^{10}$ GeV,
we can have successful leptogenesis 
even if $|s_{13}\sin\delta| = 0$
($J_{\rm CP} = 0$) and the only 
CP-violating parameter is the 
low energy Majorana phase $\alpha_{32}$.
%
%
\subsubsection{\label{sec:NHNHMajCP1}
\large{Dirac CP-Violation in $U_{\rm PMNS}$ and Leptogenesis}}
%
\indent\    
  The next question we would like to address is
under what conditions we could have a successful 
leptogenesis if the only CP-violating parameter   
is the Dirac phase in $\pmns$, i.e., 
if, e.g., $\beta_{23} = 0, \pi$,
the Majorana phase $\alpha_{32}$ 
takes a CP-conserving value and
$\sin(\alpha_{32}/2) = 0$
\footnote{
By imposing the condition
$\sin(\alpha_{32}/2) = 0$ 
we exclude the possibility of 
$\alpha_{32}$ taking the 
CP-conserving values 
$\alpha_{32} = \pi, 3\pi$,
in which case the term 
associated with the violation
of CP-symmetry due to the matrix
$R$ will dominate in
$|{\rm Im}(U^*_{\tau 2}U_{\tau 3})|$,
and correspondingly in
$|\epsilon_{\tau}|$.
If $\beta_{23} = \pi/2, 3\pi/2$,
the indicated possibility would be
avoided if $\alpha_{32}$ takes a
CP-conserving value and
$|\sin(\alpha_{32}/2)| = 1$.
},
so that $\alpha_{32} = 2\pi k$, $k=0,1,...$.
We choose for concreteness again 
{\it real} $R_{12}$ and $R_{13}$ 
which for $\alpha_{32} = 0~(2\pi)$
and $\beta_{23} = \pi~(0)$ maximise $|\epsilon_{\tau}|$
and $|Y_B|$, i.e., $|R_{12}|^2 = (1 +  
\sqrt{\deltasol}/\sqrt{\deltaatm})^{-1} \cong 0.85$,
$|R_{13}|^2 = 1 -  |R_{12}|^2 \cong 0.15$.
We will present results for
the baryon asymmetry $|Y_B|$
also for the values of 
$R_{12}$ and $R_{13}$ 
which maximise $|Y_B|$ in the case of 
$\alpha_{32} = 0~(2\pi)$
and $\beta_{23} = 0~(\pi)$,  
$|R_{12}|^2 = 0.75$ and $|R_{13}|^2 = 0.25$.
For $\delta = \pi q$, $q=0,1,2,...$, 
the CP-symmetry is not violated
and  $|\epsilon_{\tau}| = 0$.

  For the chosen CP-conserving values of 
$\alpha_{23}$ and $\beta_{23}$ 
we get from Eqs. (\ref{eq:Upara}), (\ref{Imt2t3}) 
and (\ref{tmal1}):
\beq
\label{Imt2t3DCPV}
\left | {\rm Im}\,\left (e^{i\beta_{23}}U^*_{\tau 2}U_{\tau 3}\right )\right | 
= c_{23}^2\, s_{12}\, s_{13}\, \left | \sin \delta \right |\,,
\eeq
\begin{eqnarray}
\widetilde{m_{\tau}} 
&=& 
\sqrt{\deltaatm} \left | 
\left (\frac{\deltasol}{\deltaatm} \right )^{\frac{1}{4}}\,
|R_{12}|\,\left (s_{23}c_{12} + c_{23}s_{12}s_{13}\, e^{i\delta}\right )\,
- \kappa' \, |R_{13}|\,c_{23}\,c_{13}
\right |^2\, 
\nonumber\\
&\cong& 
\frac{\sqrt{\deltaatm}}
{1+\left ( \frac{\deltaatm}{\deltasol} \right )^{\frac{1}{2}}}
\left | (c_{23}c_{13} - \kappa'\, s_{23}c_{12}) 
+ c_{23}s_{12}s_{13}\, e^{i\delta}\right |^2, 
~\kappa' \equiv e^{i(\beta_{23} + \frac{\alpha_{32}}{2})} = \pm 1\,, 
\label{tmtauDCP2}
\end{eqnarray}
%
where in Eq.~(\ref{tmtauDCP2})
we have used the values of
$|R_{12}|$ and $|R_{13}|$ specified earlier. 
Taking into account 
that $|{\rm Im}(e^{i\beta_{23}}U^*_{\tau 2}U_{\tau 3})| \cong 0.027 
|s_{13}\sin\delta|$,
we find from Eq.~(\ref{maxepstau1}):
\beq
|\epsilon_{\tau}|  \cong 2.2\times 10^{-9}\,
\left (\frac{|s_{13}\sin\delta|}{0.20}\right )
\left (\frac{\sqrt{\deltaatm}}{0.05~{\rm eV}}\right )
\left (\frac{M_1}{10^{9}~{\rm GeV}}\right )\,.
\label{maxepstau2} 
\eeq
%
Thus, the maximal asymmetry $|\epsilon_{\tau}|$ 
in the case of CP-violation due
only to the Dirac phase $\delta$ in $U_{\rm PMNS}$
is approximately by a factor of 6 smaller 
than the maximal asymmetry due to 
violation of the CP-symmetry 
by the Majorana phase
$\alpha_{32}$ of $U_{\rm PMNS}$.

   Given $\widetilde{m_{\tau}}$, one can
determine $\widetilde{m_{2}}$ from 
Eq.~(\ref{tm2MajCP1}). It follows from Eqs. 
(\ref{tmtauDCP2}) and (\ref{maxepstau2}) that 
$|Y_B(\delta)| = |Y_B(2\pi - \delta)|$. 
One has to analyze 
the cases of $\kappa' \equiv e^{i(\alpha_{32}/2 + \beta_{23})}
= + 1$ ($\alpha_{32}/2 + \beta_{23}  = 2\pi k$, $k=0,1,...$) 
and $\kappa' = - 1$ ($\alpha_{32}/2 + \beta_{23} = 
\pi (2k + 1)$, $k=0,1,2,...$) separately.  

  For $\kappa' = - 1$ we find
$\widetilde{m_{\tau}} \cong 0.25\sqrt{\deltaatm} >
\widetilde{m_2}\cong 0.05\sqrt{\deltaatm}$.
For the values of the parameters 
employed in this analysis
(chosen, in particular, 
to maximise $|\epsilon_{\tau}|$ and $|Y_B|$),
$\widetilde{m_{\tau}}$ and
$\widetilde{m_2}$ exhibit weak 
dependence on $\sin\theta_{13}$ 
(and therefore on $\delta$),
which can be neglected. 
This implies that the maximum  
of the baryon asymmetry $|Y_B|$
as a function of the Dirac phase $\delta$,
will take place at values of $\delta = (\pi/2)(2k +1)$,
$k=0,1,...$, for which  $|\epsilon_{\tau}|$
also has a maximum.
Using Eq.~(\ref{eta1}) to calculate
the relevant efficiency factors
$\eta(0.66 \widetilde{m_\tau})$ and $\eta(0.71\widetilde{m_2})$,
we get from Eq.~(\ref{BAUM1large}): 
\beq
|Y_B| \cong 2.8\times 10^{-13}\,|\sin\delta| \,
\left (\frac{s_{13}}{0.2}\right )
\left (\frac{M_1}{10^{9}~{\rm GeV}}\right )\,.
\label{YBNHDminus}
\eeq
%
The asymmetry of interest is predominantly 
in the lepton number $L_{e} + L_{\mu}$.
Thus, in order to reproduce the 
observed baryon asymmetry, taken 
to lie in the interval
$8.0\times 10^{-11} \ltap |Y_B| \ltap 9.2\times 10^{-11}$,
$s_{13}|\sin\delta|$ and $M_1$, in the case analised,
should satisfy
\beq
2.9 \ltap |\sin\delta| \,
\left (\frac{s_{13}}{0.2}\right )
\left (\frac{M_1}{10^{11}~{\rm GeV}}\right ) 
\ltap 3.3 \,. 
\label{s13M1NHD1}
\eeq
%
Given that
$s_{13}|\sin\delta|\ltap 0.2$,  
the lower bound in this inequality
can be satisfied only for
$M_1 \gtap 2.9\times 10^{11}~{\rm GeV}$.
Recalling that the flavour effects in 
leptogenesis of interest are 
fully developed for  
$M_1 \ltap 5\times 10^{11}~{\rm GeV}$,
we obtain a {\it lower bound on the values of
$|s_{13}\sin\delta|$ and $s_{13}$ for which 
we can have successful leptogenesis 
in the case considered}:
\beq
|\sin\theta_{13}\, \sin\delta| \gtap 0.11\,,~~~
\sin\theta_{13} \gtap 0.11\,.
\label{s13NHD2}
\eeq
%
The lower limit (\ref{s13NHD2})  corresponds to
\beq
|J_{\rm CP}| \gtap 2.4\times 10^{-2}\,,
\label{s13NHD3}
\eeq
%
where we have used the best fit values of
$\sin2\theta_{12}$ and $\sin2\theta_{23}$. 
Values of $s_{13}$ in the range 
given in Eq.~(\ref{s13NHD2})
can be probed in the forthcoming Double 
CHOOZ  \cite{DCHOOZ} and future reactor neutrino experiments~\cite{DayaB}.
CP-violation effects with magnitude determined
by $|J_{\rm CP}|$ satisfying  (\ref{s13NHD3})
are within the sensitivity of the next generation 
of neutrino oscillation experiments, designed to search 
for CP- or T- symmetry violations in the oscillations
\cite{machines}. Actually, since in the case 
under discussion the wash-out factor
$|\eta_B| \equiv 
|\eta(0.66 \widetilde{m_\tau}) - \eta(0.71\widetilde{m_2})|$
in the expression for $|Y_B|$ practically does not depend on
$s_{13}$ and $\delta$, while both $|Y_B| \propto |s_{13}\sin\delta|$
and $|J_{\rm CP}| \propto |s_{13}\sin\delta|$,
there is a direct relation between $|Y_B|$ and 
$|J_{\rm CP}|$ for given neutrino 
oscillation parameters, 
$R_{12}$, $R_{13}$ and $M_1$:
\beq
\frac{|Y_B|}{M_1/(10^{11}~{\rm GeV})} \cong
3.0\times 10^{-8}\,|\eta_B| \, |J_{\rm CP}|
\cong 1.3\times 10^{-9}\, |J_{\rm CP}|\,,
\label{YBJCPNHDminus}
\eeq
%
where we have used the best fit values 
of the neutrino oscillation parameters,
$|R_{12}| = 0.92$, $|R_{13}| = 0.39$ and
$\kappa' = - 1$.

  For $\kappa' = + 1$, the maximum of 
$|Y_B|$ as a function of $|R_{12}|$ ($|R_{13}|$) 
takes place in the case being investigated at
$|R_{12}| = 0.86$ ($|R_{13}|= 0.51$)
and we will employ this value
of $|R_{12}|$ ($|R_{13}|$) in the remaining
part of the current analysis.  
If $\kappa' = + 1$, a strong 
cancellation between the terms in the 
round bracket in Eq.~(\ref{tmtauDCP2}) takes place
and we have typically 
$\widetilde{m_{\tau}} \ll \widetilde{m_2}$.
Indeed, for $s_{23} = c_{23} = 1/\sqrt{2}$,
$s_{12} = 0.2$ and $\delta = \pi/2$ one finds 
$\widetilde{m_{\tau}}\cong 2.5\times 10^{-2} \sqrt{\deltaatm}$ 
and $\widetilde{m_2} \cong 0.28\sqrt{\deltaatm}$.
Using Eq.~(\ref{eta1}) to calculate
the corresponding efficiency factors,
we get from Eq.~(\ref{BAUM11large}) 
for the baryon asymmetry:
\beq
|Y_B| \cong 3.6~(0.38)\times 10^{-13}\,|\sin\delta| \,
\left (\frac{M_1}{10^{9}~{\rm GeV}}\right )\,.
\label{YBNHDplus}
\eeq
%
The numerical factor in the round brackets 
corresponds to the value of  $|R_{12}| = 0.92$ 
($|R_{13}| = 0.39$) which maximises $|\epsilon_{\tau}|$, 
but does not maximise $|Y_B|$.

  The result we get for the maximal value 
of $|Y_B|$, i.e., for 
$|R_{12}| = 0.86$ ($|R_{13}|= 0.51$),
are quite similar to those we have obtained  
in the case of $\kappa' = - 1$, 
the only difference being that now 
the generated lepton asymmetry is predominantly 
in the tau lepton charge.
The observed baryon asymmetry lying in the interval
$8.0\times 10^{-11} \ltap |Y_B| \ltap 9.2\times 10^{-11}$,
can be reproduced if
$M_1 \gtap 2.2\times 10^{11}~{\rm GeV}$.
For the flavour effects to fully develop one must have
$M_1 \ltap 5\times 10^{11}~{\rm GeV}$,
which, together with the requirement of successful
leptogenesis, implies 
\beq
|\sin\theta_{13}\, \sin\delta| \gtap 0.09\,,~~~
\sin\theta_{13} \gtap 0.09\,.
\label{s13NHD4}
\eeq
%
This lower limit corresponds to
\beq
|J_{\rm CP}| \gtap 2.0\times 10^{-2}\,,
\label{s13NHD5}
\eeq
%
where we have used again the best fit values of
$\sin2\theta_{12}$ and $\sin2\theta_{23}$. 
The ranges of values of 
$\sin\theta_{13}$ and of $|J_{\rm CP}|$
we find in the case being studied 
are also within the sensitivity 
respectively of the planned $\theta_{13}$ 
reactor neutrino experiments 
\cite{DCHOOZ,DayaB} and of
the neutrino oscillation 
experiments on CP- (T-) violation
\cite{machines}.

 It is interesting to note 
that, as it follows from Eq.~(\ref{YBNHDplus}),
for $|R_{12}| = 0.92$ and 
$|R_{13}| = 0.39$, which maximise $|\epsilon_{\tau}|$ 
but do not maximise $|Y_B|$,
the predicted baryon asymmetry $|Y_B|$
is smaller by approximately one order of magnitude
in spite of the fact that the indicated values 
of $|R_{12}|$ and $|R_{13}|$ are rather close 
to the values $|R_{12}| = 0.86$, $|R_{13}|= 0.51$
which maximise $|Y_B|$. As $|R_{12}|$ increases beyond 0.86, 
$|Y_B|$ rapidly decreases, as is also clearly seen in
Fig. \ref{MaxYBDiracMajplus.eps}.
Actually, for $|R_{12}| = 0.92$ and 
$|R_{13}| = 0.39$, the 
observed baryon asymmetry
cannot be reproduced for 
$M_1 \ltap 10^{12}~{\rm GeV}$
and $\sin\theta_{13}\leq 0.20$:
we get $|Y_B|\ltap 4\times 10^{-11}$.
Figures \ref{MaxYBDiracMajplus.eps} and 
\ref{MaxYBDiracMajminus.eps} 
show also that for real $R_{12}$ and $R_{13}$,
satisfying $|R_{12}|^2 + |R_{13}|^2 = 1$,
and CP-violation generated only
by the Dirac phase in $U_{\rm PMNS}$,   
we can have successful leptogenesis 
if both $\sin\theta_{13}$ and
$|R_{12}|$ have relatively large values:
the observed baryon asymmetry
cannot be reproduced if
$|R_{12}| \ltap 0.6$ and/or
$\sin\theta_{13} \ll 0.1$. 
 
  The results obtained in the present 
subsection are illustrated in Figs. 
\ref{MaxYBDiracMajplus.eps}
and \ref{MaxYBDiracMajminus.eps},
as well in Fig. \ref{Figs678241106.eps}. 
In Fig. \ref{Figs678241106.eps}
we show the dependence of $|Y_B|$ on
the Dirac phase $\delta$ which was varied in the interval
$[0,2\pi]$, for $s_{13} = 0.1;~0.2$, 
$M_1 = 5\times 10^{11}~{\rm GeV}$
and for $\kappa' = + 1$ 
and $\kappa' = - 1$. 
The correlation between the rephasing invariant 
$J_{\rm CP}$ which controls the magnitude 
of the CP-violation effects in neutrino oscillations
and the baryon asymmetry $Y_B$ is illustrated 
in Fig. \ref{NH2YBJalpha0s1302.eps} for 
$s_{13} = 0.2$, $M_1 = 5\times 10^{11}~{\rm GeV}$
and $\kappa' = + 1$. 
Both figures are obtained 
for real  $R_{12}$ and $R_{13}$ 
which maximise $|Y_B|$.

\vspace{0.5cm}

   In conclusion of the present subsection 
we note that one can treat in a similar way the alternative 
possibility of $\beta_{23} = \pi/2~(3\pi/2)$.
In this case we have $\epsilon_{\tau} \propto 
{\rm Re}(U^*_{\tau 2}U_{\tau 3})$.
The CP symmetry will be violated at low energies 
if both ${\rm Re}(U^*_{\tau 2}U_{\tau 3})\neq 0$
and ${\rm Im}(U^*_{\tau 2}U_{\tau 3})\neq 0$.
Maximal asymmetry $|\epsilon_{\tau}|$
is obtained for 
$\cos(\alpha_{32}/2) = \pm 1$, and, if
$s_{13}$ is non-negligible, for
$\cos (\delta - \alpha_{32}/2) = \pm 1$.
These conditions are satisfied for the
CP-conserving values of $\alpha_{32}$ and 
$\delta$: $\alpha_{32} = 0;~2\pi$, 
$\delta = 0;~\pi$. The CP-symmetry
is broken by the matrix $R$ 
($\beta_{23} = \pi/2~(3\pi/2)$).
It is easy to convince oneself
that the expression for the asymmetry 
$|\epsilon_{l}|$, $l=e,\mu,\tau$, for 
$\beta_{23} = \pi/2~{\rm or}~3\pi/2$, $\alpha_{32} = 0~(2\pi)$ 
and $\delta = 0(~\pi)$, coincides 
with the expression for the same 
asymmetry in the case respectively of 
$\beta_{23} = 0~{\rm or}~\pi$,
$\alpha_{32} = \pi~(3\pi/2)$ and 
$\delta = 0~(2\pi)$.
Although, for $\beta_{23} = \pi/2;~3\pi/2$ we also have
${\rm max}(r) \cong 1.2$,
the maximum of $r$ corresponds to 
$|R_{12}|^2 = (1 - \sqrt{\deltasol}/\sqrt{\deltaatm})^{-1} 
\cong 1.22$, $|R_{13}|^2 = |R_{12}|^2 -1 \cong 0.22$. 
Therefore the wash-out factors
$\widetilde{m_{l}}$ will differ from those 
in the case of $\beta_{23} = 0;~\pi$
and maximal asymmetry $|\epsilon_{\tau}|$.
The values of $|R_{12}|$ and $|R_{13}|$  
which maximise $|\epsilon_{\tau}|$ are
not guaranteed to maximise also 
the baryon asymmetry $|Y_B|$.
Further investigation of this case is, however,
outside the scope of the present study.

%
\subsection{\label{sec:NHIH}
\large{Inverted Hierarchical Light Neutrino Mass Spectrum}}
%
\indent\
Given the inequalities $m_3 \ll m_1 < m_2$, 
it seems natural to assume 
that the terms $\propto \sqrt{m_3}$ 
are sufficiently small and give negligible 
contributions to the lepton flavour asymmetries 
$\epsilon_{l}$
and to the wash-out mass parameters
$\widetilde{m_l}$.
In the ``one-flavour'' approximation this case 
was studied in \cite{PRST05}. It was found that 
the lepton asymmetry is strongly suppressed
by the factor $\deltasol/\deltaatm$
and leptogenesis can produce 
the observed value of the
baryon asymmetry only if the lightest
heavy Majorana neutrino $N_1$
has a relatively large mass:
$M_1 \gtap 7\times 10^{12}$ GeV.
Using Eq.~(\ref{epsa2}) and the fact 
that in the case of interest we have 
$m_{1,2} \cong \sqrt{\deltaatm}$ and
$(m_{2} - m_{1}) \cong \deltasol/(2\sqrt{\deltaatm})$,
it is not difficult to obtain the 
conditions under which the indicated terms 
$\propto \sqrt{m_3}$ in $\epsilon_{l}$ 
can be neglected. The latter depend on whether
the $R_{11}R_{12} \equiv e^{i\beta_{12}}|R_{11}R_{12}|$  
is real or purely imaginary. In the case 
of $\beta_{12}\equiv \widetilde{\beta_{11}} + \widetilde{\beta_{12}} = \pi q$, $q=0,1,2,...$, we get:
\begin{equation}
2\,\left (\frac{m_3}{\sqrt{\deltasol}}\right )^{\frac{1}{2}}\,
\left ( \frac{\deltaatm}{\deltasol} \right )^{\frac{3}{4}}\,
\frac{\left |R_{13} \right |}{\left | R_{12(11)}\right |} \ll 1\,,~~
\beta_{12} = \pi q, ~q=0,1,2,...\,.
\label{IHm3R130b120}
\end{equation}
%
Since $2(\deltaatm/\deltasol)^{\frac{3}{4}} \cong 26.4 \gg 1$,
this inequality suggests two possibilities.\\
i) Equation (\ref{IHm3R130b120}) holds and the terms
$\propto \sqrt{m_3}$ in
$\epsilon_{l}$ and $\widetilde{m_l}$ 
are indeed negligible. The simplest realisation
of this possibility corresponds to \cite{PRST05} 
setting $R_{13} = 0$, which in turn implies 
the decoupling of the heaviest RH Majorana 
neutrino $N_3$.\\
ii) The alternative possibility is 
that terms $\propto \sqrt{m_3}$ in
$\epsilon_{l}$ and $\widetilde{m_l}$
are dominant in spite of the fact that
$m_3 \ll m_1,m_2$. This would 
require the ratio in the left-hand side 
in Eq.~(\ref{IHm3R130b120}) to be much bigger than 1.
A possible simple realisation corresponds to
setting $R_{11} = 0$ or $R_{12} = 0$.
Since in the latter case 
$|\epsilon_{l}| \propto \sqrt{m_3/m_2}$,
the asymmetry $|\epsilon_{l}|$ will not 
be suppressed only if $m_3$ is 
sufficiently large. The latter condition 
is satisfied for values of 
$m_3$ in the interval $10^{-2}\sqrt{\deltasol} 
\ltap m_3 \ltap 0.5\sqrt{\deltasol}$, for which 
we still have $m_3 \ll m_{1,2}$.

 If, however, $\beta_{12} = \pi/2 (2q + 1)$, $q=0,1,2,...$,
i.e., if $R_{11}R_{12} = \pm i~|R_{11}R_{12}|$,
we obtain a very different condition:
\begin{equation}
\left (\frac{m_3}{\sqrt{\deltaatm}}\right )^{\frac{1}{2}}\,
\frac{\left |R_{13} \right |}{\left | R_{12(11)}\right |} \ll 1\,,~~
\beta_{12} = (\pi/2)(2q + 1), ~q=0,1,2,...\,.
\label{IHm3R130b12Im}
\end{equation}
%
For $m_3 \ll m_{1,2,}$, 
this condition can be naturally satisfied if 
$|R_{13}|$ is sufficiently small and, in particular,
if $|R_{13}| = 0$.

%
\subsubsection{\label{sec:NHIHR130}
\large{The case of Real $R_{11}R_{12}$
and $N_3$ Decoupling 
($R_{13} = 0$)
}}
%
\indent\  
  The terms $\propto \sqrt{m_3}$ in
$\epsilon_{l}$ and $\widetilde{m_l}$
are negligible and we get:
\begin{equation}
\epsilon_{l} \simeq 
 ~\frac{3M_1\sqrt{\deltaatm}}{32\pi v^2}
\left( \frac{\deltasol}{\deltaatm}\right)\;
\left (\frac{\deltasol}{\deltaatm} \right )^{\frac{1}{4}}\,
\frac{\left | R_{11}R_{12} \right | }
{|R_{11}|^2 + |R_{12}|^2}\,
{\rm Im}\,\left (e^{i\beta_{12}}\, U^*_{l 1}U_{l 2}\right )\,,~~~
l=e,\mu,\tau \,,
\label{epslIH1}
\end{equation}
%
where $\beta_{12}\equiv {\rm arg}(R_{11}R_{12})$.
The phase $\beta_{12}$ parametrises 
the effect of CP-violation due to the matrix $R$
in the asymmetries $\epsilon_{l}$.
In the case of CP-invariance, $\beta_{12} = 0$ or $\pi/2$
depending on whether ${\rm Im}(U^*_{l 1}U_{l 2})=0$
or ${\rm Re}(U^*_{l 1}U_{l 2}) =0$, respectively,
and $\epsilon_{l} = 0$, $l=e,\mu,\tau$.
Note that the asymmetries $\epsilon_{l}$ 
are suppressed by the factor
$\deltasol/(2\deltaatm) \cong 1.6 \times 10^{-2}$ 
with respect to the
asymmetries $\epsilon_{l}$ we have
obtained in the case of NH 
light neutrino mass spectrum.

   From Eq.~(\ref{eq:Upara}) we get 
for the quantity 
${\rm Im}(e^{i\beta_{12}}U^*_{\tau 1}U_{\tau 2})$
of interest:
\beq
\label{Imt1t2}
{\rm Im}\,\left (e^{i\beta_{12}}\,U^*_{\tau 1}U_{\tau 2}\right ) 
= - {\rm Im}\,\left [ e^{i\beta_{12}}\, e^{i\frac{\alpha_{21}}{2}}\,
\left (c_{12}s_{23} + s_{12}c_{23}s_{13}\,e^{i\delta}\right )
\left (s_{12}s_{23} - c_{12}c_{23}s_{13}\,e^{-i\delta}\right )\right ]\,, 
\eeq
%
where $\alpha_{21}$ is the Majorana phase 
which enters also into the expression for 
$\meff$ in $\betabeta$-decay.
The wash-out mass parameters 
$\widetilde{m_l}$ for the 
three lepton asymmetries read \cite{PRST05}:
\begin{equation}
\widetilde{m_l} \cong 
\sqrt{\deltaatm} \left | R_{11}\,U^*_{l 1} + 
R_{12}\,U^*_{l 2}\right |^2\,,~~l=e,\mu,\tau~.
\label{tmlIH1}
\end{equation}
%
Using the unitarity of the PMNS matrix $U$ we obtain: 
\begin{equation}
\widetilde{m_e} + \widetilde{m_\mu} + \widetilde{m_{\tau}} =
\sqrt{\deltaatm} \left ( |R_{11}|^2 + |R_{12}|^2 \right )\,.
\label{sumtlIH1}
\end{equation}
%

 In what follows we limit our discussion to 
the case of real $R_{11}$ and $R_{12}$.
In this case we have  $|R_{11}|^2 + |R_{12}|^2 = 1$, and  
$\beta_{12} = 0;~\pi$ which correspond to 
${\rm sgn}(R_{11}R_{12}) = +1;~(-1)$.
Obviously, the values of $|R_{11}| = |R_{12}| = 1/\sqrt{2}$
maximise the asymmetries $\epsilon_{l}$.
As can be shown, for generic values of the phases 
$\alpha_{21}$ and $\delta$,
these values of $|R_{11}|$ 
and $|R_{12}|$ maximise also $|Y_B|$.
Now we have 
\begin{equation}
\widetilde{m_e} + \widetilde{m_\mu} + \widetilde{m_{\tau}} =
\sqrt{\deltaatm}\,.
\label{sumtlIH2}
\end{equation}
%

 Using the best values of the neutrino oscillation parameters
$\sin^22\theta_{23}$, $\sin^2\theta_{12}$,
$\deltasol$ and $\deltaatm$, given in Eq.~(\ref{th12th23}) 
and the upper limit $s_{13} \leq 0.2$,
we get maximal $|\epsilon_{\tau}|$ and $|Y_B|$ 
as can be shown, for $\alpha_{21} = \pi/2$, 
$\delta = \pi$ and $\beta_{12} = 0$: 
\beq
|\epsilon_{\tau}|  \cong 1.5\times 10^{-10}\,
\left (\frac{\sqrt{\deltaatm}}{0.05~{\rm eV}}\right )
\left (\frac{M_1}{10^{9}~{\rm GeV}}\right )\,,
\label{maxepstIH1}
\eeq
\beq
|Y_B| \cong 2.2\times 10^{-14}
\left (\frac{\sqrt{\deltaatm}}{0.05~{\rm eV}}\right )
\left (\frac{M_1}{10^{9}~{\rm GeV}}\right )\,.
\label{maxYBIH1}
\eeq
%
Clearly, for $M_1 \ltap 10^{12}~{\rm GeV}$
for which the flavour effects in leptogenesis 
can be substantial, it is impossible to 
reproduce the observed baryon asymmetry
in the case of IH light neutrino mass spectrum, 
$R_{13} = 0$ and real $R_{11}$ and $R_{12}$,
$|R_{11}|^2 + |R_{12}|^2 = 1$. 
The main reason for this result
lies in the fact that the lepton asymmetries 
$\epsilon_{l}$ are suppressed by the factor
$\deltasol/(2\deltaatm) \cong 1.6 \times 10^{-2}$,
while the wash-out effects are rather large owning to
the constraint (\ref{sumtlIH2}).

%
\subsubsection{\label{sec:NHIHR130LYB}
\large{Generating $Y_B$ Compatible 
with the Observations
}}
%

\indent\ Consider next the case of 
$\beta_{12} = (\pi/2)(2q + 1)$, $q=0,1,2,...$.
Now the product $R_{11}R_{12}$ is purely imaginary:
$R_{11}R_{12} = i \kappa |R_{11}R_{12}|$,
$i \kappa \equiv e^{i\beta_{12}}$, $\kappa = \pm 1$. 
We shall assume further that condition
(\ref{IHm3R130b12Im}) holds and thus the terms 
$\propto \sqrt{m_3}$ in $\epsilon_{l}$ and $\widetilde{m_l}$
are negligible. A simple realisation of this 
scenario corresponds to $R_{13} = 0$ ($N_3$ decoupling),
and to $|R_{11}|^2 - |R_{12}|^2 = 1$.
Under the indicated conditions we have:
\begin{equation}
|\epsilon_{\tau }| \simeq 
 ~\frac{3M_1\sqrt{\deltaatm}}{16\pi v^2}
\frac{2\left | R_{11}R_{12} \right | }
{|R_{11}|^2 + |R_{12}|^2}\,
\left | {\rm Re}\,\left (U^*_{\tau 1}U_{\tau 2}\right ) \right |\,,~~~
l=e,\mu,\tau \,,
\label{epslIH2}
\end{equation}
%
where 
\beq
\label{Ret1t2}
\left | {\rm Re}\,\left (U^*_{\tau 1}U_{\tau 2}\right )\right | 
\cong 
\left | c_{12}s_{12}s^2_{23}\,\cos \frac{\alpha_{21}}{2} -
c_{23}s_{23}s_{13}\left [c^2_{12}\,\cos \left(\frac{\alpha_{21}}{2} -\delta
\right)
- s^2_{12}\, \cos \left(\frac{\alpha_{21}}{2}+\delta\right)\right ]\right |\,,
\eeq
%
and we have neglected the term $\propto s_{13}^2c_{12}s_{12}c^2_{23}$
in the expression for $|{\rm Re}(U^*_{\tau 1}U_{\tau 2})|$.
The maximal value of the factor 
$r' \equiv 2|R_{11}R_{12}|/(|R_{11}|^2 + |R_{12}|^2)$ 
is $r' = 1$ and is reached for $|R_{11}|^2 \gg 1$.
However, even for $|R_{11}|^2 = 1.5$ and $|R_{12}|^2 = 0.5$, 
$r'$ is rather close to its maximal value: $r' = \sqrt{3}/2 \cong 0.87$.
In the case of purely Majorana or Dirac CP-violation
from the PMNS matrix we obtain:  
\begin{eqnarray}
\label{Ret1t21}
\left | {\rm Re}\,\left (U^*_{\tau 1}U_{\tau 2}\right )\right | 
&\cong&  
s_{23}\, \left (c_{12}\, s_{12}\, s_{23} \pm
s_{13}\,c_{23}\,\cos2\theta_{12}\right )\,
\left | \cos \frac{\alpha_{21}}{2} \right |\,,~\delta = 0,\pi,2\pi\,,\\ 
\left | {\rm Re}\,\left (U^*_{\tau 1}U_{\tau 2}\right )\right |
&\cong& s_{13}\,c_{23}\,s_{23}\left | \sin \delta \right |\,,~~
\cos \frac{\alpha_{21}}{2}=0\,.
\end{eqnarray}
%

{\bf {\it A. Majorana CP-Violation from $U_{\rm PMNS}$}}\\

   In what follows we first set $s_{13} = 0$.
The CP-symmetry is violated by the Majorana phase
$\alpha_{21}$ only. Both $|\epsilon_{\tau }|$ and $|Y_B|$ 
vanish for $\alpha_{21} = \pi (2q + 1)$, $q=0,1,...$.
The wash-out parameters 
$\widetilde{m_\tau}$ and $\widetilde{m_2}$ are given by
\begin{equation}
\widetilde{m_\tau} \cong 
\sqrt{\deltaatm} \left [ s^2_{12}\, s^2_{23}\, |R_{11}|^2 + 
c^2_{12}\, s^2_{23}\, |R_{12}|^2  + 
2\, \kappa\, c_{12}\, s_{12}\, s^2_{23}\, |R_{11}R_{12}|\,
\sin\frac{\alpha_{21}}{2} \right ]\,,
\label{tmtIH2}
\end{equation}
%
\begin{equation}
\widetilde{m_2} = 
\sqrt{\deltaatm} \left ( |R_{11}|^2 + |R_{12}|^2 \right ) - 
\widetilde{m_{\tau}}\,.
\label{mt2IH2}
\end{equation}
%
Obviously, we have $|\epsilon_{\tau }(\alpha_{21})| = 
|\epsilon_{\tau }(2\pi - \alpha_{21})|$,
$\widetilde{m_\tau}(\alpha_{21}) =
 \widetilde{m_\tau}(2\pi - \alpha_{21})$,
and therefore $|Y_B(\alpha_{21})| = 
|Y_B(2\pi - \alpha_{21})|$.
Numerical calculations we have performed show
that the maximal value of the baryon asymmetry 
$|Y_B|$ in the case under discussion 
corresponds to $|R_{11}|^2 \cong 1.1 - 1.4$ 
(see Fig. \ref{Figs1011241106.eps}) 
and we will use
these values of $|R_{11}|^2$ and the corresponding
values of $|R_{12}|^2 = 0.1 - 0.4$, 
in our further analysis. 
The results we obtain for  $|Y_B|$ 
depend strongly on whether $\kappa = + 1$ or
$\kappa = - 1$. 

  For $\kappa = + 1$ and $|R_{11}|^2 \cong 1.1 - 1.4$,
one has $\widetilde{m_{\tau}},\widetilde{m_2} 
\gg 2\times 10^{-3}$ eV, which corresponds to a 
strong wash-out regime. The baryon asymmetry
is maximal for $\alpha_{21} = 2\pi q$, $q=0,1,...$,
for which both $|\epsilon_{\tau }|$
and the 
efficiency factor $|\eta_B| \equiv
|\eta(0.66 \widetilde{m_\tau}) - \eta(0.71\widetilde{m_2})|$
are maximal 
\footnote{Note that for $\alpha_{21} = 2\pi q$, $q=0,1,...$, and
$s_{13} = 0$ we have 
${\rm Im} (U^*_{\tau 1}U_{\tau 2}) = 0$ and 
$J_{\rm CP} = 0$. Correspondingly, the CP-invariance is not 
violated by $\alpha_{21}$ and $\delta$; 
it is violated at high energies by the matrix $R$.}
($|\cos(\alpha_{21}/2)| = 1$).
Numerical studies show that the
absolute maximum of $|Y_B|$ 
for $\kappa = + 1$ is reached at 
$|R_{11}|^2 \cong 1.1$ 
(Figs. \ref{Figs1011241106.eps} and \ref{IHs130bl02ralpha21plus.eps}).

  In the case of $\kappa = - 1$ and values of $|R_{11}|^2$
indicated above, which maximise $|Y_B|$,
there is a strong partial compensation between 
the three terms in $\widetilde{m_\tau}$. 
As a consequence, the 
efficiency factor $|\eta_B|$ and 
the asymmetry $|Y_B|$ are 
for, e.g., $\alpha_{21} = \pi/2$ 
and the values of $|R_{11}|^2$ which maximise 
$|Y_B|$, approximately by a factor of 5 
bigger 
than the values they have
in the case of  $\kappa = + 1$. Moreover,
the maximum of $|Y_B|$ takes place
at $|R_{11}|^2 \cong 1.4$ 
and $\alpha_{21} \cong 2\pi/3;~4\pi/3$
(Figs. \ref{Figs1011241106.eps} and 
\ref{IHs130bl02ralpha21minus.eps}),
rather than at $|R_{11}|^2 \cong 1.1$ and
$\alpha_{21} = 2\pi q$, $q=0,1,...$.
In both cases of $\kappa = + 1$ and $\kappa = - 1$ 
we have $|Y_B| = 0$ for $\alpha_{21} = \pi (2q +1)$. 
 
 We get for the maximal baryon asymmetry 
(i.e., at $|R_{11}|^2 \cong 1.1$ and
$\alpha_{21} = 0;~2\pi$ for $\kappa = + 1$, 
and at $|R_{11}|^2 \cong 1.4$ 
and $\alpha_{21} \cong 2\pi/3;~4\pi/3$ for $\kappa = - 1$, 
see Figs. \ref{Figs1011241106.eps} - 
\ref{IHs130bl02ralpha21minus.eps}):
\beq
|Y_B| \cong 1.5~(0.5)\times 10^{-12}
\left (\frac{\sqrt{\deltaatm}}{0.05~{\rm eV}}\right )
\left (\frac{M_1}{10^{9}~{\rm GeV}}\right )\,,~~~
\kappa = - 1~(\kappa = + 1)\,.
\label{YBIH1}
\eeq    
%
It follows from the preceding analysis that 
the observed value of the baryon asymmetry
$|Y_B| \gtap 8\times 10^{-11}$ can be 
reproduced in the case being studied 
for $M_1 \gtap 5.3\times 10^{10}~{\rm GeV}$ if   
$\kappa = - 1$, and for   
$M_1 \gtap 1.6\times 10^{11}~{\rm GeV}$
when $\kappa = + 1$. 
Since both the baryon asymmetry  
$|Y_B|$ and the effective Majorana mass in 
$\betabeta$-decay, $\meff$ depend on 
the Majorana phase $\alpha_{21}$, for given values 
of the other parameters there exists a direct 
correlation between the values of  $|Y_B|$ and $\meff$.
The latter is illustrated in Fig. \ref{IHneutrinoless}.\\

{\bf {\it B. Dirac CP-Violation from $U_{\rm PMNS}$ 
($\alpha_{21} = \pi$)
}}\\

  One can treat in a similar manner the case 
of  $s_{13} \neq 0$ and CP-violation 
generated only by the Dirac phase $\delta$ in 
$U_{\rm PMNS}$, $\alpha_{21} = \pi (2k+1)$, $k=0,1,...$. 
We have $e^{i\alpha_{21}/2} = i\kappa'$, where 
$\kappa' = 1$ for $k=0,2,4,...$, and $\kappa' = -1$
if $k=1,3,...$. It follows from eqs. (\ref{epslIH2}) 
and (\ref{Ret1t21}) that in this case the asymmetry 
$|\epsilon_{\tau }| \propto |s_{13}\sin\delta|$
and one can expect the baryon asymmetry 
$|Y_B|$ to be suppressed by $s_{13}$. 
Obviously, we have 
$|\epsilon_{\tau }|=0$ and $|Y_B| = 0$ for 
$\delta = 0;~\pi;~2\pi$.  
For the wash-out mass parameters 
$\widetilde{m_{\tau}}$ and $\widetilde{m_{2}}$ we get:
\begin{equation}
\widetilde{m_\tau} \cong 
\sqrt{\deltaatm}\, \left |s_{23}\, 
\left (s_{12}\, |R_{11}| - \kappa\, \kappa'\, c_{12}\, |R_{12}|\right ) 
- c_{23}\, s_{13}\,e^{-i\delta}\,
\left (c_{12}\, |R_{11}| + \kappa\, \kappa'\, s_{12}\, |R_{12}|\right )
\right |^2\,,
\label{tmtIHD1}
\end{equation}
%
and 
$\widetilde{m_2} = 
\sqrt{\deltaatm}(|R_{11}|^2 + |R_{12}|^2) - \widetilde{m_{\tau}}$.
The two possibilities $\kappa \kappa' = - 1$ 
and $\kappa \kappa'= + 1$ lead to drastically 
different results.

 For $\kappa \kappa' = - 1$, the term 
with the factor  $s_{23}$ in the expression 
for $\widetilde{m_{\tau}}$ in Eq.~(\ref{tmtIHD1}) 
gives the dominant contribution 
and determines the magnitude of 
$\widetilde{m_{\tau}}$.
For, e.g., $\delta = \pi/2$ 
(for which $|\epsilon_{\tau }|$ is maximal),
the maximum of the baryon asymmetry
$|Y_B|$ takes place at $|R_{11}| \cong 1.05$
($|R_{12}| \cong 0.32$).
For these values of $\delta$ and $|R_{11}|$, both 
wash-out mass parameters satisfy 
$\widetilde{m_{\tau,2}} \gtap 2\times 10^{-2}$ eV.
Thus, the baryon asymmetry is generated in 
the strong wash-out regime.
Correspondingly, the efficiency factor 
$|\eta_B|$ is relatively small,
$|\eta_B| \ltap 6\times 10^{-3}$.
The observed value of the baryon 
asymmetry $|Y_B| \cong (8.0 - 9.2)\times 10^{-11}$
can be reproduced only if 
$s_{13} \cong 0.2$ and 
$M_1 \cong (4 - 5)\times 10^{11}~{\rm GeV}$;
for  $s_{13} \cong 0.1$, this requires a value of 
$M_1 \cong 8\times 10^{11}~{\rm GeV}$.

   We obtain completely different results 
in the case of $\kappa \kappa' = + 1$.
In this case there can be a deep mutual 
compensation  between the two terms in the 
bracket multiplied by 
$s_{23}$ in the right-hand side of Eq.~(\ref{tmtIHD1}) and we can have 
$\widetilde{m_{\tau}} \cong 
(1.5 - 2.0)\times 10^{-3}$ eV 
for the values of $|R_{11}|$ 
($|R_{12}|$) which maximise 
the baryon asymmetry $|Y_B|$. 
Correspondingly, $|Y_B|$ can be 
generated in the weak wash-out 
regime and can be much larger 
than in the case of $\kappa \kappa' = - 1$.

  More specifically, 
the maximum of $|Y_B|$ with respect 
to the Dirac phase $\delta$ and 
$|R_{11}|$ ($|R_{12}|$)
takes place approximately at 
(or relatively close to) $\delta \cong \pi/2;~3\pi/2$
for any $s_{13} \ltap 0.2$ of 
interest, and at $|R_{11}| \cong 1.30;~1.60$ 
($|R_{12}|\cong 0.83;~1.25$)
for $s_{13} = 0.20;~0.10$, respectively 
(Figs. \ref{Fig12IHDalphapi040107R}
and \ref{Fig13IHDalphapi040107Delt}); 
for $s_{13} \ltap 0.02$ 
it is located at $|R_{11}| 
\cong 1.07$ ($|R_{12}| \cong 0.38$). 
We have used again the best fit values of
$\deltaatm$, $\sin2\theta_{12}$ and $\sin2\theta_{23}$ 
to obtain the location of the maxima in
$|R_{11}|$. In the case of 
$s_{13} = 0.10$ there exists a second 
local maximum of $|Y_B|$ at 
$|R_{11}| \cong 1.15$ ($|R_{12}| \cong 0.57$), 
at which $|Y_B|$ is approximately by a 
factor of 1.4 smaller than at the 
absolute maximum 
at $|R_{11}| \cong 1.60$ (see 
Fig. \ref{Fig12IHDalphapi040107R}). 
The positions of the indicated maxima 
of $|Y_B|$ are determined essentially 
by the position of the absolute maximum
of the efficiency factor  
$|\eta_B| \cong
|\eta(0.66 \widetilde{m_\tau}) - \eta(0.71\widetilde{m_2})|$.
The latter corresponds approximately to 
$\widetilde{m_\tau} \cong 1.5 - 1.8 \times 10^{-3}$ eV
and negligible 
\footnote{Indeed, since $|R_{11}|^2 + |R_{12}|^2 > 1$,
we have $\widetilde{m_2} > 5.0 \times 10^{-2}$ eV
and therefore  $|\eta(0.71\widetilde{m_2})| < 2\times 10^{-3}$.}
$|\eta(0.71\widetilde{m_2})|$. 
At the maximum,
$|\eta_B| \cong 6.7\times 10^{-2}$.
In the case of $s_{13} \cong 0.20$, as can be shown,  
$\widetilde{m_\tau} \gtap 2.3 \times 10^{-3}$ eV and
the minimal value of $\widetilde{m_\tau}$ 
corresponds to negligible 
$s^2_{23}(s_{12}|R_{11}| - c_{12}|R_{12}|)^2 \cong 0$.
The latter condition is fulfilled approximately for 
$|R_{11}|^2 \cong c^2_{12}/\cos2\theta_{12} \cong 1.75$ and
$|R_{12}|^2 \cong s^2_{12}/\cos 2\theta_{12} \cong 0.75$,
which is very close to the value 
$|R_{11}|^2 \cong 1.69$ obtained by 
numerical calculations. 
Thus, for $s_{13} \cong 0.20$ we have for 
$\widetilde{m_\tau}$ which maximises 
$|\eta_B|$ ($|Y_B|$):  
$\widetilde{m_\tau} \cong 
\sqrt{\deltaatm}\, c^2_{23}s^2_{13} 
(c_{12}|R_{11}| + s_{12}|R_{12})^2 
\cong (c^2_{23}s^2_{13}/\cos 2\theta_{12})\sqrt{\deltaatm}$.

   In the case of $s_{13} \cong 0.10$, both terms 
in Eq.~(\ref{tmtIHD1}) ``conspire'' to 
produce $\widetilde{m_\tau} \cong 1.75 \times 10^{-3}$ eV, 
and thus maximal $|\eta_B|$ (and $|Y_B|$), 
at $|R_{11} \cong 1.60$. The second local 
maximum of $|Y_B|$ at $|R_{11}| \cong 1.15$ 
corresponds to $\widetilde{m_\tau} \cong 10^{-3}$ eV.
Again both terms in Eq.~(\ref{tmtIHD1}) 
contribute, the term 
$\propto c^2_{23}s^2_{13}$ being approximately 
by a factor 1.5 smaller than the 
term $\propto s^2_{23}$. 
In contrast, the local minimum of $|Y_B|$ 
at $|R_{11}| \cong 1.28$ 
(Fig. \ref{Fig12IHDalphapi040107R})
is associated with 
$\widetilde{m_\tau} \cong 6\times 10^{-4}$ eV.
In this case the contribution 
of the term $\propto s^2_{23}$ in 
$\widetilde{m_\tau}$ is negligible.

 Finally, for $s_{13} \ltap 0.02$, the term 
with the factor $\propto c_{23}s_{13}$
in the expression for $\widetilde{m_\tau}$, 
Eq.~(\ref{tmtIHD1}), plays no role in the 
determination of the maxima of $|\eta_B|$ and 
$|Y_B|$. Thus, in this case, in particular,
$|Y_B|$ depends on $s_{13}$ and $\delta$ only 
through $\epsilon_{\tau }$ and we have  
$|Y_B| \propto s_{13}|\sin\delta|$.  

  For the maximal value of the 
baryon asymmetry $|Y_B|$ for $s_{13} = 0.20~(0.10)$
and $\delta = \pi/2$ ($|R_{11}| \cong 1.30~(1.60)$) 
we get (Fig. \ref{Fig12IHDalphapi040107R}):
\beq
|Y_B| \cong 1.7~(1.0)\times 10^{-12}
\left (\frac{\sqrt{\deltaatm}}{0.05~{\rm eV}}\right )
\left (\frac{M_1}{10^{9}~{\rm GeV}}\right )\,,~~
\delta = \frac{\pi}{2}\,,~~s_{13}=0.20~(0.10)\,.
\label{YBIHD1}
\eeq    
%
It follows from the above result that if, e.g., 
$s_{13} = 0.10$, we can obtain
baryon asymmetry $|Y_B| \gtap 8\times 10^{-11}$ 
compatible with the observations 
for $M_1\gtap 8\times 10^{10}~{\rm GeV}$.

  What is the minimal value of $s_{13}|\sin \delta|$ 
for which we can have successful leptogenesis?
For sufficiently small values of $s_{13}$ 
we get maximal $|\eta_B| \cong 6.5\times 10^{-2}$ 
($\widetilde{m_\tau} \cong 1.75 \times 10^{-3}$ eV)
at $|R_{11}| \cong 1.07$ independently of the value 
of $s_{13}$ (see the discussion preceding Eq.~(\ref{YBIHD1})). 
The baryon asymmetry at the maximum is given by:
\beq
|Y_B| \cong 8.1\times 10^{-12}\, s_{13}\, |\sin\delta|\,
\left (\frac{\sqrt{\deltaatm}}{0.05~{\rm eV}}\right )
\left (\frac{M_1}{10^{9}~{\rm GeV}}\right )\,.
\label{YBIHD2}
\eeq    
%
Thus, for $M_1 \ltap 5\times 10^{11}~{\rm GeV}$, 
we can have successful leptogenesis and get 
$|Y_B|\cong (8.0 - 9.2)\times 10^{-11}$ provided 
\beq
|\sin\theta_{13}\, \sin\delta| \gtap 0.02\,,~~~
\sin\theta_{13} \gtap 0.02\,.
\label{s13IHD1}
\eeq
%
The preceding lower limit corresponds to
\beq
|J_{\rm CP}| \gtap 4.6\times 10^{-3}\,.
\label{JCPIHD1}
\eeq
%
Values of $\sin\theta_{13}$ 
and of $|J_{\rm CP}|$
as small as 0.02 and $4.6\times 10^{-3}$, 
respectively,
can be probed in neutrino oscillation 
experiments at neutrino factories
\cite{machines}.

   The results in the case under 
discussion are illustrated
in Figs. \ref{Fig12IHDalphapi040107R}
and \ref{Fig13IHDalphapi040107Delt}.
In Fig. \ref{Fig12IHDalphapi040107R}
we show the baryon asymmetry $Y_B$ 
as a function of $|R_{11}|$  for $\delta = \pi/2$,  
while Fig. \ref{Fig13IHDalphapi040107Delt}
exhibits the dependence of 
$|Y_B|$ on the Dirac phase $\delta$ 
for the values of $|R_{11}|$ from the interval 
$|R_{11}|\cong (1.05 - 1.7)$, which maximise 
$|Y_B|$. The results presented in both figures 
are for $s_{13}=0.1;~0.2$, $\alpha_{21} = \pi$ 
($\kappa' = + 1$), 
$\kappa = + 1$, $\kappa = - 1$ and
$M_1 = 2\times 10^{11}$ GeV.
In Fig. \ref{Fig14YBJCPIHDalphapi040107} we show
the correlation between the 
rephasing invariant $J_{\rm CP}$ (in blue) 
and the baryon asymmetry $Y_{\rm B}$
in the case under discussion,
(CP-violation due to the Dirac phase $\delta$
$\alpha_{21} = \pi$) for
$s_{13} = 0.2$, $M_1 = 2\times 10^{11}$ GeV, 
$\kappa = + 1$ and $|R_{11}| = 1.3 $.
The Dirac phase $\delta$ is varied in the interval
[0,2$\pi$].\\ 

{\bf {\it C. CP-Violation due the Dirac Phase in $U_{\rm PMNS}$ 
and $R$ ($\alpha_{21} = 0;~2\pi$)
}}\\

   We will consider next briefly the 
``mixed'' case of CP-violation corresponding to
$\alpha_{21}= 2\pi k$, $k=0,1,...$, 
$\delta$ taking values in the interval 
[0,2$\pi$], and purely imaginary
$R_{11}R_{12}$, $R_{11}R_{12} 
 = i\kappa |R_{11}R_{12}|$, $\kappa \pm 1$.
For $\delta = 0;~\pi$, this case provides an 
interesting example of 
breaking of CP-symmetry due to 
$R_{11}R_{12}$ being purely imaginary.

   It proves convenient to write
$e^{i\alpha_{21}/2} = \kappa'$, where $\kappa' = \pm 1$.
We have in the case under discussion:
\beq
\label{Ret1t2DR1}
\left | {\rm Re}\,\left (U^*_{\tau 1}U_{\tau 2}\right )\right |
\cong \left | c_{12}\, s_{12}\, s^2_{23} -
c_{23}\, s_{23}\, \cos 2\theta_{12}\, s_{13}\, \cos\delta \right |\,.
\eeq
%
Obviously, the term $\propto s_{13}\cos\delta$
plays a sub-dominant role in the asymmetry 
$|\epsilon_{\tau }|$.
The wash-out mass parameter $\widetilde{m_\tau}$ reads:
\begin{eqnarray}
\widetilde{m_\tau} &=&
\sqrt{\deltaatm} \left [ s^2_{23}\, \left ( s^2_{12}\, |R_{11}|^2 + 
c^2_{12}\,|R_{12}|^2\right ) + 
c^2_{23}\, s^2_{13}\,\left (c^2_{12}\, |R_{11}|^2 + 
s^2_{12}\,|R_{12}|^2\right ) \right . \nonumber \\  
&-& \left .  2\, c_{23}\, s_{23}\, s_{13}\, 
\left (c_{12}\, s_{12}\,\cos\delta +
\kappa\, \kappa'\,|R_{11}R_{12}|\,\sin\delta \right ) \right ]\,.
\label{tmtIHDR1}
\end{eqnarray}
%
It follows from eqs. (\ref{epslIH2}), 
(\ref{Ret1t2DR1}) and (\ref{tmtIHDR1}) 
that $|\epsilon_{\tau }(\delta)| = 
|\epsilon_{\tau }(2\pi - \delta)|$,
$\widetilde{m_\tau}(\delta,\kappa\kappa') =  
\widetilde{m_\tau}(2\pi -\delta,-\kappa\kappa')$, and 
therefore $|Y_B(\delta,\kappa\kappa')| = 
Y_B(2\pi -\delta,-\kappa\kappa')|$.

   For $s_{13} \ltap 0.01$, the terms 
$\propto s_{13}$ and $\propto s^2_{13}$  
in the expressions for $|\epsilon_{\tau }|$
and $\widetilde{m_\tau}$ can be neglected.  
The CP-violation effects are practically 
due to the purely imaginary $R_{11}R_{12}$:
the baryon asymmetry $Y_B$ does not depend on 
$s_{13}$ and $\delta$.
However, even in this case it is possible to have 
successful leptogenesis. The baryon asymmetry
$Y_B$ can be generated in a regime which is 
intermediate between 
the weak and strong wash-out ones:
$\widetilde{m_\tau} \gtap 8\times 10^{-3}$ eV.
The asymmetry $|Y_B|$ has two very similar (in magnitude) 
maxima at $|R_{11}| \cong 1.05;~1.12$. At these 
maxima we obtain: 
\beq
|Y_B| \cong 5.2~(5.1)\times 10^{-13}\,
\left (\frac{\sqrt{\deltaatm}}{0.05~{\rm eV}}\right )
\left (\frac{M_1}{10^{9}~{\rm GeV}}\right )\,,~~~s_{13}=0,~~ 
|R_{11}| \cong 1.05~(1.12)\,.
\label{YBIHDR1}
\eeq    
%
Thus, one can have $|Y_B| = (8.0 - 9.2)\times 10^{-11}$
for $M_1 \gtap 1.6\times 10^{11}~{\rm GeV}$.

    If $s_{13} \gg 0.01$, the effects of 
the Dirac CP-violating phase $\delta$ are 
non-negligible; for $s_{13} \cong (0.1 - 0.2)$,
they are substantial for any value of $\delta$. 
The magnitude of the baryon asymmetry 
$|Y_B|$ depends on the value of $s_{13}$ and,
through the wash-out mass 
parameters $\widetilde{m_\tau}$ and $\widetilde{m_2}$,
on the sign factor $\kappa \kappa'$.
In the case of $\delta = \pi/2$, the maximum of 
$|Y_B|$ is located approximately at 
$|R_{11}| \cong (1.05 - 1.10)$ 
(Fig. \ref{Figs1516IH0KandinskiR11}).
For $\kappa \kappa' = + 1$,
the baryon asymmetry at the maximum 
is by a factor $\sim 2$ larger than 
that for $\kappa \kappa' = - 1$.
We can have successful leptogenesis 
in the case of $s_{13} = 0.1~(0.2)$   
for $M_1 \gtap 1.2~(1.0)\times 10^{11}~{\rm GeV}$.
These results are illustrated in 
Figs. \ref{Figs1516IH0KandinskiR11} and
\ref{Figs1718IH0KandinskiDelta}.
In Fig. \ref{Fig17YBJCPIH0DR} we show
the correlation between the 
rephasing invariant $J_{\rm CP}$ 
and the baryon asymmetry $Y_{\rm B}$
for  $\alpha_{21} = 0$, $s_{13} = 0.2$, $\kappa = + 1$,
$|R_{11}| = 1.05$ and $M_1 = 2\times 10^{11}$ GeV.
The Dirac phase $\delta$ is varied in 
the interval [0,2$\pi$]. 

\vspace{0.5cm}

   As we have indicated at the beginning of this 
subsection, the observed value of the baryon asymmetry
can be reproduced in the case of 
real $R_{1j}$, $j=1,2,3$,
if the terms $\propto \sqrt{m_3}$ in
$\epsilon_{l}$ and $\widetilde{m_l}$
are dominant in spite of the fact that
$m_3 \ll m_1,m_2$. A simple 
realisation of this possibility corresponds to
having $R_{11} = 0$ or $R_{12} = 0$, and 
sufficiently large $m_3$ still obeying
the inequality $m_3 \ll m_{1,2}$.
The latter conditions can be satisfied 
for $m_3$ possessing a value in the interval
$10^{-2}\sqrt{\deltasol} 
\ltap m_3 \ltap 0.5\sqrt{\deltasol}$.
A more detailed discussion 
of this case will be presented elsewhere.

%
\subsection{\label{sec:NHQD}
\large{Quasi-Degenerate Light Neutrino Mass Spectrum}}
%
 
We turn now to the QD spectrum, 
for which $m_1 \simeq m_2 \simeq m_3 \gg
\sqrt{\deltasol}, \sqrt{\deltaatm}$ and 
we take as conventional lower limit on the masses 
$m_1 \gtap 0.1$~eV.
In this case the contribution of all the masses 
to the CP-asymmetry $\epsilon_\tau$ needs 
to be taken into account. 
We study the two cases in which 
the products of the parameters in $R$
entering in the asymmetry,
$R_{1i} R_{1j}$, are either real or 
purely imaginary.
This corresponds to having 
$\beta_{ij} \equiv \widetilde{\beta_{1i} }+\widetilde{ \beta_{1j} }= 2 k \pi/2, (2 k+ 1) \pi/2$,
$k=0,1,....$, respectively.
We consider first the case of $R_{1i} R_{1j}$ real
and equal to $\pm |R_{1i} R_{1j}|$.
The CP-asymmetry can be written as:
\begin{eqnarray}
|\epsilon_\tau | = & \hspace{-3truecm}
\displaystyle\frac{3 M_1}{16 \pi v^2} \displaystyle\frac{1}{\sum_i |R_{1i}|^2} 
\left| (m_2 - m_1) |R_{11} R_{12}| e^{i  
\beta_{12}}
\mathrm{Im} \Big(U_{\tau 1}^\ast U_{\tau 2} \Big) \right. \nonumber \\ 
 & 
\hspace{-2truecm} \left.
+ (m_3 - m_1) |R_{11} R_{13}| e^{i 
\beta_{13}}
\mathrm{Im} \Big(U_{\tau 1}^\ast U_{\tau 3} \Big)  + (m_3 - m_2) |R_{12} R_{13}| e^{i 
\beta_{23}}
\mathrm{Im} \Big(U_{\tau 2}^\ast U_{\tau 3} \Big) \right| ~, \\
  \simeq &  \hspace{-4truecm}
 \displaystyle\frac{3 M_1}{16 \pi v^2} \frac{1}{\sum_i |R_{1i}|^2 }
  \left| \frac{\deltasol}{2 m_1} |R_{11} R_{12}| e^{i 
  \beta_{12}}
\mathrm{Im} \Big(U_{\tau 1}^\ast U_{\tau 2} \Big)  \right. \nonumber \\ 
 & 
\left.
 \pm \ds\frac{\deltaatm}{2 m_1}  |R_{11} R_{13}| e^{i 
 \beta_{13}}
\mathrm{Im} \Big(U_{\tau 1}^\ast U_{\tau 3} \Big)
\pm \ds\frac{\deltaatm}{2 m_1} |R_{12} R_{13}| e^{i 
\beta_{23}}
\mathrm{Im} \Big(U_{\tau 2}^\ast U_{\tau 3} \Big) \right| ~,
\label{epsilonQDapprox}
\end{eqnarray}
where we have neglected terms of order $\deltasol / \deltaatm$.
The signs $\pm$ in Eq.~(\ref{epsilonQDapprox}) refer to the 
quasi-degenerate spectrum with a normal or inverted hierarchy.
This information could not be obtained in experiments
which are sensitive to the overall neutrino mass scale, 
as neutrinoless double beta decay or direct neutrino mass searches.
However, 
by exploiting matter effects, 
long baseline neutrino oscillation 
and atmospheric neutrino experiments
might be able to establish if the spectrum
is with normal or inverted hierarchy.
Neglecting the terms proportional to $\deltasol$, 
we can further simplify the expression in 
Eq.~(\ref{epsilonQDapprox}):
\begin{eqnarray}
|\epsilon_\tau|   \simeq  & \hspace{-7truecm}
 \displaystyle\frac{3 M_1 m_1}{16 \pi v^2} \frac{1}{\sum_i |R_{1i}|^2 } \frac{\deltaatm}{2 m_1^2}
s_{23} c_{23} c_{13} |R_{13}| \nonumber \\
& \hspace{-2truecm} \times  \left| |R_{11}| \!  \Big( \!  s_{12} \sin  \!   \ds\frac{\alpha_{31}}{2}
 \!   -  \!  c_{12} \ds\frac{c_{23}}{s_{23}} s_{13} \sin  \!  \Big( \!   \ds\frac{\alpha_{31}}{2}  \!   -  \!  \delta \!   \Big) \! \Big) 
  \!  \pm  \!   |R_{12}| \Big(  \!   \!\!  - c_{12}  \sin \!   \ds\frac{\alpha_{32}}{2}  \!  -  \!  s_{12} \ds\frac{c_{23}}{s_{23}} s_{13} 
\sin  \!  \Big(  \!   \ds\frac{\alpha_{32}}{2}  \!  -  \!   \delta \Big)  \!    \Big)   \!   \right| ,
\end{eqnarray}
where $\pm$ refer to the case of $\widetilde{\beta_{11}} = \widetilde{\beta_{12}}$ and
$\widetilde{\beta_{12}} = k \pi + \widetilde{\beta_{11}}$, $k=0,1,...$, respectively.
Notice that, in general, the asymmetry is suppressed by $\deltaatm/m_1^2$.
For comparison with the NH case studied in Section~\ref{sec:NHNH},
we can have a mild suppression of order 
$ (\deltaatm)^{3/4} / 
\big( m_1 (\deltasol)^{1/4} \big) \sim 0.2 \ (1 \ \mathrm{eV} / m_1)$
for large values of $m_1$.
If $R_{13}$ is negligible ($N_3$ decoupling),
the dominant contribution is proportional to
$\deltasol$ which amounts to an additional suppression factor
of $\deltasol/\deltaatm \sim 0.03$ \cite{PRST05}.
Typically, we get a CP-asymmetry of order
\begin{eqnarray}
|\epsilon_\tau| & \sim &
1.2 \times 10^{-6} \ds\frac{M_1}{10^{11} ~ \mathrm{GeV}} 
\ds\frac{0.1~\mathrm{eV}}{m_1} |R_{13} R_{11}|  \nonumber \\
& &\hspace{-1truecm} \times \left| 0.55 \sin \ds\frac{\alpha_{31}}{2}  - 0.17 \ds\frac{s_{13}}{0.2}  \sin \big( \ds\frac{\alpha_{31}}{2} -\delta \big) 
-   0.84  \sin  \ds\frac{\alpha_{32}}{2}   - 0.11 \ds\frac{s_{13}}{0.2}
\sin  \Big(    \ds\frac{\alpha_{32}}{2}   -    \delta   \Big) \right| ,
\end{eqnarray}
where we have taken for definiteness 
$R_{12} = R_{11}$ and $\sum_i |R_{1i}|^2 =1$.
The asymmetry decreases linearly with $m_1$ and we have evaluated
it for the minimal value of $m_1$ which is allowed for the QD spectrum.
Notice that, as far as $\alpha_{31}, \alpha_{32}$ is not too small, 
$\sin \frac{\alpha_{31}}{2} \gg 0.1 \ (s_{13}/0.2)$,
the contribution of the $\delta$ phase will be subdominant.
The phases $\alpha_{21}$ and $\alpha_{31}$ 
enter in the effective Majorana mass parameter
for neutrinoless double beta decay.
For the QD light neutrino mass spectrum,
we have  
\begin{equation}
\meff = m_1 \left| c_{12}^2 c_{13}^2 + s_{12}^2 c_{13}^2 e^{i \alpha_{21}}
+ s_{13}^2 e^{ i (\alpha_{31} -2  \delta ) }\right|~.
\end{equation}
The dependence on the phase difference $(\alpha_{31}-2 \delta)$
is suppressed by $s_{13}^2$ and can be neglected. 
In principle, the phase $\alpha_{21}$ can be measured in future
neutrinoless double beta decay experiments
if a sufficient precision will be achieved in the measurement
of $\meff$ and of the masses, and if the problem of the computation
of nuclear matrix elements will be solved~\cite{PPW}.
The prospects of measuring the other Majorana phase
$\alpha_{31}$ are far beyond the sensitivities of the 
present and future planned experiments.

In the case of $R_{1i} R_{1 j}$
purely imaginary, i.e. 
$\widetilde{\beta_{1i}} + \widetilde{\beta_{1j}}= (2 k + 1 ) \pi/2$, $k=0,1,...$,
in general, there is no suppression as the masses
enter in the CP-asymmetry via the combination $m_i + m_j$.
We get 
\begin{eqnarray}
|\epsilon_\tau |
= & \hspace{-10truecm} \displaystyle\frac{3 M_1 m_1}{8 \pi v^2} \frac{1}{\sum_i |R_{1i}|^2} \nonumber \\
 & \times \Big| \pm |R_{11} R_{12} | \mathrm{Re} \big( U_{\tau 1}^\ast U_{\tau 2} \big) 
\pm |R_{11} R_{13} | \mathrm{Re} \big( U_{\tau 1}^\ast U_{\tau 3} \big) 
\pm |R_{12} R_{13} | \mathrm{Re} \big( U_{\tau 2}^\ast U_{\tau 3} \big) 
\Big| ~,
\label{epsilonRim}
\end{eqnarray}
where the $+ (-)$ refer to $ \beta_{ij} = \pi/2 \, (3 \pi/2)$.
The expression in Eq.~(\ref{epsilonRim}) is rather lengthy but can be 
simplified if we neglect $\theta_{13}$. 
In this case it reads:
\begin{eqnarray}
|\epsilon_\tau |
=  &  \hspace{-11truecm} \displaystyle\frac{3 M_1 m_1}{8 \pi v^2} \frac{1}{\sum_i |R_{1i}|^2}
 \nonumber \\
 &
\hspace{-1.5truecm} \times \left| \pm |R_{11} R_{12}| (\! - s_{12} c_{12} s_{23}^2  \cos \ds\frac{\alpha_{21}}{2} \!)
\pm  |R_{11} R_{13}| (\! s_{12} s_{23} c_{23}  \cos \ds\frac{\alpha_{31}}{2} \! )
\pm |R_{12} R_{13}| (\! - c_{12} s_{23} c_{23}  \cos \ds\frac{\alpha_{32}}{2} \!) \right|\, .
\end{eqnarray}
As we have neglected the terms $\propto s_{13}$,
there is no dependence on the phase $\delta$. However,
both Majorana phases enter in the 
expression for $\epsilon_\tau$.

Let's now turn to the wash-out factors.
Using the unitarity condition on $U$, we find:
\begin{equation}
\widetilde{m_2} + \widetilde{m_\tau} = m_1 \ \sum_k |R_{1k}|^2.
\label{washoutQD}
\end{equation}
In the case of real $R_{1k}$, for instance,
this implies \cite{PRST05} 
$\widetilde{m_2} + \widetilde{m_\tau} = m_1$.
Therefore, we can expect that the wash-out mass parameters 
will typically be much larger than $3\times 10^{-3}$ eV,
leading to a strong suppression of the baryon asymmetry.
More specifically, $\widetilde{m_\tau}$ 
is given by:
\begin{equation}
\widetilde{m_\tau} = 
m_1 \left|R_{11} U_{\tau 1}^\ast + R_{12} U_{\tau 2}^\ast 
+ R_{13} U_{\tau 3}^\ast \right|^2 ~.
\end{equation}

 We will study first the case of real $R_{ij}$.
 Taking, for example, $R_{11} = R_{12} = R_{13}$,
we obtain $\widetilde{m_\tau} = 0.053~ m_1$,
for $\alpha_{21}=\pi$ and $\alpha_{31}=\pi$.
In this case the strong wash-out regime formulas 
apply and we have
\begin{equation}
\eta(390/589 \, \widetilde{m_\tau}) = 
3.3 \times 10^{-2} \, \Big(\frac{0.1 ~ \mathrm{eV}}{m_1}\Big)^{1.16}~,
\end{equation}
resulting in a baryon asymmetry
which is substantially smaller than the
observed one.

  A larger 
efficiency factor can be achieved
in the regime interpolating between
the ones of strong and weak wash-out effects. 
However, it should be noticed that 
$\eta(\widetilde{m_\tau}) $ and $\epsilon_\tau$
depend on the same parameters and they cannot be maximized 
independently. 
In fact, for real $R_{ij}$, we can rewrite
the CP-asymmetry as:
\begin{eqnarray}
|\epsilon_\tau | & \simeq & \displaystyle\frac{3 M_1 m_1}{32 \pi v^2} \frac{\deltaatm}{m_1^2 \sum_i |R_{1i}|^2}
\left| \mathrm{Im} (R_{13} U_{\tau 3}  \sum_\beta R_{1 \beta} U_{\tau \beta}^\ast ) \right| ~,
\nonumber \\
& \leq & \displaystyle\frac{3 M_1 m_1}{32 \pi v^2} \frac{\deltaatm}{m_1^2 \sum_i |R_{1i}|^2}
\ \sqrt{\frac{\widetilde{m_\tau}}{m_1}} \ |R_{13} U_{\tau 3}| ~.
\end{eqnarray}
Therefore, while the 
efficiency factor $\eta(\widetilde{m_\tau})$
increases when $\widetilde{m_\tau}$ decreases
as the wash-out regime changes from strong to weak, 
$|\epsilon_\tau|$ decreases accordingly.
In the strong wash-out regime in the $\tau$ flavour
and $\widetilde{m_\tau} \ll m_1$,
the baryon asymmetry has an upper bound:
\begin{equation}
|Y_B| \leq \frac{12}{37 g_\ast} \frac{ 3 M_1 m_1}{32 \pi v^2}
\frac{\deltaatm}{m_1^2} |R_{13}| c_{23} 
\sqrt{\frac{\widetilde{m_\tau}}{m_1}} \left( \frac{0.2 \times 10^{-3} \ \mathrm{eV}} {\widetilde{m_\tau}}
\frac{589}{390} \right)^{1.16} ~.
\end{equation}
Conversely, for weak wash-out in the $\tau$ flavour,
we have
\begin{equation}
|Y_B| \leq \frac{12}{37 g_\ast} \frac{ 3 M_1 m_1}{32 \pi v^2}
\frac{\deltaatm}{m_1^2} |R_{13}| c_{23}
\sqrt{\frac{\widetilde{m_\tau}}{m_1}} \left( \frac{\widetilde{m_\tau}} {0.2 \times 10^{-3} \ \mathrm{eV}}
\frac{390}{589} \right)~.
\end{equation}
%
Thus, the maximal baryon asymmetry is obtained
for intermediate wash-out effect. For definiteness we 
can estimate the upper bound on $Y_B$ at 
$\widetilde{m_\tau} \simeq 3 \times 10^{-3}$~eV.
For the smallest allowed value of $m_1$
for the QD spectrum, $m_1 = 0.1$~eV, we get:
\begin{equation}
|Y_B| \ltap 5 \times 10^{-11} |R_{13}| \frac{M_1}{10^{11} \ \mathrm{GeV}} ~.
\end{equation}
Requiring that $\widetilde{m_\tau} \ll m_1$ imposes 
a fine tuning on the values of the parameters $R_{1i}$.
For simplicity, we search for the solution of the equation
$\widetilde{m_\tau}  \sim 3 \times 10^{-3} \ \mathrm{eV} \simeq 0$, 
which corresponds to:
\begin{eqnarray}
\label{eqab} R_{11} |U_{\tau1}| - R_{12} |U_{\tau 2}| \cos \big( \frac{\alpha_{21}}{2} \big) + R_{13} |U_{\tau 3}| \cos  \big( \frac{\alpha_{31}}{2} \big) \simeq 0 ~, \\
\label{eqac}
R_{12} |U_{\tau 2}| \sin \big( \frac{\alpha_{21}}{2} \big) - R_{13} |U_{\tau 3}| \sin  \big( \frac{\alpha_{31}}{2} \big) \simeq 0 ~.
\end{eqnarray}
If $\alpha_{21} \, (\alpha_{31}) \, [\alpha_{32}]=0$, we have that 
$R_{13} \, (R_{12}) \, [R_{11}] =0$ as well.
Otherwise, the solution of Eqs.~
(\ref{eqab}) and (\ref{eqac}) is given by:
\begin{equation}
R_{11} = R_{13}  \frac{c_{23}}{s_{12} s_{23}} \frac{\sin{\alpha_{32} / 2}}{\sin{\alpha_{21} / 2}} 
\ \ \ \mathrm{and} \ \ \ 
R_{12} = R_{13} \frac{c_{23}}{c_{12} s_{23}} \frac{\sin{\alpha_{31} / 2}}{\sin{\alpha_{21} / 2}} ~.
\end{equation}

Another possibility consists in having
$\widetilde{m_\tau} \simeq m_1$ and small $\widetilde{m_2}$.
However, also in this case the baryon asymmetry
is suppressed due to the values of the CP-violating phases
required to have $\widetilde{m_2}\ll m_1$,
namely, $\alpha_{21}, \alpha_{31} \sim 0, \pi$.
In conclusion, for real $R$, 
it might be possible to reproduce 
the observed baryon asymmetry,
but only for relatively 
large values of $M_1 \gg 10^{11}$~GeV
and small values of $m_1$. 
A careful and detailed analysis should 
be performed on a case by case basis.
A measurement of $m_1$ in the upper end of the range of allowed
values $[0.1 \, \mathrm{eV}, 2.3 \, \mathrm{eV}]$
in the quasi-degenerate spectrum
would strongly disfavour, if not rule out, the possibility
of having leptogenesis due uniquely to the low energy CP-violating phases,
for hierarchical RH neutrinos and real $R$.

   For $R_{1i} R_{1j} = \pm i |R_{1i} R_{1j}|$, the CP-asymmetry
is enhanced by a factor $4 m_1^2 / \deltaatm$. 
Also in this case an upper bound on $|Y_{\rm B}|$
can be derived, and it depends on $\widetilde{m_\tau}$.
A sufficiently large baryon asymmetry might be obtained
for relatively large values of $M_1$.
A more detailed analysis of this case is 
beyond the scope of the present study.

%
\section{Baryon Asymmetry from Low Energy CP-Violating 
Dirac and Majorana Phases in $\pmns$: the Case of 
Quasi-Degenerate RH Neutrinos}
%
\noindent
In this Section we extend our previous findings to the case in which the
RH neutrinos are  quasi-degenerate in mass, $M_1\simeq M_2\simeq M_3$
and the matrix $R$ is real. We therefore consider the case
in which the CP parities of the heavy and light Majorana neutrinos
are such that $\rho_i^N=\rho^\nu_j=1$ for all $i,j=1,2,3$. In such a case,
indeed, 
CP invariance  corresponds to having all the elements of the matrix $R$ real,
see Eq.~(\ref{RCPinv1}),  and $\delta=\alpha_{21}=\alpha_{31}=0$. 
The degenerate
pattern for the RH neutrino masses may arise if, for instance, 
 there is a slightly broken SO(3) symmetry in the
RH sector. The baryon asymmetry receives a contribution from the decay of all
three RH neutrinos. 
The CP asymmetry in a given flavour $l$ generated by the decay of the 
RH neutrino 
$N_i$ ($i=1,2,3$) is dominated by the one-loop self energy contribution
\cite{oneloop} and reads

\begin{eqnarray}
\epsilon_i^l &=&-\sum_{j\neq i}\,\frac{M_i}{M_j}\,
\frac{\Gamma_j}{M_j}\,S_{ij}\, I_{ij}^l\, ,\,\,\,
\Gamma_j=\frac{\lambda\lambda^\dagger_{jj} M_j}{8\pi}\, ,\,\,
\left(\lambda\lambda^\dagger\right)_{ii}=\frac{M_i}{v^2} \sum_\ell \, m_\ell\, 
R^2_{i \ell}\, , 
\nonumber\\
S_{ij} &=& \frac{M_j^2\Delta M_{ij}^2}{\left(\Delta M^2_{ij}\right)^2+
M_i^2\Gamma_j^2}\, ,\,\, \Delta M_{ij}^2=M_j^2-M_i^2\, ,\nonumber\\
I_{ij}^l &=& \frac{1}{\left(\lambda\lambda^\dagger\right)_{ii}}
\frac{1}{\left(\lambda\lambda^\dagger\right)_{jj}}\,\frac{M_i M_j}{v^4}
\sum_\ell \left(R_{i\ell} R_{j\ell} m_\ell\right)\sum_{t\, s }
\sqrt{m_t m_s} R_{it}R_{js}
{\rm Im}\left( U_{l s}U_{l t}^*\right)\, .
\label{QDRH1}
\end{eqnarray}
Notice, in particular,
that $I^l_{ij}=-I^l_{ji}$ and $S_{ij}=-S_{ji}$. 
The CP asymmetry $\epsilon_i^l$ is resonantly enhanced when
$\Gamma_j=\Delta M^2_{ij}/M_i$. At the resonance 

\begin{equation}
S_{ij} \simeq  \frac{M_i}{2\,\Gamma_j}\simeq \frac{M_j}{2\,\Gamma_j}\, ,
\,\,\,
\epsilon_{ij}^l \simeq  -\frac{1}{2} \sum_{j\neq i}\,I_{ij}^l\, .
\end{equation}
The  washing out of a given flavour $l$ 
is now operated by the $\Delta L=1$ scatterings involving all  three
RH neutrinos. Therefore, the parameter $\widetilde{m_{l}}$ is given by

\begin{equation}
\widetilde{m_{l}}\simeq 
\sum_{j}\frac{\left|\lambda_{jl}\right|^2 v}{M_j}\simeq
\frac{\left(\lambda^\dagger\lambda\right)_{ll} v}{M_1}
=\sum_{\ell}\, m_\ell\, \left|U_{l\ell}\right|^2\, ,
\end{equation} 
where we have set the nearly equal masses of the RH neutrinos approximately 
equal to
$M_1$.
If the resonance operates for all three RH neutrinos, the CP asymmetry
in the flavour $l$ is 

\begin{equation}
\label{sum}
\epsilon_l =2\,\sum_{i<j}\epsilon^l_{ij}=-\,\sum_{i<j}\, 
I^l_{ij}=2
\frac{\sum_{i<j}\sum_{\ell t s } R_{i\ell} R_{j\ell} R_{it}\,R_{js}
m_\ell \sqrt{m_s m_t}
{\rm Im}\left(U_{l s}U_{l t}^*\right)}{\left(\sum_p \, m_p\, 
R^2_{i p}\right)\left(\sum_q \, m_q\, 
R^2_{j q}\right)}\, .
\end{equation}
It does  not depend upon the mass of the RH neutrinos
(if the running of the parameters is neglected). 
The elements of the matrix $R$ can be 
parametrized by introducing a real antisymmetric 
matrix $A$ \cite{PPY03}
\begin{equation}
R\equiv e^A={\bf 1}+\frac{1-\cos r}{r^2}\, A^2+
\frac{\sin r}{r}\,A\,,~~r=\sqrt{A_{12}^2+A_{23}^2+A_{13}^2}\, ,
\end{equation}
%
where ${\bf 1}$ is the $3\times 3$ unity matrix and $A_{ij}=-A_{ji}$ are the
elements of the matrix $A$. In the limiting case where the RH neutrinos are
exactly degenerate, one can perform an orthogonal rotation on the
RH neutrino states which leaves the mass matrix of the RH neutrinos
proportional to the unity matrix and  defines a physically equivalent
reparametrization  of the RH neutrinos. This amounts to saying that
for $M_1=M_2=M_3$, the real matrix $R$ can be set equal to the
unity matrix (or $A_{12}=A_{23}=A_{13}=0$). The flavour
asymmetries in Eq.~(\ref{QDRH1}) vanish if $R={\bf 1}$ which reflects the fact
that no baryon asymmetry can be obtained in the exactly degenerate case. 
On the other hand, if the
degeneracy is slightly broken, the elements of the matrix $A$ are expected to
be tiny, but not all vanishing and the baryon asymmetry is typically 
different from zero.
We are now ready to study all three possible cases for the spectrum
of the light neutrinos. We will restrict ourselves to the case in which
the resonant condition
is not satisfied for all couples of  RH neutrinos and work under the condition
that $M_1\simeq M_2\lsim M_3$; 
the flavour asymmetries 
are generated resonantly  only by  the decays of  $N_1$ and $N_2$ and we will
set $I_{13}\simeq I_{23}\simeq 0$.

We find
\begin{eqnarray}
\epsilon_l &\simeq & -\sum_{I<j}\left[R_{i3} R_{j3}
\left(\Delta m^2_{\rm A}\right)^{1/2}-
R_{i1} R_{j1}\left(\Delta m^2_\odot\right)^{1/2}\right]\nonumber\\
&\times & \left[
\left(R_{i3}R_{j2}-R_{i2}R_{j3}\right)
\left(\Delta m^2_\odot\right)^{1/4}\left(\Delta m^2_{\rm A}\right)^{1/4}
{\rm Im} \left(U_{l 2}
U_{l 3}^*\right)\right]\nonumber\\
&\times &\left[\left(\Delta m^2_\odot\right)^{1/2}
R_{i2}^2+\left(\Delta m^2_{\rm A}\right)^{1/2}
R_{i3}^2\right]^{-1}\nonumber\\
&\times &\left[\left(\Delta m^2_\odot\right)^{1/2}
R_{j2}^2+\left(\Delta m^2_{\rm A}\right)^{1/2}
R_{j3}^2\right]^{-1}\, ,\nonumber\\
\widetilde{m_l}&\simeq & 
\left(\Delta m^2_\odot\right)^{1/2}\left|U_{l 2}\right|^2
+\left(\Delta m^2_{\rm A}\right)^{1/2} \left|U_{l 3}\right|^2\, 
\end{eqnarray}
in the normal hierarchical case;  

\begin{eqnarray}
\epsilon_l &\simeq & -\sum_{i<j}\left[-R_{i3} R_{j3}
\left(\Delta m^2_{\rm A}\right)^{1/2}-
R_{i1} R_{j1}\frac{\left(\Delta m^2_\odot\right)}{2\left(
\Delta m^2_{\rm A}\right)^{1/2}}\right]\nonumber\\
&\times & \left[\left(R_{i 2}R_{j1}-R_{i1}R_{j2}\right)
\left(\Delta m^2_{\rm A}\right)^{1/2} {\rm Im }\left(U_{l 1}
U_{l 2}^*\right)\right]\nonumber\\
&\times &\left[\left(\Delta m^2_{\rm A}\right)^{1/2} 
R_{i1}^2+\left(\Delta m^2_{\rm A}\right)^{1/2}
R_{i2}^2\right]^{-1}\nonumber\\
&\times &\left[\left(\Delta m^2_{\rm A}\right)^{1/2} 
R_{j1}^2+\left(\Delta m^2_{\rm A}\right)^{1/2}
R_{j2}^2\right]^{-1}\, ,\nonumber\\
\widetilde{m_l}&=& \left(\Delta m^2_{\rm A}\right)^{1/2}
 \left|U_{l 1}\right|^2+
\left(\Delta m^2_{\rm A}\right)^{1/2}\left|U_{l 2}\right|^2\, 
\end{eqnarray}
in the inverted hierarchical case and, finally, 

\begin{eqnarray}
\epsilon_l &\simeq & -\frac{1}{2\,m^2}
\sum_{i<j}\frac{1}{2\,m^2}\left[-R_{i3} R_{j3}
\left(\Delta m^2_{\rm A}\right)-
R_{i1} R_{j1}\left(\Delta m^2_\odot\right)\right]\nonumber\\
&\times & \left[\left(R_{i 2}R_{j1}-R_{i1}R_{j2}\right)
 {\rm Im }\left(U_{l 1}
U_{l 2}^*\right)\right.\nonumber\\
&+&
\left(R_{i 3}R_{j2}-R_{i2}R_{j3}\right)
{\rm Im} \left(U_{l 2}
U_{l 3}^*\right)\nonumber\\
&+&
\left.\left(R_{i 3}R_{j1}-R_{i1}R_{j3}\right){\rm Im} \left(U_{l 1}
U_{l 3}^*\right)\right]\, ,\nonumber\\
\widetilde{m_l}&\simeq & m\, 
\end{eqnarray}
in the degenerate case. In this latter case, 
since the washing-out factors
are   approximately same, the expressions
for the baryon asymmetries may be simplified if $R$ is real.
Indeed, the total asymmetry $\epsilon_1$ vanishes
and,  if $(10^9\lsim M_1\lsim 10^{12})$ GeV, 
the flavour asymmetry $\epsilon_2=-\epsilon_\tau$ while
$\widetilde{m_2}\simeq 2m\simeq 2\widetilde{m_\tau}$. 
One finds

\begin{equation}
Y_B \simeq  - \frac{12}{37} \frac{222}{417}\,Y_\tau \,.
\end{equation}
In Fig. \ref{RHdegeneratelightQDBYmnu.eps}, 
we show the correlation of the baryon 
asymmetry with the effective Majorana mass 
in neutrinoless double beta  decay for the case of quasi-degenerate
RH neutrinos and QD spectrum of light neutrinos.  
A number of projects aim to reach a sensitivity to $\left|\langle
m_\nu\rangle\right|\sim\left(0.01-0.05\right)$ eV \cite{aa} and can certainly
probe the region of values of $\left|\langle
m_\nu\rangle\right|$ for successfull  baryon asymmetry 
from the PMNS phases only. In particular, a direct information
on the Majorana phase $\alpha_{21}$ may come from the  measurement
of $\langle
m_\nu\rangle$, $m$, and $\sin^2 2\theta_{12}$,
\begin{equation}
\sin^2\frac{\alpha_{21}}{2}\simeq \left(1-\frac{\left|\langle
m_\nu\rangle\right|^2}{m^2}\right)\frac{1}{\sin^2 2\theta_{12}}\, , 
\end{equation}
and might tell us if enough baryon asymmetry may be generated uniquely from
the PMNS Majorana phases.

%
\section{Extension to the MSSM}
%

\noindent
The extension to our findings to the supersymmetric version of the SM, the
so-called Minimal Supersymmetric Standard Model (MSSM) is rather
straightforward \cite{antusch}. One has to consider the presence of the 
supersymmetric partners of the RH heavy neutrinos, the so-called
sneutrinos $\widetilde{N_i}$ ($i=1,2,3)$, which  also
give a contribution to the flavour asymmetries, and of the 
supersymmetric partners of the lepton doublets, the so-called slepton
doublets.
Since the effects of 
supersymmetry breaking may be safely neglected, the  flavour
CP asymmetries in the MSSM are twice those in the SM and double is also
the possible channels by which a lepton flavour asymmetry
is reproduced. However, the $\Delta L=1$ scatterings washing out the
asymmetries are also doubled and the number of relativistic
degrees of freedom is almost twice the  one for  the SM case. As a result,
introducing new degrees of freedom and interactions
does not appreciably change the flavour asymmetries 
with respect to the values obtained within the SM. 

There are  however  two other and important  differences with respect to the
SM case. First, in the MSSM, the flavour-independent formulae can only 
be applied for temperatures larger 
than $(1+\tan^2 \beta)\times 10^{12}$ GeV, where $\tan\beta$ indicates the
ratio of the vacuum expectation values of the two Higgs fields
of the MSSM. Indeed, the 
squared charged lepton Yukawa couplings in the MSSM are multiplied 
by $(1+\tan^2 \beta)$. Consequently, 
charged $\mu$ and $\tau$ lepton Yukawa couplings are in thermal equilibrium 
for
 $(1+\tan^2 \beta)\times 10^5 \: \mbox{GeV} 
\ll T \ll (1+\tan^2 \beta)\times 10^{9} \: \mbox{GeV}$  
and all flavours in the Boltzmann equations are to be 
treated separately. For 
$(1+\tan^2 \beta)\times 10^9 \: \mbox{GeV} \ll T \ll (1+\tan^2 \beta)
\times 10^{12} \: \mbox{GeV}$, only the $\tau$ Yukawa coupling 
is in equilibrium and only the $\tau$ flavour is treated separately 
in the Boltzmann equations, while the $e$ and $\mu$ flavours 
are indistinguishable.
This implies that the range of the RH (s)neutrino masses where
flavour is relevant in leptogenesis 
 is greater than the one in the SM by the factor
$(1+\tan^2 \beta)$ which is large even for moderate values
of $\tan\beta$. 
As a consequence, the lower bounds given in 
Eqs. (\ref{s13NHD2}) - (\ref{s13NHD3}) change approximately to 
\beq
|\sin\theta_{13}\, \sin\delta|,\sin\theta_{13} \gtap 
0.11\, (1+\tan^2 \beta)^{-1}\,,~~~
\label{s13NHD2SUSY}
\eeq
%
\beq
|J_{\rm CP}| \gtap 2.4\times 10^{-2}\,(1+\tan^2 \beta)^{-1}\,.
\label{s13NHD3SUSY}
\eeq
%
The lower bounds in Eqs. (\ref{s13NHD4}) - (\ref{s13NHD5}) 
and in Eqs. (\ref{s13IHD1}) - (\ref{JCPIHD1})
change in a similar way. 
The shift of the range of the 
heavy Majorana neutrino masses,
in which the lepton flavour effects are significant,
to larger values has important implications also 
if the spectrum of the RH (s)neutrinos 
is hierarchical and the 
light neutrinos possess inverted hierachical (IH) 
spectrum.  This is 
illustrated in Fig. \ref{RHhierarchicalIHsusy.eps}, where we plot
the baryon asymmetry versus the quantity $J_{\rm CP}$ for the
IH spectrum of light neutrinos in the 
supersymmetric case for a given set of parameters. 
Let us recall that for real $R$ matrix elements 
$R_{11}$ and $R_{12}$, it was impossible to obtain 
baryon asymmetry compatible with the observations
in the corresponding non-SUSY case.
The reader should be also 
warned that, for values of $\tan\beta\gsim 30$, 
radiative corrections to the physical 
neutrino parameters should be
accounted for \cite{rad}. 

Secondly, the relation between the baryon asymmetry and the
lepton flavour asymmetries has to be modified to account for the
presence of two Higgs fields. 
Between 
$(1+\tan^2 \beta)\times 10^9$ and $(1+\tan^2 \beta)\times 10^{12}$ GeV, 
the relation is  
\begin{equation}
Y^\mathrm{MSSM}_{B} \simeq  -\frac{10}{31 g_*} \left( 
\hat{\epsilon}_2\eta\left(\frac{541}{761}\widetilde{m_2}\right)
 + \hat{\epsilon}_\tau\eta\left(\frac{494}{761}
\widetilde{m_\tau}\right) \right),
\end{equation}
where the hat superscripts
indicates that the flavour lepton asymmetries are computed including 
leptons and sleptons. Notice that if the spectrum
of RH (s)neutrinos is quasi-degenerate as well as that  of the light
neutrinos, the wash-out factor
are also the approximately same and the expressions
for the baryon asymmetries may be simpliflied if $R$ is real.
Indeed, the total asymmetry $\epsilon_1$ vanishes
and one of the flavour asymmetries may be expressed in terms of the
others. Under these circumstances, one finds
\begin{equation}
Y^\mathrm{MSSM}_B  \simeq -\frac{10}{31}\frac{447}{988} 
Y^\mathrm{MSSM}_{\tau}\, .
\end{equation}
%

 Finally,  in the case of supersymmetric leptogenesis one should
also face the problem arising from the so-called gravitino bound. The latter
is posed by the possible overproduction of gravitinos during the reheating
stage after inflation. Being only gravitationally coupled to 
the SM particles, gravitinos
may decay very late jeopardising the successfull predictions of  
Big Bang nucleosynthesis.
This does not happen, however, if gravitinos are not efficiently generated 
during reheating, that is if the  reheating temperature $T_{RH}$
 is bounded from above, $T_{RH}\lsim 10^{10}$ GeV \cite{grav}. 
The severe bound on the reheating temperature makes the generation of the
RH neutrinos problematic (for a complete study in the one-flavour case
see  \cite{lept}), if the latter are a few times 
heavier than the reheating temperature, rendering the 
thermal leptogenesis scenario   difficult. There are, though, two possible ways out to this problem. First,
leptogenesis might occur in a non-thermal way, that is the RH neutrinos
might be generated not through thermal scatterings, 
but by other mechanisms, e.g.
at the  preheating stage \cite{preh}. 
Alternatively, and maybe more interestingly, as we have previously 
seen if the two lightest RH neutrinos are quasi-degenerate in mass, 
the final baryon asymmetry does not depend upon their common mass. 
The latter, therefore,  might be smaller than the 
largest possible reheating temperature   and thermal leptogenesis might 
take place without any limitation from the gravitino bound.
%
\section{Conclusions}
%
%
\noindent
In this paper we have systematically investigated the 
connection between the leptogenesis and the low 
energy CP-violation
in the lepton (neutrino) sector. 
Our study was stimulated by
the recent progress 
in the understanding of the 
importance of lepton flavour 
effects in leptogenesis.
It lead to the realization that 
these effects can play 
a crucial role in the leptogenesis scenario, 
both from the quantitative and 
the qualitative point of view. 
When the lepton flavour effects 
are taken into account, 
the final baryon asymmetry is the sum of 
three different contributions 
given by the CP asymmetries 
generated in each flavour (lepton charge), 
properly weighted by the corresponding
wash-out  factor.  
In the one-flavour approximation, which 
holds only if leptogenesis is taking place 
at a temperature higher than about $10^{12}$ GeV,  
the final baryon asymmetry is proportional 
to the total baryon CP asymmetry (summed over the 
three flavours) and weighted by a single 
wash-out factor (obtained by summing the  wash-out 
factor of the three lepton flavours).

  There are many differences 
between the predictions for the
baryon asymmetry $Y_{\rm B}$ obtained in
the one-flavour approximation 
and in the case when the flavour effects 
are accounted for. The baryon asymmetry $Y_{\rm B}$, 
derived in the one-flavour approximation, for instance,
vanishes if the light neutrinos are degenerate in mass.
Correspondingly, $Y_{\rm B}$ has to be proportional 
to a difference of masses of the light 
neutrinos. In the ``flavour'' case this suppression 
can be absent even when the leptogenesis CP violation 
is due entirely to the low energy phases 
in the PMNS matrix. 
However, the most significant difference is that
in the one flavour approximation there is no direct 
connection between the leptogenesis CP-violating
parameters and the CP-violating parameters - 
Dirac and Majorana phases, 
present in the lepton (neutrino) sector. 
In particular, a possible future observation of 
CP violation in neutrino oscillations
would not automatically imply,
within the ``one-flavour'' leptogenesis scenario,
the existence of a baryon asymmetry.
In the ``flavoured'' treatment of leptogenesis, however, 
this conclusion does not universally  hold and the
observation of CP violation in the lepton (neutrino) 
sector would generically imply a nonvanishing baryon 
asymmetry. Including the effects of lepton flavour, 
therefore, allows to build a new bridge between 
the CP violation in leptogenesis and
the observables depending on the CP-violating 
Dirac and Majorana phases 
in the PMNS neutrino mixing matrix,
such as the CP violating rephasing invariant $J_{\rm CP}$
which controls the magnitude of 
CP-violation effects in neutrino
oscillations, the effective Majorana mass $\meff$
in neutrinoless double beta decay, etc.
The study of such a connection has been the main 
subject of our paper. 

  We have first derived 
the constraints the requirement of
CP-invariance imposes on the neutrino    
Yukawa couplings $\lambda$ and on the elements of 
the complex orthogonal matrix $R$ appearing in the
``orthogonal'' parametrisation of $\lambda$. 
The CP-parities of the light and heavy 
Majorana neutrinos, which take the 
values $\pm i$, play a special role 
in these constraints.
The CP-invariance constraints 
are useful for understanding 
the source of CP violation generating 
the CP asymmetries in the
heavy Majorana neutrino decays.
One example is the case of 
{\it real matrix $R$ and 
specific CP-conserving 
values of the Majorana and Dirac phases 
in the PMNS neutrino mixing matrix}, 
which corresponds to violation
of CP-symmetry at high energy 
in leptogenesis, leading to the 
generation of non-zero baryon asymmetry. 
The indicated constraints 
help to clarify also 
under which conditions 
the leptogenesis CP-asymmetries 
are due entirely to the low energy
CP-violating phases of the PMNS matrix.

 Taking into account the lepton 
flavour effects in leptogenesis,
we have subsequently investigated in detail the
possibility that the CP-violation necessary 
for the generation of the baryon asymmetry of the Universe
is due exclusively to the Dirac and/or Majorana CP-violating 
phases in the PMNS matrix, and thus is directly related
to the low energy CP-violation in the lepton sector
(e.g., in neutrino oscillations, etc.).    
We have derived results for two types of spectrum 
of the heavy RH Majorana neutrinos: 
i) hierarchical, in which the lightest
RH neutrino $N_1$ is much lighter than the other 
two RH neutrinos, and ii) 
quasi-degenerate, in which the two lightest 
RH neutrinos $N_{1,2}$ are almost degenerate in mass
and have masses which are smaller than the mass of the
third one. For each of the two cases, we have presented  
predictions for the baryon asymmetry
for three types of spectra of 
the light Majorana neutrinos:
normal hierarchical (NH),
inverted hierarchical (IH) and
quasi-degenerate (QD).
In all numerical calculations we have 
used the best fit values of the solar and 
atmospheric neutrino oscillation parameters,
$\deltaatm$, $\deltasol$, $\sin^2\theta_{23}$ 
and $\sin^2\theta_{12}$,
given in Section 2.

\begin{figure}[ht]
  \centerline{
  {\includegraphics[width=10truecm,height=7.5cm]{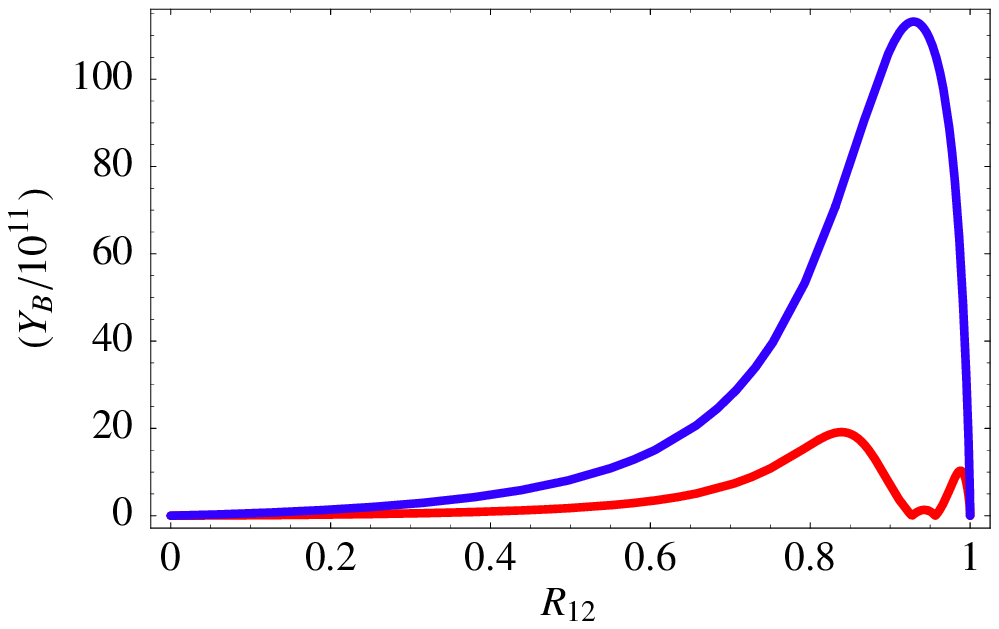}}}
  \caption{ 
The baryon asymmetry $Y_B$ as a function of
$R_{12}$ in the case of real $R_{12}$ and $R_{13}$,
${\rm sign}\left(R_{12}R_{13}\right) = +1$ ($\beta_{23} = 0$),
$R^2_{12} + R^2_{13} = 1$, $s_{13}=0.20$,
hierarchical RH neutrinos and NH light neutrino mass spectrum
and a) Majorana CP-violation (blue line), $\delta=0$ and $\alpha_{32} = \pi/2$
($\kappa = +1$), and
b) Dirac CP-violation (red line), $\delta=\pi/2$ and $\alpha_{32} = 0$ 
($\kappa' = +1$),
for $M_1 = 5\times 10^{11}$ GeV.
The neutrino oscillation parameters
$\deltasol$, $\sin^2\theta_{12}$, $\deltaatm$ and 
$\sin^22\theta_{23}$ are fixed 
at their best fit values.
}
\label{MaxYBDiracMajplus.eps}
\end{figure}

\begin{figure}[ht]
  \centerline{
{\includegraphics[width=10truecm,height=7.5cm]{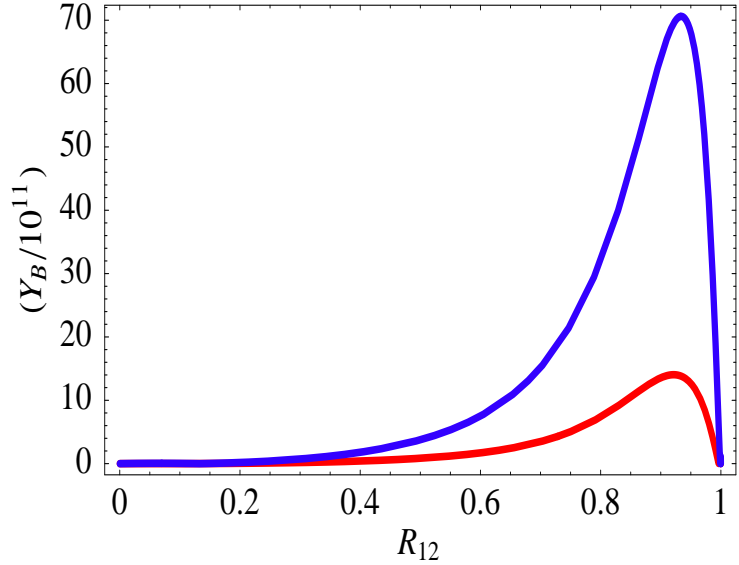}}}
  \caption{ 
The same as in Fig. \ref{MaxYBDiracMajplus.eps} but for
${\rm sign}\left(R_{12}R_{13}\right) = -1$ ($\beta_{23} = \pi$)
and
a) Majorana CP-violation (blue line), 
$\delta=0$ and $\alpha_{32} = \pi/2$ ($\kappa = -1$), and
b) Dirac CP-violation (red line), $\delta=\pi/2$ and $\alpha_{32} = 0$
($\kappa' = -1$).
}
\label{MaxYBDiracMajminus.eps}
\end{figure}

\begin{figure}[ht]
  \centerline{
   {\includegraphics[width=10truecm,height=8cm]{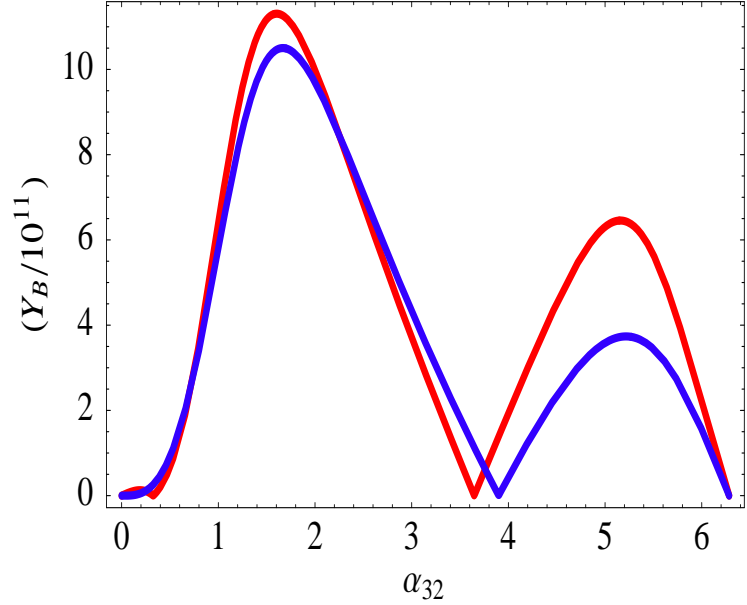}}}
  \caption{ 
The baryon asymmetry as a function of the 
Majorana phase $\alpha_{32}$ varying in the interval 
$\alpha_{32} = [0,2\pi]$ in the case of 
Majorana CP-violation, 
hierarchical RH neutrinos and NH light neutrino mass spectrum,  
for $\delta = 0$, real $R_{12}$ and $R_{13}$,
$|R_{12}| =0.92$, $|R_{13}| =0.39$, 
${\rm sgn}(R_{12}R_{13}) = + 1$
($\beta_{23} = 0$, $\kappa = + 1$),
$M_1 = 5\times 10^{10}$ GeV, 
and two values of  $s_{13}$:
$s_{13}=0$ (blue line) and $0.2$ (red line).
}
\label{Fig4alp02pis130200plus.eps}
\end{figure}

\begin{figure}[ht]
  \centerline{
   {\includegraphics[width=10truecm,height=8cm]{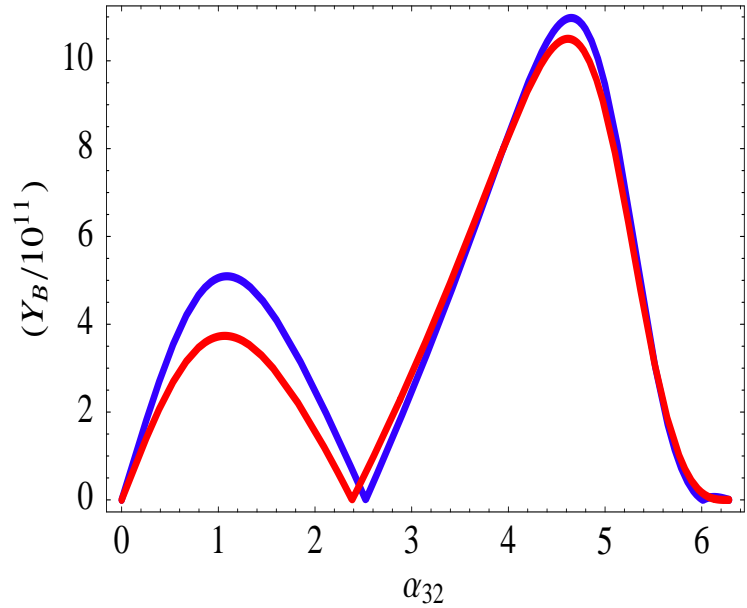}}}
  \caption{ 
The same as in Fig. \ref{Fig4alp02pis130200plus.eps}
but for real $R_{12}$ and $R_{13}$ 
having opposite signes, 
${\rm sgn}(R_{12}R_{13}) = - 1$
($\beta_{23} = \pi$, $\kappa = - 1$),
$|R_{12}| = 0.92$,  $|R_{13}| =0.39$, 
and two values of $s_{13}$: 
$s_{13}=0$ (red line) and $s_{13}=0.1$ (blue line).
}
\label{Fig4balp02piminus.eps}
\end{figure}

\begin{figure}
  \centerline{
 {\includegraphics[width=10truecm,height=8cm]{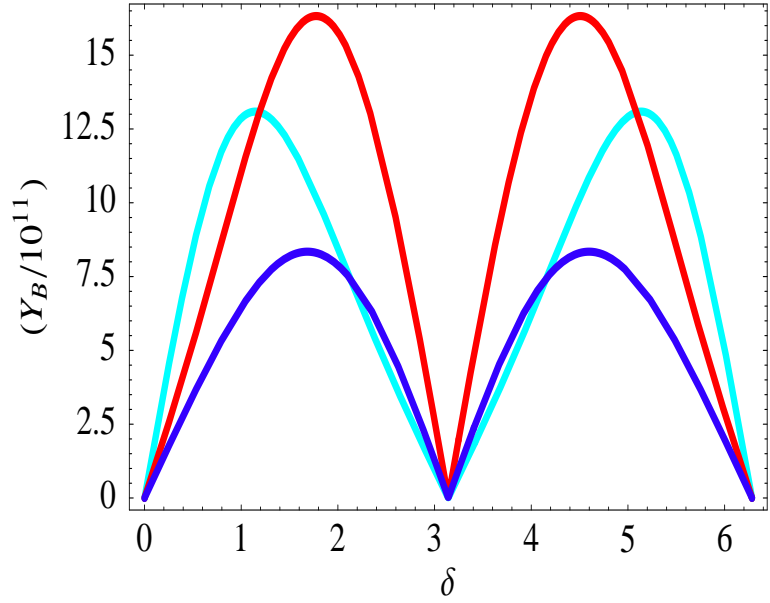}}}
  \caption{ 
  The baryon asymmetry $|Y_B|$ as a 
function of the Dirac phase $\delta$ 
varying in the interval $\delta = [0,2\pi]$ in the case of 
Dirac CP-violation, $\alpha_{32}=0;~2\pi $, 
hierarchical RH neutrinos and 
NH light neutrino mass spectrum, 
for $M_1 = 5\times 10^{11}$ GeV,
real $R_{12}$ and $R_{13}$ satisfying
$|R_{12}|^2 + |R_{13}|^2 = 1$,
$|R_{12}| =0.86$, $|R_{13}| =0.51$, 
${\rm sign}\left(R_{12}R_{13}\right)=+1$, and
for i) $\alpha_{32}=0$ ($\kappa' = + 1$),
$s_{13}=0.2$ (red line) and    
$s_{13}=0.1$ (dark blue line), 
ii) $\alpha_{32}= 2\pi$ ($\kappa' = - 1$),
$s_{13}=0.2$ (light blue line).
}
 \label{Figs678241106.eps}
 \end{figure}

\begin{figure}
  \centerline{
{\includegraphics[width=10truecm,height=8cm]{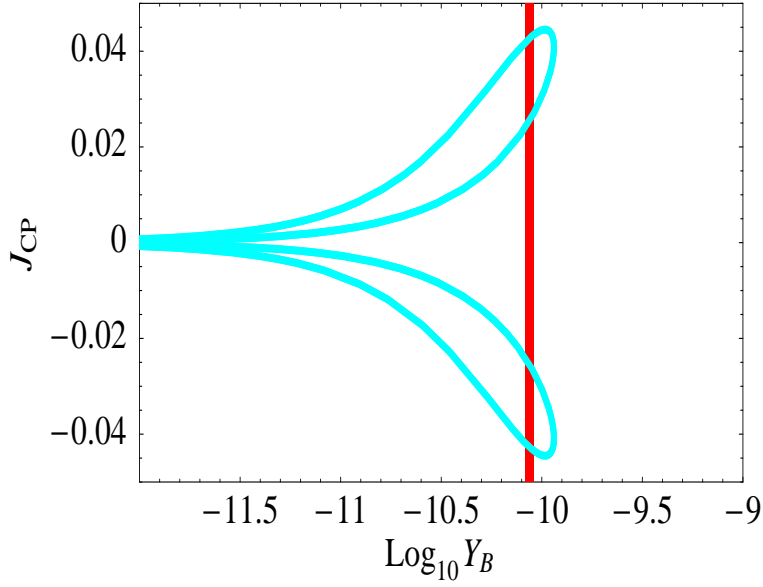}}}
  \caption{ 
The correlation between the 
rephasing invariant $J_{\rm CP}$ (in blue) 
and the baryon asymmetry $Y_{\rm B}$
 when varying the Dirac phase
$\delta = [0,2\pi]$, in  
the case of hierarchical RH neutrinos and 
NH light neutrino mass spectrum and for 
$s_{13}=0.2$, $\alpha_{32}=0~(2\pi)$, $|R_{12}| =0.86$,
$|R_{13}| = 0.51$, 
${\rm sign}\left(R_{12}R_{13}\right)= + 1~(-1)$
($\beta_{23} = 0~(\pi)$, $\kappa' = + 1$), 
$M_1 = 5\times 10^{11}$ GeV . 
The red region denotes the 
$2\sigma$ allowed range of $Y_{\rm B}$.
}
\label{NH2YBJalpha0s1302.eps}
\end{figure}

 For hierarchical RH neutrino mass spectrum, 
the lepton flavour effects are relevant in 
leptogenesis for $M_1\ltap \times 10^{12}$ GeV.
The predicted baryon asymmetry $Y_{\rm B}$ 
depends linearly on $M_1$. In order to reproduce 
the observed baryon asymmetry, 
generically values of 
$M_1\gtap 3\times 10^{10}$ GeV are 
required. We have shown that if the light 
neutrinos have a NH spectrum,  
the requisite baryon asymmetry 
can be produced  
if  the only source of CP violation is 
either the Majorana phases or the Dirac phase
in the PMNS matrix, $U_{\rm PMNS}$ 
(Figs. \ref{MaxYBDiracMajplus.eps} - 
\ref{NH2YBJalpha0s1302.eps}).  
When the only CP-violating parameter is the 
low energy Majorana phase 
$\alpha_{32} \equiv \alpha_{31} - \alpha_{21}$,
we can have successful leptogenesis
as long as $|\sin(\alpha_{32}/2)|$ is 
not exceedingly small 
and $M_1 \gtap 3.5\times 10^{10}$ GeV.
If  the only source of CP violation 
is the Dirac phase $\delta$ in $U_{\rm PMNS}$,
the observed baryon asymmetry can be reproduced
provided $M_1 \gtap 2\times 10^{11}$ GeV
and $|\sin\theta_{13}\sin\delta|\gtap 0.1$,
$\theta_{13}$ being the CHOOZ angle.
This condition leads to the inequality 
$\sin\theta_{13}\gtap 0.1$ and to 
the following lower bound on the CP-violating
rephasing invariant $J_{\rm CP}$,
associated with the Dirac phase in the PMNS matrix:
$|J_{\rm CP}| \gtap 2.0\times 10^{-2}$.
Values of $\sin\theta_{13} \gtap 0.1$  
can be probed in the forthcoming Double 
CHOOZ and Daya Bay reactor neutrino experiments.
CP-violation effects with magnitude determined
by $|J_{\rm CP}| \gtap 2.0\times 10^{-2}$ 
are within the sensitivity of the next generation 
of neutrino oscillation experiments, designed to search 
for CP- or T- symmetry violation in the oscillations.
Moreover, since in this case both 
$|Y_{\rm B}| \propto |\sin\theta_{13}\sin\delta|$
and $|J_{\rm CP}| \propto |\sin\theta_{13}\sin\delta|$,
given the other parameters on which 
$|Y_{\rm B}|$ and $|J_{\rm CP}|$ depend, 
there exists a correlation between the rephasing invariant 
$J_{\rm CP}$, which controls the magnitude 
of the CP-violation effects in neutrino oscillations,
and the baryon asymmetry $Y_B$ 
(Fig. \ref{NH2YBJalpha0s1302.eps}).

\begin{figure}
  \centerline{
   {\includegraphics[width=10truecm,height=8cm]{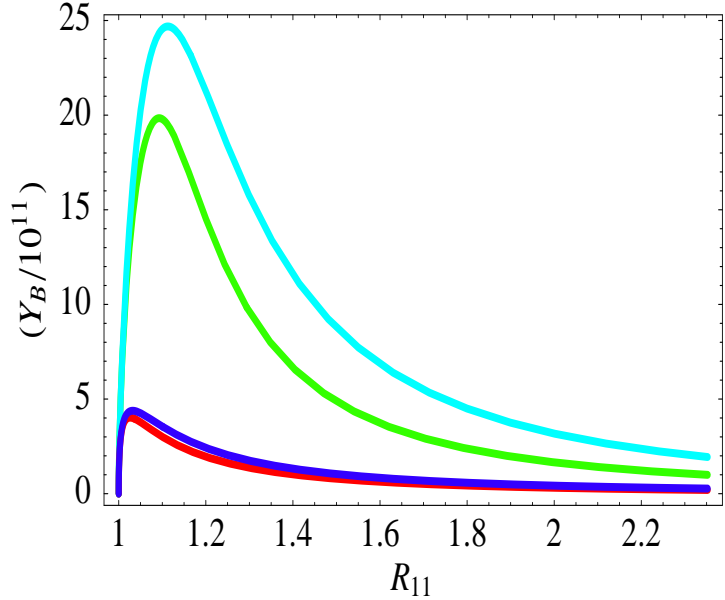}}}
  \caption{ 
The baryon asymmetry $Y_B$ as a function of
$|R_{11}|$ in the case of 
hierarchical RH neutrinos and 
IH light neutrino mass spectrum, 
Majorana CP-violation, $\delta=0$ and $\alpha_{32} = \pi/2$,
$M_1 = 2\times 10^{11}$ GeV,
purely imaginary $R_{11}R_{12} = i\kappa |R_{11}R_{12}|$
and $\kappa = + 1$ (dark blue and red lines),
$\kappa = - 1$ (light blue and green lines),
$|R_{12}|^2 - |R_{13}|^2 = 1$,
and for
$s_{13}=0.2$ (green and red lines) and $s_{13}=0$ (light and dark blue lines).
}
\label{Figs1011241106.eps}
\end{figure}

\begin{figure}
  \centerline{
   {\includegraphics[width=10truecm,height=8cm]{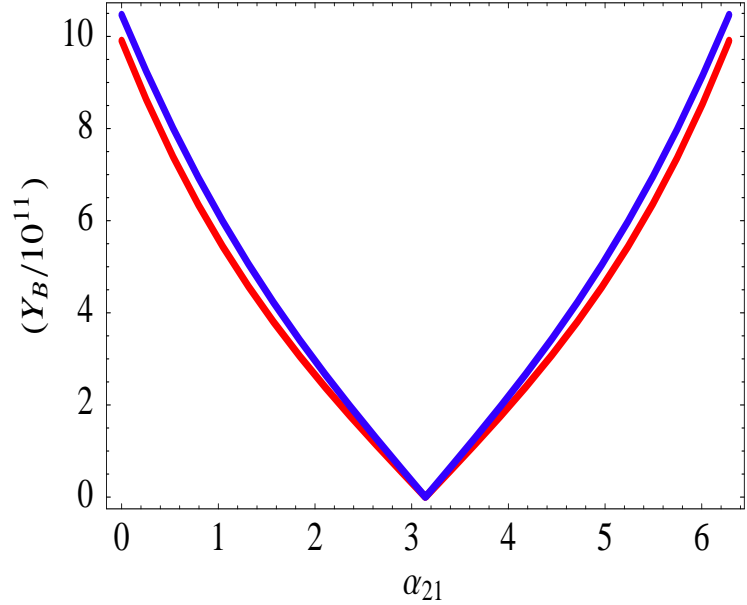}}}
  \caption{ 
The baryon asymmetry as a function of the 
Majorana phase $\alpha_{32}$ varying in the interval 
$\alpha_{32} = [0,2\pi]$ in the case of 
hierarchical RH neutrinos and 
IH light neutrino mass spectrum,  
Majorana CP-violation,
$\delta = 0$, purely imaginary
$R_{11}R_{12} = i\kappa |R_{11}R_{12}|$,
$\kappa = + 1$
($\beta_{12} = \pi/2$), $|R_{11}|^2 - |R_{12}|^2 = 1$,
$|R_{11}| = 1.05$, and for
$M_1 = 2\times 10^{11}$ GeV, 
and two values of  $s_{13}$:
$s_{13}=0$ (blue line) and $0.2$ (red line).
}
\label{IHs130bl02ralpha21plus.eps}
\end{figure}

\begin{figure}
  \centerline{
{\includegraphics[width=10truecm,height=8cm]{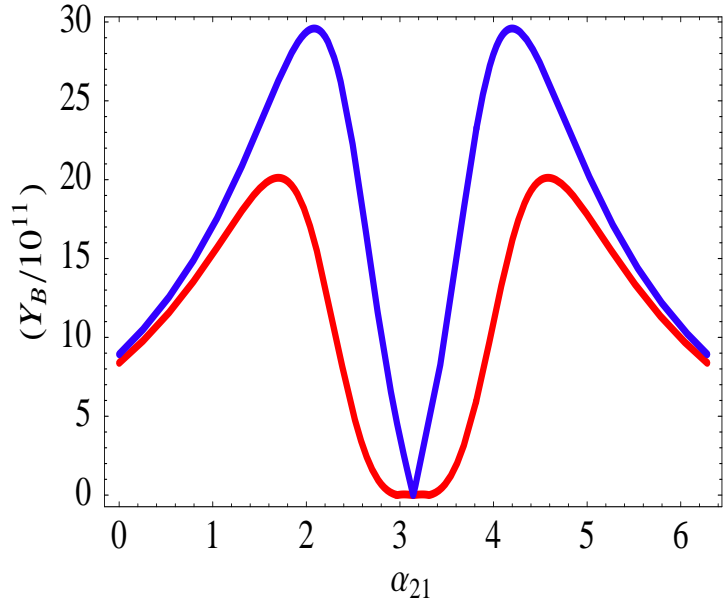}}}
  \caption{ 
The same as in 
Fig. \ref{IHs130bl02ralpha21plus.eps}, but for 
$\kappa = - 1$ ($\beta_{12} = 3\pi/2$) and 
$|R_{11}| = 1.2$.
}
\label{IHs130bl02ralpha21minus.eps}
\end{figure}

\begin{figure}
  \centerline{
{\includegraphics[width=10truecm,height=8cm]{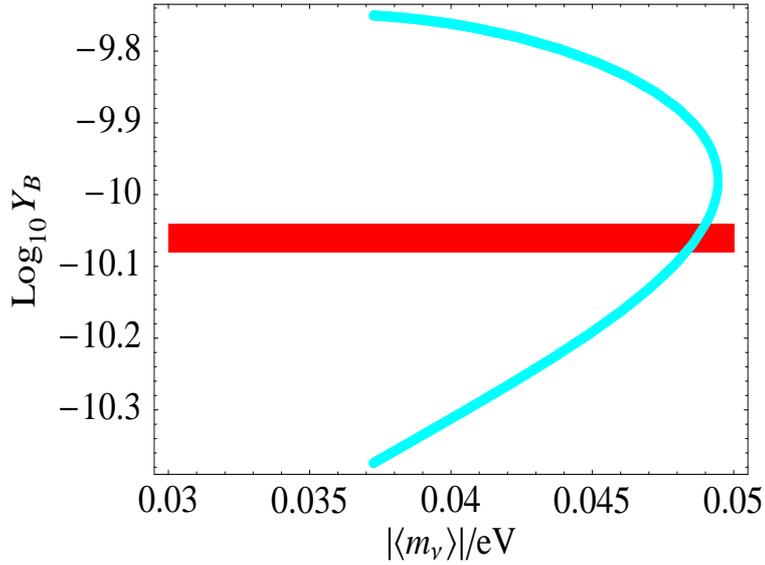}}}
  \caption{ 
The baryon asymmetry $|Y_B|$ versus the effective
Majorana mass in neutrinoless double beta decay,
$\meff$, in the case of Majorana CP-violation, 
hierarchical RH neutrinos and IH light neutrino mass spectrum,  
for $\delta = 0$, $s_{13} = 0$, purely imaginary
$R_{11}R_{12} = i\kappa |R_{11}R_{12}|$,
$\kappa = + 1$
($\beta_{12} = \pi/2$), $|R_{11}|^2 - |R_{12}|^2 = 1$,
$|R_{11}| = 1.05$ and $M_1 = 2\times 10^{11}$ GeV.
The Majorana phase $\alpha_{21}$ is varied in the interval
$[-\pi/2, \pi/2]$. 
}
\label{IHneutrinoless}
\end{figure}

\begin{figure}
  \centerline{
{\includegraphics[width=10truecm,height=8cm]{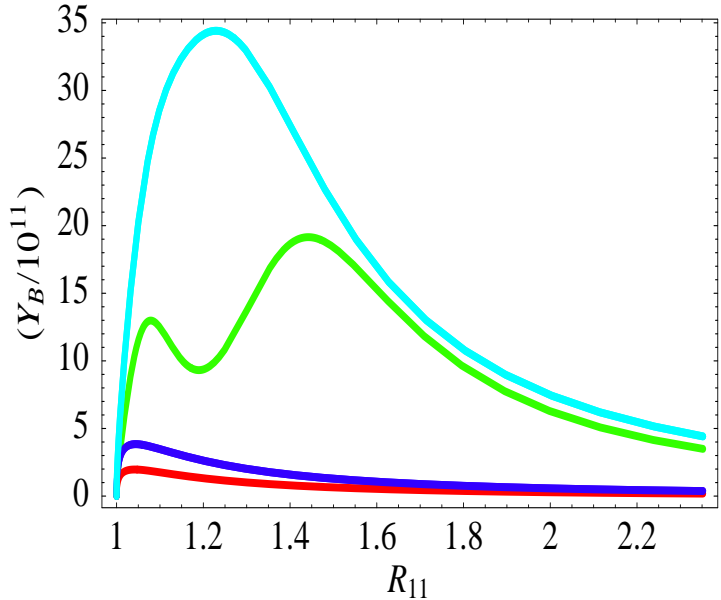}}}
  \caption{ 
The baryon asymmetry $|Y_B|$ as a function of
$|R_{11}|$ in the case of hierarchical RH neutrinos 
and IH light neutrino mass spectrum and 
Dirac CP-violation, $\delta = \pi/2$,
$\alpha_{21} = \pi $ ($\kappa' = + 1$), 
purely imaginary 
$R_{11}R_{12} = i\kappa |R_{11}R_{12}|$
($|R_{12}|^2 - |R_{13}|^2 = 1$),
for $\kappa = - 1$ (red and dark blue lines) and
$\kappa = + 1$ (light blue and green lines),
$s_{13}=0.2$ (light blue and dark blue lines) and
$s_{13}=0.1$ (green and red lines),
and $M_1 = 2\times 10^{11}$ GeV.
}
\label{Fig12IHDalphapi040107R}
\end{figure}

\begin{figure}
  \centerline{
 {\includegraphics[width=10truecm,height=7cm]{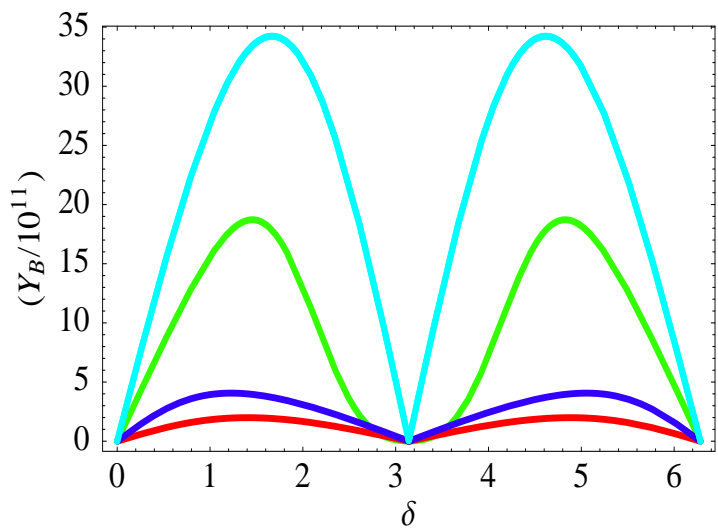}}}
  \caption{ 
The asymmetry $|Y_B|$ as a function 
of the Dirac phase $\delta$ 
in the case of hierarchical RH neutrinos,  
IH light neutrino mass spectrum, 
Dirac CP-violation,  
$\alpha_{21} = \pi$ ($\kappa' = + 1$),
$R_{11}R_{12} = i~\kappa~|R_{11}R_{12}|$ 
($|R_{11}|^2 - |R_{12}|^2 = 1$),
$\kappa = - 1$ (red and dark blue lines),
$\kappa = + 1$ (light blue and green lines), 
for $M_1 = 2\times 10^{11}$ GeV,
and  $s_{13} = 0.1$ (red and green lines)
and $s_{13} = 0.2$ (dark blue and light blue lines). 
Values of $|R_{11}|$, 
which maximise $|Y_B|$ 
have been used: $|R_{11}| = 1.05$ 
in the case of $\kappa = - 1$, 
and $|R_{11}| = 1.3~(1.6)$  
for $\kappa = + 1$ and $s_{13} = 0.2~(0.1)$.
}
\label{Fig13IHDalphapi040107Delt}
\end{figure}

\begin{figure}
  \centerline{
{\includegraphics[width=10truecm,height=8cm]{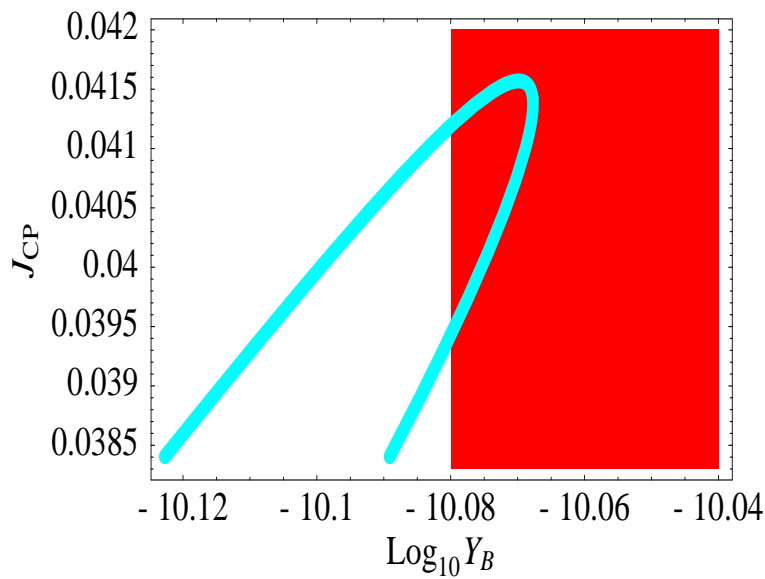}}}
  \caption{ 
The correlation between the 
rephasing invariant $J_{\rm CP}$ (in blue) 
and the asymmetry $Y_{\rm B}$
in the case of hierarchical RH neutrinos, 
IH light neutrino mass spectrum, 
Dirac CP-violation, $\alpha_{21}=\pi$ ($\kappa' = + 1$),
$R_{11}R_{12} = i~\kappa~|R_{11}R_{12}|$ 
($|R_{11}|^2 - |R_{12}|^2 = 1$),
$\kappa = + 1$,
and for $s_{13}=0.2$, 
$M_1 = 5\times 10^{10}$ GeV
and $|R_{11}| = 1.3$. 
The Dirac phase $\delta$ is varied in the interval 
$[0,2\pi]$. The red region denotes the 
$2\sigma$ allowed range of $Y_{\rm B}$.
}
\label{Fig14YBJCPIHDalphapi040107}
\end{figure}

\begin{figure}
  \centerline{
{\includegraphics[width=10truecm,height=8cm]{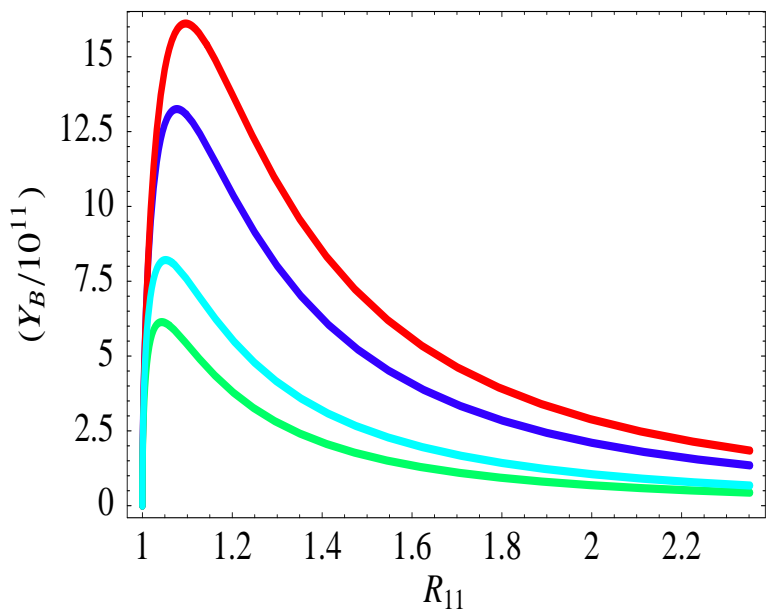}}}
  \caption{ 
The baryon asymmetry $Y_B$ as a function of
$|R_{11}|$ in the case of hierarchical RH neutrinos 
and IH light neutrino mass spectrum, 
$\delta = \pi/2$,
$\alpha_{21} = 0$, and 
purely imaginary $R_{11}R_{12} = i\kappa |R_{11}R_{12}|$
satisfying $|R_{12}|^2 - |R_{13}|^2 = 1$,
for  $\kappa = + 1$ (red and dark blue lines),
$\kappa = - 1$ (light blue and green lines),
$M_1 = 2\times 10^{11}$ GeV
and two values of  $s_{13}$:
$s_{13}=0.2$ (red and light blue lines) 
and $s_{13}=0.1$ (dark blue and green lines).
}
\label{Figs1516IH0KandinskiR11}
\end{figure}

\begin{figure}
  \centerline{
{\includegraphics[width=10truecm,height=8cm]{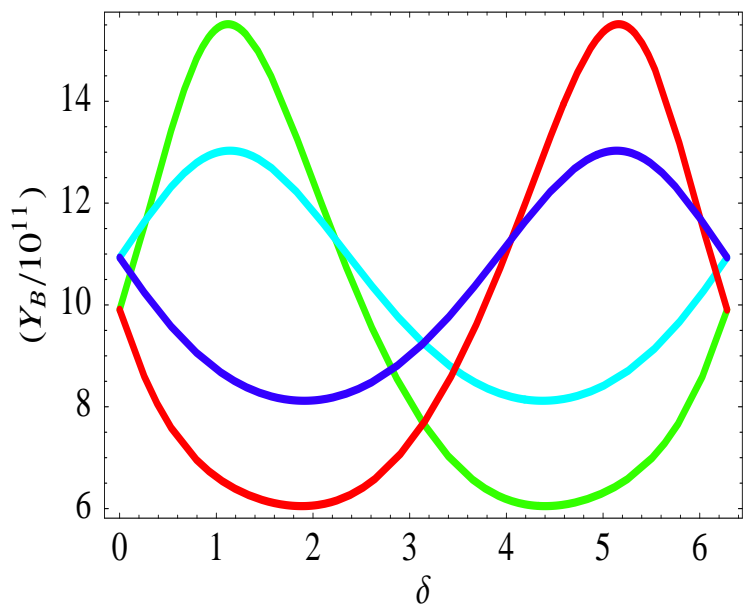}}}
  \caption{ 
The baryon asymmetry as a function of the Dirac phase $\delta$ 
in the case of hierarchical RH neutrinos and 
IH light neutrino mass spectrum, for 
$\alpha_{21} = 0$,
$M_1 = 2\times 10^{11}$ GeV, purely imaginary 
$R_{11}R_{12} = i~\kappa~|R_{11}R_{12}|$ 
($|R_{11}|^2 - |R_{12}|^2 = 1$),
with $\kappa = + 1$ (green and light blue lines )
and $\kappa = - 1$ (red and dark blue lines), 
and for $|R_{11}| = 1.05$
and two values of $s_{13}$:
$s_{13} = 0.1$ (light and dark blue lines),  
$s_{13} = 0.2$ (red and green lines). 
}
\label{Figs1718IH0KandinskiDelta}
\end{figure}

\begin{figure}
  \centerline{
{\includegraphics[width=10truecm,height=8cm]{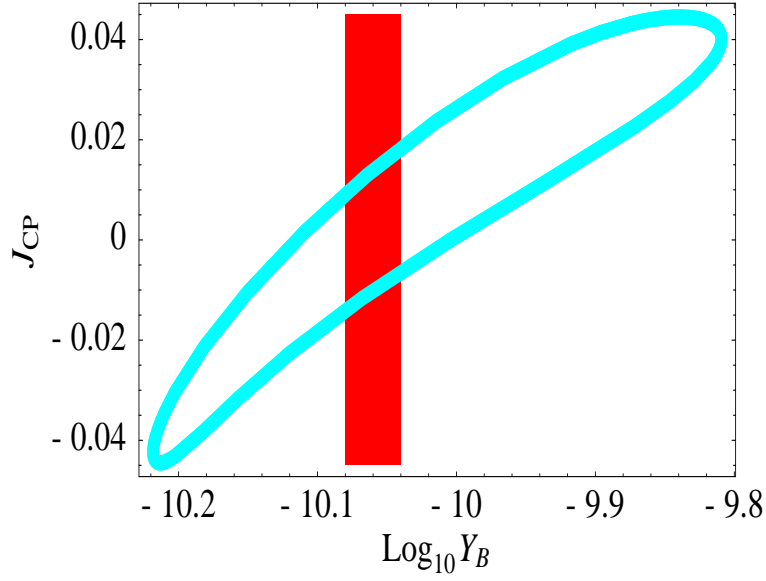}}}
  \caption{ 
The correlation between the 
rephasing invariant $J_{\rm CP}$ (in blue) 
and the baryon asymmetry $Y_{\rm B}$
in the case of hierarchical RH neutrinos and 
IH light neutrino mass spectrum for 
$\alpha_{21} = 0$, $s_{13} = 0.2$,
$M_1 = 2\times 10^{11}$ GeV, 
$R_{11}R_{12} = i~\kappa~|R_{11}R_{12}|$ 
($|R_{11}|^2 - |R_{12}|^2 = 1$),
and for $\kappa = + 1$ and $|R_{11}| = 1.05$.  
The Dirac phase $\delta$ is varied in the interval
$[0,2\pi]$. The red region denotes the 
$2\sigma$ allowed range of $Y_{\rm B}$.
}
\label{Fig17YBJCPIH0DR}
\end{figure}

  In the case of IH light neutrino mass spectrum 
and negligible lightest neutrino mass $m_3$,
the observed baryon asymmetry 
cannot be reproduced if the 
product of the elements of the matrix $R$, 
$R_{11}R_{12}$, is purely real: the generated baryon
asymmetry is generically small,
being suppressed by the additional 
factor $\deltasol/\deltaatm$.
However, if $R_{11}R_{12}$ is purely imaginary 
(and CP-conserving), a sufficiently large 
baryon asymmetry  compatible with the 
observations can be obtained  
both when the only source of CP-violation 
is the Majorana phase $\alpha_{21}$, 
or the Dirac phase $\delta$, in $U_{\rm PMNS}$
(Figs. \ref{Figs1011241106.eps} - 
\ref{Fig14YBJCPIHDalphapi040107}).  
In the case of Majorana CP-violation, 
depending on the ${\rm sgn}({\rm Im}(R_{11}R_{12}))$,
values of $M_1 \gtap 5\times 10^{10}~{\rm GeV}$ 
or somewhat larger (e.g., $M_1 \gtap 
1.6\times 10^{11}~{\rm GeV}$) are required.
Since both the baryon asymmetry  
$|Y_B|$ and the effective Majorana mass in 
$\betabeta$-decay, $\meff$, depend  on 
the Majorana phase $\alpha_{21}$, 
for given values of the other parameters 
there exists a direct correlation between the values of  
$|Y_B|$ and $\meff$ (Fig. \ref{IHneutrinoless}).
We have shown that one can have successful leptogenesis
in the case under discussion also if  
$s_{13} \neq 0$ and the CP-violation is 
generated only by  the Dirac phase 
$\delta$ in $U_{\rm PMNS}$
(Figs. \ref{Fig12IHDalphapi040107R} and 
\ref{Fig13IHDalphapi040107Delt}).
For $M_1 \ltap 5\times 10^{11}$ GeV,
the observed baryon asymmetry 
can be reproduced if
$|\sin\theta_{13}\sin\delta|\gtap 0.02$.
This requirement implies that we should 
have also $\sin\theta_{13}\gtap 0.02$
and $|J_{\rm CP}| \gtap 4.6\times 10^{-3}$.
Values of $\sin\theta_{13}$ and of $|J_{\rm CP}|$
as small as 0.02 and $4.6\times 10^{-3}$, 
respectively, can be probed in neutrino 
oscillation experiments at neutrino factories.
There exists a correlation between 
the rephasing invariant $J_{\rm CP}$ 
and the baryon asymmetry $Y_B$ in this case 
as well (Fig. \ref{Fig14YBJCPIHDalphapi040107}).

  The analysis we have performed showed that
if the light neutrinos have QD spectrum,
the baryon asymmetry is generically too 
small mainly due to the large wash-out 
suppression factor. 

\begin{figure}
  \centerline{
{\includegraphics[width=10truecm,height=8cm]{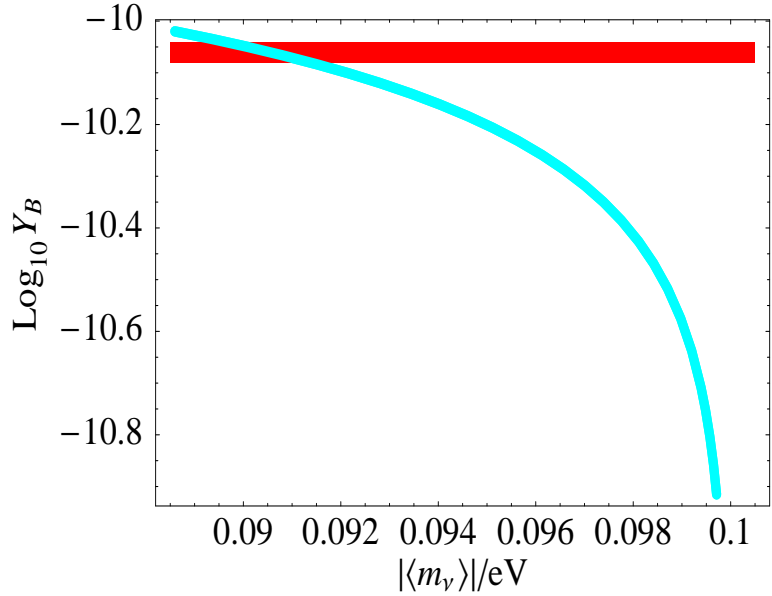}}}
  \caption{ 
The baryon asymmetry versus the $\betabeta$-decay 
effective Majorana mass $\left|\langle m_\nu\rangle \right|$ 
in the case of QD heavy RH neutrinos and QD 
light neutrino mass spectrum, and  for
$\delta=\pi/3$, $s_{13}=0.01$,
$M_1 =   10^{10}$ GeV and $m=0.1$ eV.
The Majorana phase $\alpha_{32}$ is varied 
in the interval $[0,\pi/3]$.
}
\label{RHdegeneratelightQDBYmnu.eps}
\end{figure}

\begin{figure}
  \centerline{
{\includegraphics[width=10truecm,height=8cm]{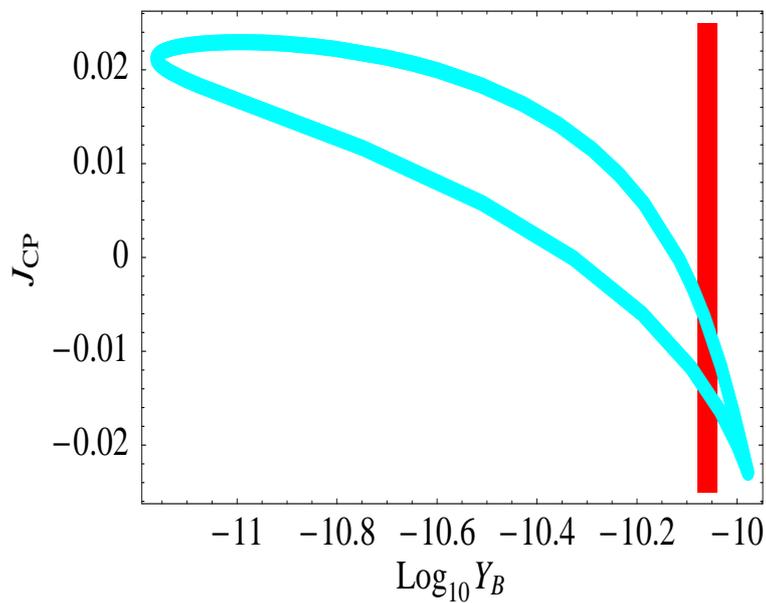}}}
  \caption{ 
  The rephasing invariant $J_{\rm CP}$ versus
the baryon asymmetry in the case of supersymmetric 
hierarchical RH neutrinos and IH light neutrino mass spectrum and 
for $\alpha_{32}=\pi/4$,
$s_{13}=0.1$,  $R_{12} =0.86$,
$R_{13} =0.5$, ${\rm sign}\left(R_{12}R_{13}\right)=+1$ and 
$M_1 =  6\times  10^{12}$ GeV.
The Dirac phase $\delta$ is varied in the interval [0,$2\pi$]. 
}
\label{RHhierarchicalIHsusy.eps}
\end{figure}

 For heavy RH Majorana neutrinos with
QD spectrum, leptogenesis takes place 
through a resonance effect. The main new feature 
is that the final baryon 
asymmetry does not depend 
on the mass of the RH neutrinos. 
This property is crucial, 
allowing the generation of a 
sufficiently large baryon asymmetry even 
in the case of QD light 
neutrino mass spectrum.
In the latter case the predicted 
baryon asymmetry is correlated with the 
effective Majorana mass in the neutrinoless 
double beta decay  
(Fig. \ref{RHdegeneratelightQDBYmnu.eps}).

 Finally, we have discussed 
how the results on leptogenesis we have obtained 
will be modified in the minimal 
supersymmetric extension of the 
Standard Theory (MSSM) with right-handed Majorana 
neutrinos and see-saw mechanism of neutrino mass 
generation. We have noticed, in particular,
that for hierarchical heavy Majorana 
neutrino mass spectrum, 
the range of the lightest RH neutrino mass
$M_1$ for which the 
lepton flavour effects are relevant in leptogenesis 
is greater than the one in the non-supersymmetric case 
by the factor $(1+\tan^2 \beta)$, 
$M_1 \ltap (1+\tan^2 \beta)\times 10^{12}$ GeV,
$\tan\beta$ being the
ratio of the vacuum expectation values 
of the two Higgs fields of the MSSM. 
This can have important implications
especially in the cases when the 
generation of the baryon asymmetry in the
non-supersymmetric case is strongly suppressed.
We have also stressed that a quasi-degenerate spectrum 
of the heavy RH Majorana neutrinos 
is welcome in the case of 
supersymmetric leptogenesis since it
renders the gravitino bound harmless. 

  The results obtained in the present article 
underline the importance of understanding the 
status of the CP-symmetry in the lepton sector 
and, correspondingly, of the 
experiments aiming to measure the CHOOZ angle
$\theta_{13}$ and of the experimental 
searches for Dirac and/or Majorana leptonic 
CP-violation at low energies.

{\bf Acknowledgements.}
It is a pleasure to thank P. Di Bari for useful discussions.
This work was supported in part by the Italian MIUR and INFN
programs on ``Fisica Astroparticellare'' (S.T.P.) 
and by the the European Network of Theoretical 
Astroparticle Physics  ILIAS/N6 under the contract RII3-CT-2004-506222
(S.P. and S.T.P.). S.P. would like to thank the PH-TH Unit 
at CERN where this work was started.

\newpage

\end{document}